\documentclass[prd,aps,superscriptaddress,showpacs,preprintnumbers,nobibnotes,nofootinbib,amsmath,amssymb,floatfix,letterpaper,twocolumn,tighten]{revtex4}

\usepackage{dcolumn} 
\usepackage{graphicx}
\usepackage{epsfig}
\usepackage{color}

\def\parens#1{\left( #1 \right)}
\def\bracks#1{\left[ #1 \right]}

\def\ev#1{\langle #1 \rangle}

\def\ket#1{| #1\rangle}

\def\matel#1#2#3{\langle#1|#2|#3\rangle}

\newcommand{\<}{\langle}
\renewcommand{\>}{\rangle}

\def\del{\nabla}

\newcommand{\id}{\hbox{1$\!\!$1}}

\def\req#1{Eq.~(\ref{#1})}
\def\reqn#1{Equation~(\ref{#1})}
\def\reqs#1#2{Eqs.~(\ref{#1}) and~(\ref{#2})}

\def\reft#1{Table~\ref{#1}}
\def\refts#1#2{Tables~\ref{#1} and~\ref{#2}}

\def\refig#1{Fig.~\ref{#1}}
\def\refigure#1{Figure~\ref{#1}}
\def\refigs#1#2{Figs.~\ref{#1} and \ref{#2}}
\def\refigsdash#1#2{Figs.~\ref{#1}-\ref{#2}}

\def\rsec#1{Section~\ref{#1}}

\def\ben{\begin{equation}}
\def\een{\end{equation}}

\def\beq{\begin{equation}}
\def\eeq{\end{equation}}

\def\benn{\begin{equation*}}
\def\eenn{\end{equation*}}

\def\beqa{\begin{eqnarray}}
\def\eeqa{\end{eqnarray}}
\def\nn{\nonumber}

\def\bea{\begin{eqnarray*}}
\def\eea{\end{eqnarray*}}

\def\bitem{\begin{itemize}}
\def\eitem{\end{itemize}}

\def\benum{\begin{enumerate}}
\def\eenum{\end{enumerate}}

\def\btabu{\begin{tabular}}
\def\etabu{\end{tabular}}

\def\brtabu{\begin{ruledtabular}}
\def\ertabu{\end{ruledtabular}}

\def\btab{\begin{table}}
\def\etab{\end{table}}

\def\btabstar{\begin{table*}}
\def\etabstar{\end{table*}}

\def\bfig{\begin{figure}}
\def\efig{\end{figure}}

\def\barr{\begin{array}}
\def\earr{\end{array}}

\def\bverb{\begin{verbatim}}
\def\everb{\end{verbatim}}

\def\Tr#1{\mathrm{Tr}\left(#1\right)}

\def\Tra{\mathrm{Tr}}
\def\Tr{\mathrm{Tr}\,}

\def\defined{\equiv}

\def\fr#1#2{\,\frac{#1}{#2}\,}
\def\v{\vec}

\def\sr{\sqrt}
\def\x{\times}

\def\approx{\simeq}

\def\set#1{\{#1\}}

\def\cO{\mathcal{O}}

\def\cA{\mathcal{A}}
\def\cB{\mathcal{B}}
\def\cC{\mathcal{C}}
\def\cO{\mathcal{O}}

\def\muhat{\hat{\mu}}

\def\a{\alpha}
\def\b{\beta}
\def\d{\delta}

\def\g{\gamma}
\def\G{\Gamma}
\def\e{\epsilon}

\def\ph{\varphi}

\def\r{\rho}

\def\c{\chi}
\def\y{\psi}

\def\ybar{\bar{\psi}}

\def\MeV{\mathrm{MeV}}
\def\GeV{\mathrm{GeV}}

\def\bo#1{{\bf #1}}

\newcommand{\rr}[1]{\let\temp=\\\raggedright#1\let\\=\temp}

\begin{document}
\bibliographystyle{apsrev}
\preprint{BU-HEP 05-11, CPT-2005/P.045}

\renewcommand{\thefootnote}{\fnssymbol{footnote}}

\title{
Light hadron and diquark spectroscopy in quenched QCD \\with overlap quarks 
on a large lattice
}

\author{R.~Babich}
\affiliation{Department of Physics, Boston University, Boston, MA}
\author{F.~Berruto}
\affiliation{Brookhaven National Laboratory, Upton, NY}
\author{N.~Garron}
\affiliation{DESY, Platanenallee 6, 15738 Zeuthen, Germany}
\author{C.~Hoelbling}
\affiliation{Department of Physics, Bergische Universit\"at Wuppertal, Germany}
\author{J.~Howard}
\affiliation{Department of Physics, Boston University, Boston, MA}
\author{L.~Lellouch}
\affiliation{Centre de Physique Th\'eorique\footnote[1]{UMR 6207 du CNRS et des universit\'es 
d'Aix-Marseille I, II et du Sud Toulon-Var, affili\'ee \`a la FRUMAM.}, Marseille, France}
\author{C.~Rebbi}
\affiliation{Department of Physics, Boston University, Boston, MA}
\author{N.~Shoresh}
\affiliation{Harvard University, Cambridge, MA}

\date{September 12, 2005}

\renewcommand{\thefootnote}{\arabic{footnote}}

\begin{center}
\begin{abstract}
A simulation of quenched QCD with the overlap Dirac operator has been
completed using 100 Wilson gauge configurations at $\b=6$ on an $18^3
\x 64$ lattice and at $\b=5.85$ on a $14^3 \x 48$ lattice, both in
Landau gauge. We present results 
for light meson and baryon masses, meson final state ``wave functions,'' 
and other observables, as well as some details on the numerical
techniques that were used. Our results indicate that scaling
violations, if any, are small. We also present an analysis of diquark correlations 
using the quark propagators generated in our simulation.

\end{abstract}
\end{center}


\maketitle

\section{Introduction}
\label{sec-Intro}
Preserving chiral symmetry presented a notorious difficulty for
the early formulations of lattice Quantum Chromodynamics (QCD).
The problem stemmed from the fact that an ultralocal discretization
of the Dirac equation must either abandon full chiral
symmetry or introduce extra fermionic degrees of freedom
(doublers), with associated no-go theorems for lattice 
fermions~\cite{Karsten:1980wd,Nielsen:1980rz}.
It was only about a decade ago that seminal work by Kaplan,
Shamir, Neuberger, Narayanan, and others provided a way out of the impasse,
through the introduction of domain-wall \cite{Kaplan:1992bt,Shamir:1993zy} and
overlap \cite{Narayanan:1994gw,Narayanan:1993sk,Neuberger:1997fp}
formulations of lattice fermions. These two closely related discretizations of the Dirac equation
avoid the no-go theorems by forfeiting ultralocality, while
retaining locality,
and satisfy an identity, the Ginsparg-Wilson relation \cite{Ginsparg:1981bj},
which allows one to extend to the lattice the chiral symmetry
of continuum QCD~\cite{Luscher:1998pq}.  Unfortunately, the satisfactory properties
of the new lattice formulations come with a heavy computational cost,
since they entail either extension of the lattice in a fifth
dimension or very demanding large matrix manipulations.  It is therefore
important to subject these novel discretizations to the test of
lattice QCD simulations on systems of realistically large size,
in order to explore the adequacy of the necessary numerical techniques
and to validate the good properties that are expected to follow
from the preservation of chiral symmetry.  In this paper, we present
the results of such an investigation, where we used the overlap
Dirac operator to simulate quenched QCD on two lattices, of size
$18^3 \times 64$ and $14^3 \times 48$.

One early simulation of quenched QCD with overlap quarks was performed in
\cite{Giusti:2001pk,Giusti:2001yw}.  Pioneering investigations of QCD 
with the
overlap fermion discretization were also presented in
\cite{Edwards:1998wx,Hernandez:1999cu,DeGrand:2001ie,Liu:2000dw,
Hernandez:2001yn,Hernandez:2001hq,Chiu:2002eh,
Garron:2003cb,Mathur:2003zf,Chen:2003im,Chiu:2003iw,DeGrand:2003in}.  In the work
of \cite{Giusti:2001pk,Giusti:2001yw}, quenched QCD was simulated
with the Wilson gauge action at $\beta=6$ on a lattice of size $16^3 \times 32$.
An expansion into fractions, after projection of a small number
of low lying eigenvectors, was used to calculate the overlap Dirac
operator.  It was shown that the available computational methods
could produce accurate results for the quark propagators in
a reasonable amount of time with moderately powerful computer
resources (capable of 10 to 20 Gflops sustained).
Satisfactory results were obtained on the pseudoscalar spectrum,
strange quark mass, quark condensate, renormalization constants,
and a few other
observables, and the good chiral behavior of the theory was verified.
It also became apparent, however, that the extent of the lattice
in the temporal direction was too small for a meaningful calculation
of other parameters of the hadron spectrum, such as vector meson
and baryon masses, as well as for a reliable estimate of important
matrix elements.  We decided therefore to extend the scope of the
investigation by considering a bigger lattice, with a time extent
twice as large.  The spatial extent was also slightly increased,
with the final choice of lattice size, $18^3 \times 64$, motivated
by a careful assessment of the computer resources we could rely upon.
After completing the calculation of the quark propagators on the
$18^3 \times 64$ lattice, we also simulated
a system of size $14^3 \times 48$, with correspondingly coarser
lattice spacing, in order to check scaling.

For the calculations described in this paper we simulated
quenched QCD with Wilson gauge action at $\beta=6$ on the
$18^3 \times 64$ lattice and $\beta=5.85$ on the $14^3 \times 48$
lattice.  The corresponding values of the lattice spacing are
$a^{-1} = 2.12\,{\rm GeV}$ and $a^{-1} = 1.61\,{\rm GeV}$, on the basis of the Sommer scale
defined by $r_0^2F(r_0)=1.65$, $r_0=0.5\,{\rm
fm}$~\cite{Guagnelli:1998ud,Necco:2001xg}. (The determination
of $a^{-1}$ from the Sommer scale 
in quenched QCD is quite accurate and we can consider the 
corresponding statistical error negligible with respect to the 
other statistical errors in this work.)
Thus, the two lattices are of approximately the same physical volume.
For both lattices, we generated 100 gauge configurations
using the multihit Metropolis algorithm, with 11,000 lattice
upgrades for equilibration and 10,000 lattice upgrades between
subsequent configurations.  For each configuration,
we performed a gauge transformation to Landau gauge.
We then calculated
the quark propagators with source at the origin and all 12
source color-spin combinations for bare quark mass
\beq
am\in\{0.03,0.04,0.06,0.08,0.1,0.25,0.5,0.75\}
\eeq
on the finer $18^3 \times 64$ lattice, and 
\beq
am\in\{0.03,0.04,0.053,0.08,0.106,0.132,0.33,0.66,0.99\}
\eeq
on the coarser $14^3 \times 48$ lattice.  The calculation of the quark
propagators was done with a conjugate-gradient multimass
solver.  Technical details of our calculation, as well as
relevant formulae, are presented in Appendix~\ref{sec-Overlap}.
We should mention here, however, some computational
considerations which informed the choices we had to make for our
investigation.

The limitations of the quenched approximation are well-known, and forefront
calculations today tend to incorporate dynamical quarks.
A simulation with dynamical overlap quarks on a lattice
similar to the one we studied would have required, however,
computational resources at least two orders of magnitude
larger than we had available.  Moreover, an efficient calculation
of quark propagators with the overlap Dirac operator is
a necessary prerequisite for overlap simulations with dynamical quarks.
We concluded that a quenched calculation would be the best
option at present, in that it would allow us to consider
a reasonably large lattice and separate the computational
problems associated with the calculation of the overlap
operator from those of the dynamical fermion feedback.
An alternative would have been to perform mixed action
calculations, i.e.~to use overlap valence quarks
in the background of gauge configurations generated
with some other type of dynamical quarks, for example those
made available by the MILC collaboration (see http://qcd.nersc.gov).
While that might have been computationally feasible and may be a good
strategy for future calculations,
we decided not to proceed in that manner at this stage
in order to avoid the entanglement of two different sets of
computational effects as well as for a more technical reason,
on which we shall now elaborate.

The overlap Dirac operator for a massless quark is given by
\beq
D=\fr{\r}{a}\parens{1+\gamma_5 H(\r)\fr{1}{\sr{H(\r)^2}}}\;,
\label{ovop}
\eeq
where $H(\r)$ stands for the Hermitian Wilson-Dirac operator
with mass $-\r /a$, namely
\beqa
H(\r)
&=&\gamma_5 D_W(\rho) \nn\\
&=& \fr{\g_5}{2} \sum_\mu [
\g_\mu(\del_\mu-\del_\mu^\dag)+ a\del_\mu^\dag \del_\mu]
- \fr{\r \g_5 }{a}
\eeqa
with the forward lattice covariant derivative implicitly defined by
\beq
\del_\mu\y(x) = \fr{1}{a}[U_\mu(x)\y(x+a\muhat)-\y(x)]\;.
\eeq
In terms of $D$, the overlap Dirac operator for a quark of
mass $m$ is then given by $[1-(am)/(2\r)]D+m$. (See
Appendix~\ref{sec-Overlap} for more details on the overlap Dirac operator.)

In the calculation of the quark propagators, which is
based on a conjugate-gradient iterative procedure, at each step one must
apply the overlap operator to the current iterate.
\reqn{ovop} shows that this requires evaluating
the action of $(H^2)^{-1/2}$ on a quark field, for which in turn
one must use some suitable approximation to the inverse square
root.  Implementing such an
approximation, with the required degree of numerical accuracy,
becomes more and more demanding the larger the condition
number of $H^2$, i.e.~the larger the value of the ratio between its largest and smallest
eigenvalues.  Since the largest eigenvalue of $H^2$ is bounded,
in practice the condition number of the matrix depends on its
lowest eigenvalue, or, equivalently, on the gap in the eigenvalues
of $H$ around 0.  This gap in turn depends on the parameter
$\rho$ (which must be carefully chosen) as well as, loosely
speaking, on the ``amount of disorder'' of the gauge configuration.
This became quite apparent in our study, where the calculation of
the quark propagators on the smaller $14^3 \times 48$ lattice
turned out to be computationally more challenging than
the calculation on the $18^3 \times 64$ lattice, because of the increased ``roughness'' of the
background gauge configurations at the smaller value of $\beta$.
A similar dependence is encountered in the domain-wall formulation,
where the so called ``residual mass,'' which measures the
deviation from chiral symmetry induced by the truncation
in the fifth dimension, is seen to depend on the smoothness
of the gauge configuration \cite{Blum:2000kn}.
Given the fact that the gap in the eigenvalues of $H$ can be
substantially affected by the choice of
action and by the presence or absence of dynamical fermions,
we thought that it would be useful to establish a benchmark by 
performing a quenched simulation with the most traditional
action, namely the Wilson gauge action.  We thought that
demonstrating that the overlap formulation is amenable to
precise calculations on large lattices in this context would
be an important step toward the use of more elaborate gauge
actions and/or mixed action calculations.

Another choice we had to make concerned the type of
quark sources to use.  Calculations of the lowest masses
in the hadron spectrum are made more precise by the use
of extended sources.  However, we also wanted to take advantage of
our quark propagators for the study of non-perturbative
renormalization and the evaluation of selected matrix elements,
both of which required point-like sources.  Since the
high cost of evaluating overlap quark propagators made it impossible
for us to use both source terms, we had to adopt point-like sources.
Of course, this did not prevent us from using extended sink operators,
and indeed some of the results we present
here have been obtained with extended sinks.  Finally, the non-perturbative
renormalization techniques required that the gauge field be
brought to Landau gauge prior to evaluating the propagators.
We have implemented this gauge fixing for all our gauge configurations
(of course, after they have been generated),
an additional advantage being that this allowed us to calculate
quark-antiquark correlation functions in the final state as well
as diquark propagators.

Novel results with respect to the calculations presented in
\cite{Giusti:2001pk,Giusti:2001yw} have been obtained for
the vector meson spectrum and decay constants, for the baryon
spectrum, for correlations among quarks and antiquarks in the final
state (which we will refer to as final state ``wave functions''),
scaling, quark and diquark propagators, non-perturbative renormalization,
and meson matrix elements of selected four-quark operators.
In this paper, we present our results for the meson spectrum and
decay constants, meson final state wave functions, the baryon spectrum,
scaling from $\beta=5.85$ to $\beta=6$, and quark and diquark
propagators.
We will present our results for renormalization constants and matrix
elements in a companion paper.  We hope to present a more detailed
analysis of heavy quark states, as well as a study of diquark
correlations within baryons, in future publications.

In \rsec{sec-LightMesObs}, we present our results for light meson observables on the
finer $18^3 \x 64$ lattice.  
We follow in \rsec{sec-Scaling} with our results for the coarser 
$14^3 \x 48$ lattice, as well as a scaling comparison between the two
lattices.  As discussed in Appendix~\ref{sec-Overlap}, we chose $\rho=1.4$ at $\beta=6.0$ and 
$\rho=1.6$ at $\beta=5.85$ \cite{Hernandez:1998et,Hernandez:1999cu} so as to optimize the
locality properties of our overlap operator.  That choice induces a
lattice spacing dependence which would have to be parameterized in
order to allow for a continuum extrapolation.  Since our goal is only
an estimate of possible discretization errors, we do not pursue such an approach here.
In \rsec{sec-LightBaryons}, we present our results for the baryon
spectrum on the $18^3 \x 64$ lattice as well as scaling between the two
lattices. Finally, in \rsec{sec-DiquarkCorr}, we present our
analysis of diquark correlation functions and diquark spectra. In the
Appendices, we give background information on the overlap Dirac
operator, present details of our simulation techniques,
and present data tables for meson and baryon spectra and other observables.

\section{Light Meson Observables on the $18^3 \times 64$ Lattice}
\label{sec-LightMesObs}
\subsection{Correlators}

We evaluated meson correlators for point source, point sink
operators, as well as point source, extended sink operators.
As explained in the introduction, computational limitations did not 
allow us to calculate correlators with extended sources.

We first consider correlators with point sink operators.
We only consider connected diagrams, so without loss of generality
we take quark and antiquark of different flavors $f_1$ and $f_2$.  
The zero-momentum meson correlator is given by
\beq
G_{AB}(t) = \ev{\,\sum_{\v{x}} \Tra\left[S^{f_2}(0;\v{x},t)\,
\G_A\g_5\,(S^{f_1}(0;\v{x},t))^\dagger\,\g_5\G_B\right]}\;,
\eeq
where $S(0;\v{x},t)$ is the Euclidean quark propagator 
and $\G_A$ and $\G_B$ are the Dirac $\g$-matrix combinations
associated with the meson states $A$ and $B$.  
An example of such a meson correlator is shown in
\refig{fig-PPcorr}.

\bfig[t]
\includegraphics*[width=8.3cm,clip]{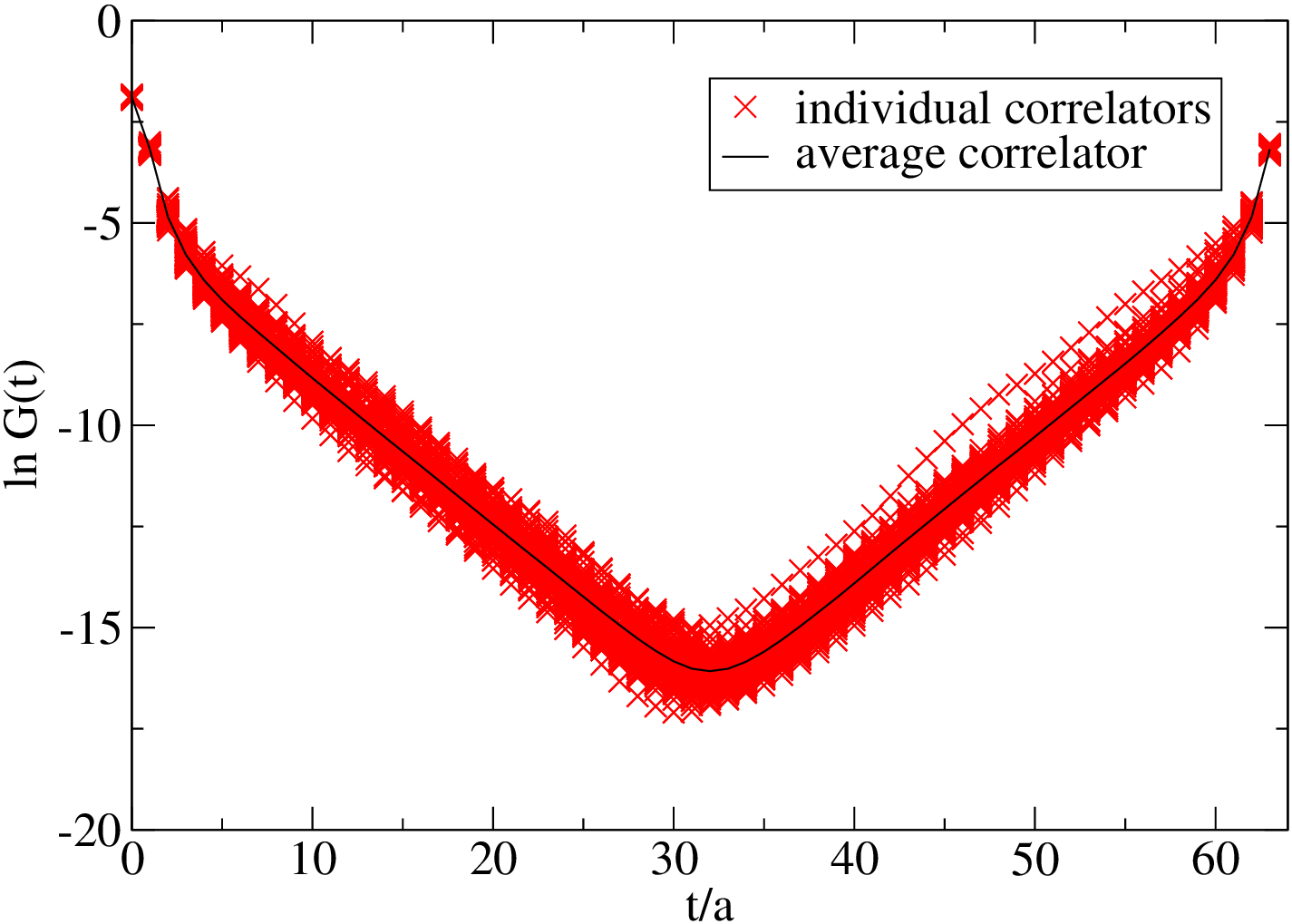}
\caption{Time dependence of a $PP$ correlator, for quark masses
  $am_1=0.08$, $am_2 = 0.10$.}
\label{fig-PPcorr}
\efig

To obtain ground state meson masses, the zero-momentum meson
correlators, which are even under time-reversal,
were fit in an appropriate fitting window to the usual functional form
\beq
G(t) = \fr{Z}{M}\,e^{-MT/2}\cosh\bracks{M\parens{\fr{T}{2}-t}}\;,
\label{eq-mesoncorrff}
\eeq
where $M$ is the meson mass and $T$ is the extent of the lattice in
time, and where we will refer to $Z$ as the correlator matrix element.

For mesons with quarks of non-degenerate mass, we combined the results
obtained by an interchange of the two quark masses.
In order to increase statistics, we folded the meson correlators over the
midpoint $T/2$ and fitted to data below that point.  Concerning
statistical errors, the data that enter into the fit 
(the meson correlators at different values of time) are highly correlated, 
but the amount of data is not sufficient for an evaluation of cross-correlations
that would be precise enough for incorporation into the fitting procedure.  Thus, we
instead used the bootstrap technique.  We found that the error estimates became stable
when the number of bootstrap samples $n_B$ reached a value of approximately
200 and used $n_B=300$ in our calculations.  We used the bootstrap
method for the estimate of the statistical errors in most of the results 
presented in this paper.

Typically, effective mass plots are used to determine the best fitting
window $\{t_{min},t_{max}\}$ for correlators.  Examples of such
effective mass plots are shown in \refig{fig-em_08}.  Using a cosh 
fit for the meson correlators allowed a consistent $t_{max}$ of $T/2$.
In order to pinpoint the best value for $t_{min}$, we considered scans
over different values of $t_{min}$ for a fixed $t_{max}$.  An example
of such data is shown in \refig{fig-tmin0810}.  The value of
$t_{min}$ chosen was the smallest value (consistent with the errors)
before the clear effect of higher states caused the mass prediction to
rise.  Consideration of the $\c^2$ value from the fit also was used to
confirm the choice.  If possible, a single value of $t_{min}$ was
chosen for all quark mass combinations.

\bfig[tp]
\includegraphics*[width=8.3cm,clip]{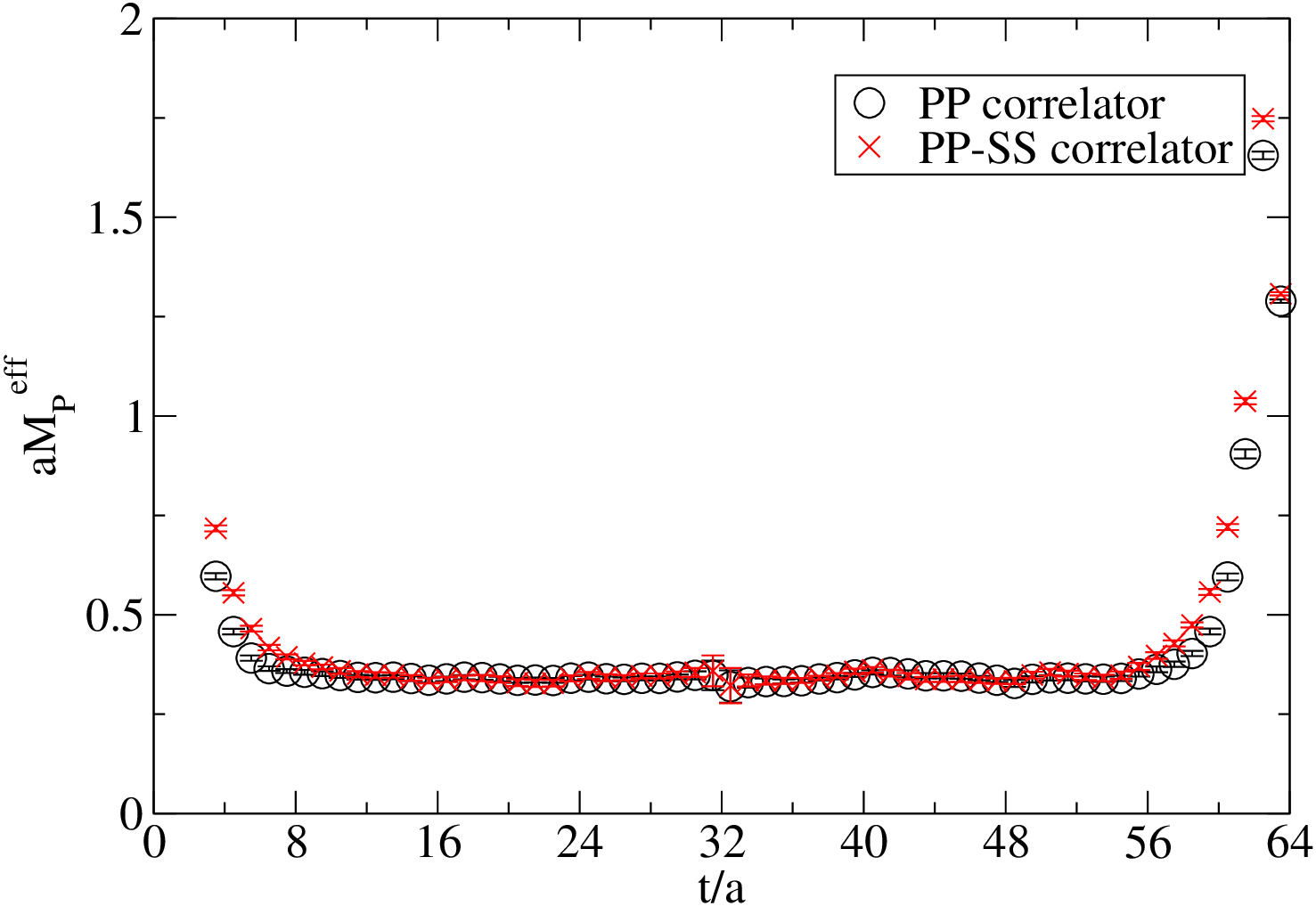}
\caption{\label{fig-em_08}Effective mass plateau for
different pseudoscalar correlators at $am=0.08$.}
\efig

\bfig[tp]
\includegraphics*[width=8.3cm,clip]{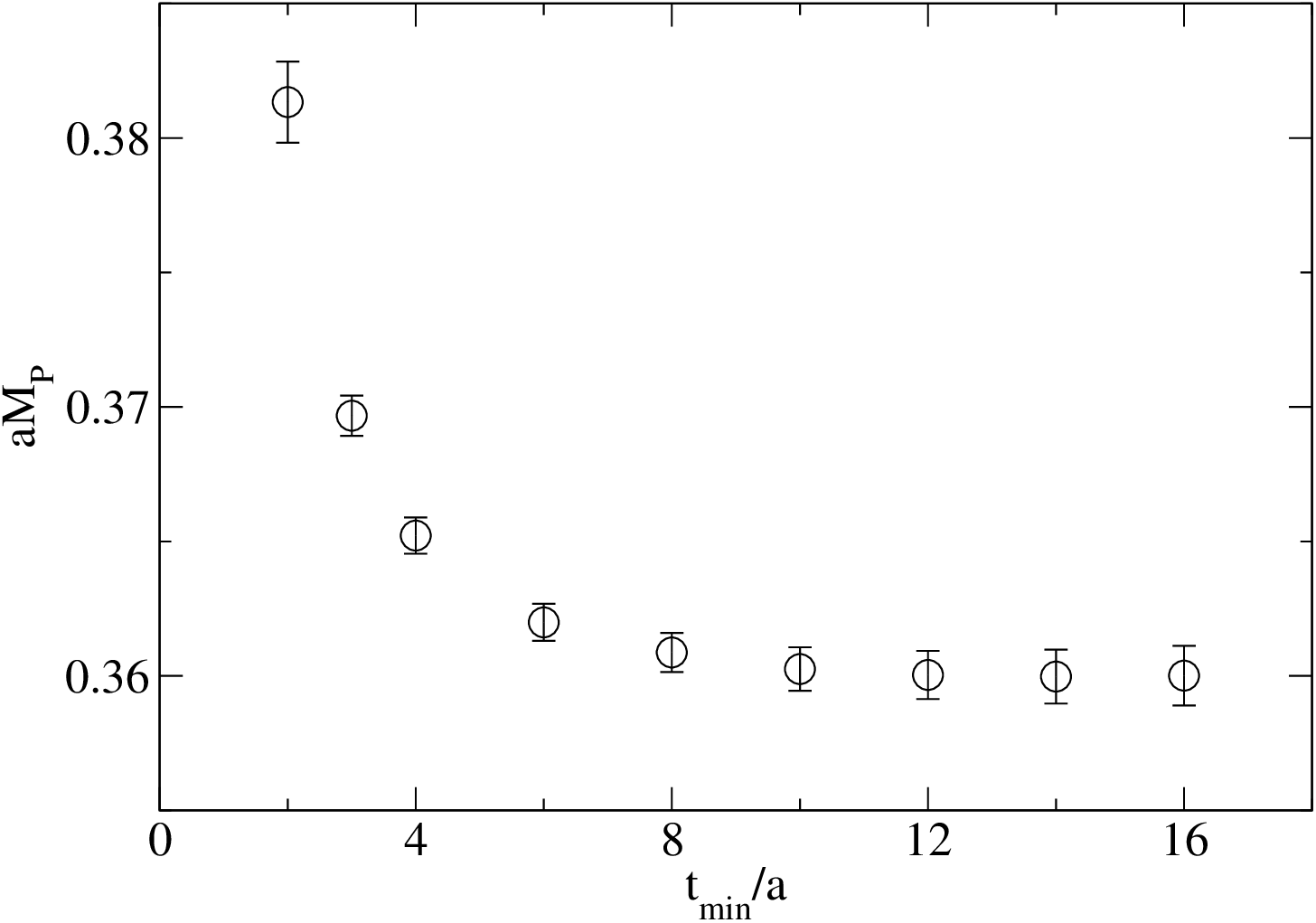}
\caption{Scanning for optimum $t_{min}$ for $PP$ mass, 
with $am_1=0.08$, $am_2 = 0.10$, and $t_{max}=32$.}
\label{fig-tmin0810}
\efig

\subsection{Meson spectra}
\label{subsec-MesonSpectra}

The effective mass data and fitting window scans
in \refigs{fig-em_08}{fig-tmin0810} suggested use of a fit
range $12\le t/a \le 32$ in order to extract the pseudoscalar mass.  For
the vector mass, the fitting range was $8 \le t/a \le 32$.
We illustrate in \refig{fig-PPall} our results for the pseudoscalar 
spectrum for all possible input quark mass combinations.   
In the quenched approximation, the correlator $G_{PP}(t)$ receives
contributions proportional to $1/m^2$ and $1/m$ from
chiral zero modes that are not suppressed by the fermionic
determinant.  These unsuppressed contributions, which should vanish in
the infinite volume limit, could be sizable at finite
volume~\cite{Hernandez:1999cu,Blum:2000kn}.
A method of handling these quenching artifacts is to consider the
difference of pseudoscalar and scalar meson correlators
\beq
G_{PP-SS}(t) = G_{PP}(t) - G_{SS}(t)\;,
\eeq
since, by chirality, the quenching artifacts cancel in the
difference~\cite{Blum:2000kn}.
The results for the pseudoscalar masses obtained with this correlator
are also included in \refig{fig-PPall}.  With our large lattice we 
do not see any significant difference between the results obtained
with $PP$ and $PP-SS$ correlators.  Thus, in the remainder of this paper,
for what concerns the pseudoscalar masses and matrix elements, we will
only consider the $PP$ correlators, except where otherwise noted.
For economy of figures, we also show in \refig{fig-PPall} the results
for correlators with extended sink operators, which will be discussed 
in section~\ref{subsec-ExtSinkOps}.

\bfig[tp]
\includegraphics*[width=8.3cm,clip]{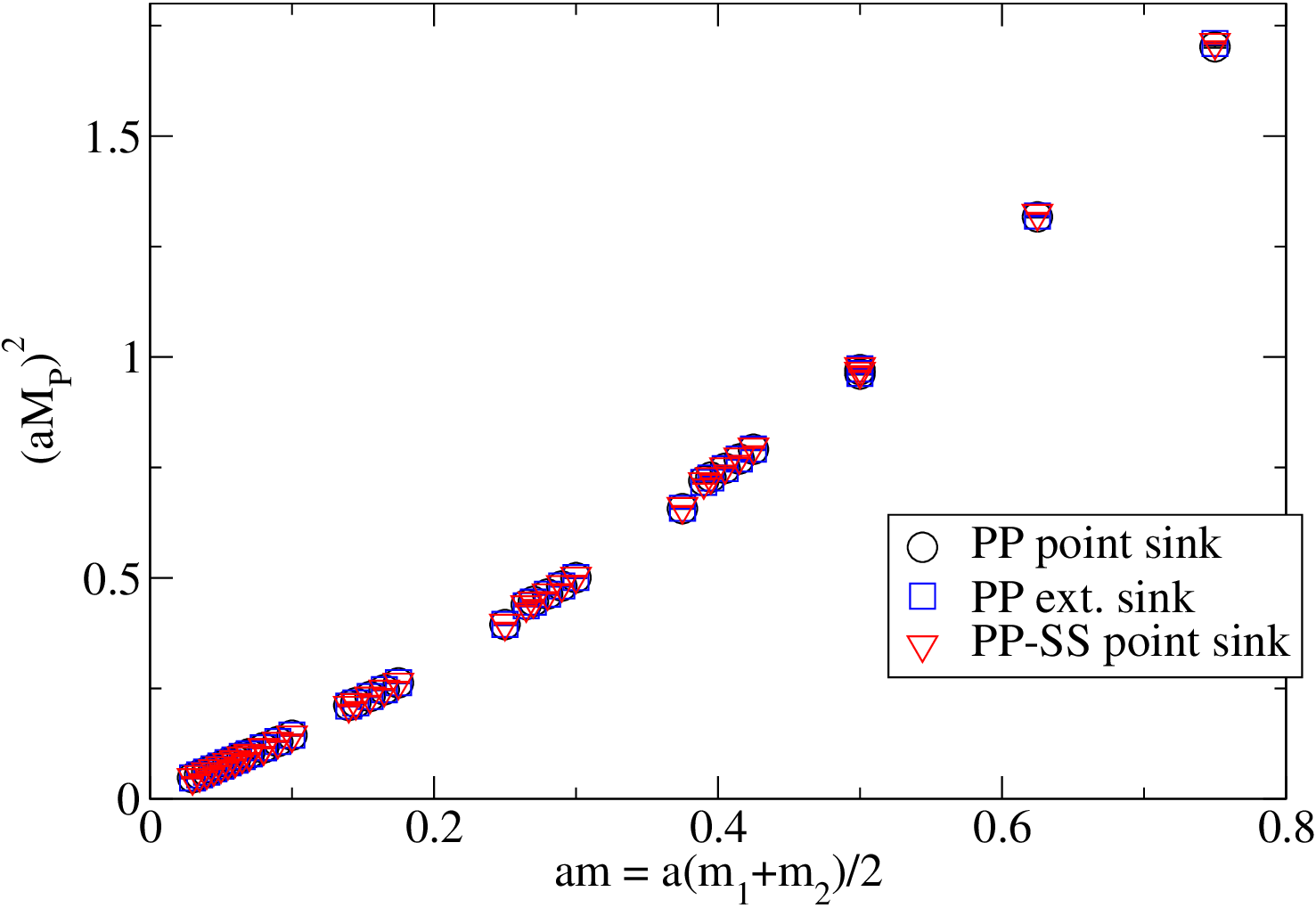}
\caption{Pseudoscalar meson spectrum for point and extended sinks.}
\label{fig-PPall}
\efig

\bfig[tp]
\includegraphics*[width=8.3cm,clip]{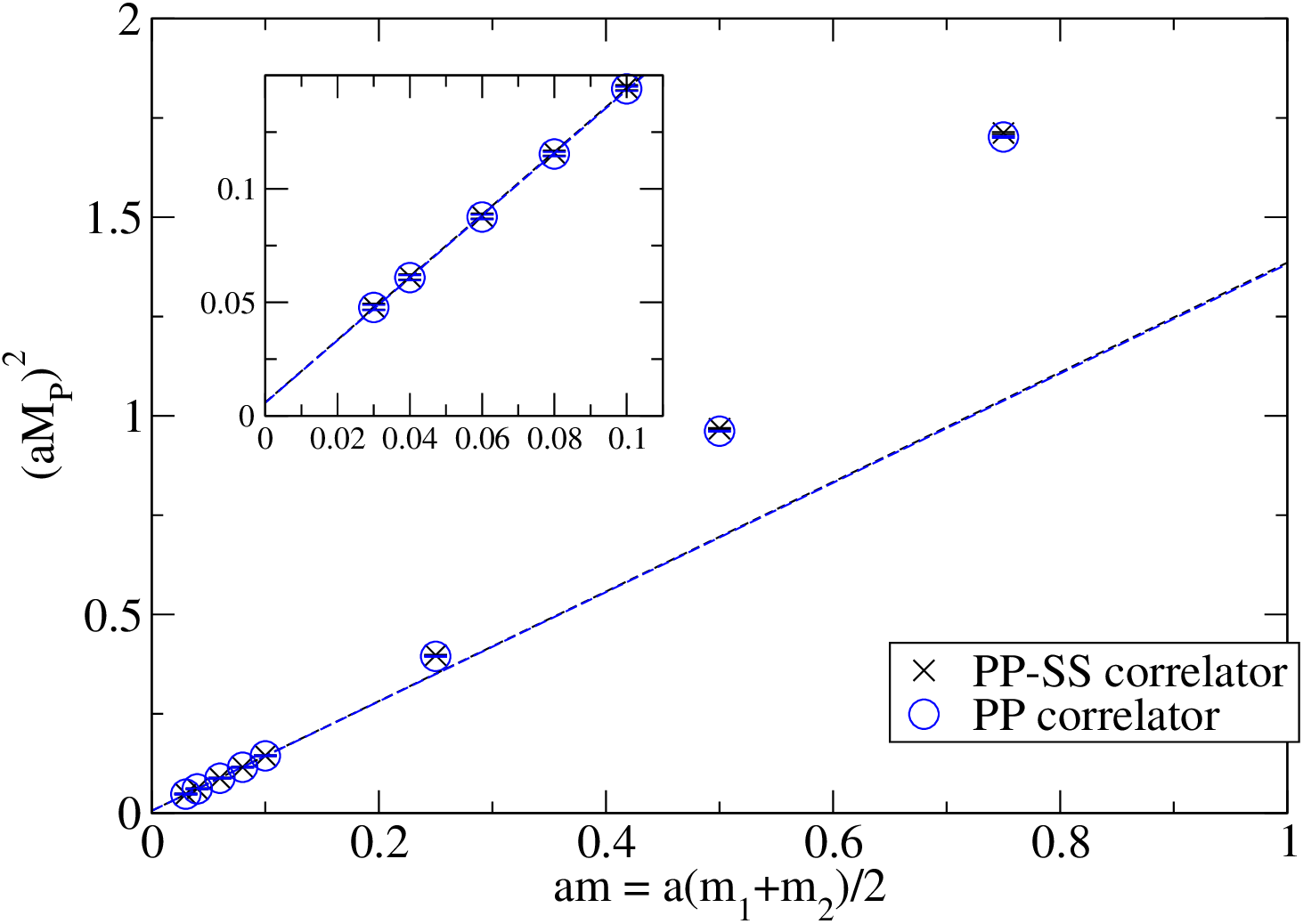}
\caption{Chiral behavior of meson masses.  The solid line
 corresponds to the best fit to \req{eq-mpfit} in the interval
 $am\le 0.1$.}
\label{fig-mqmp2}
\efig

The chiral behavior of the pseudoscalar spectrum with degenerate
quarks is illustrated in \refig{fig-mqmp2} for the $PP$
and $PP-SS$ correlators.  The lightest data point corresponds
roughly to a kaon composed of degenerate quarks of mass $am=0.03\sim
am_s/2$, where $am=a(m_1+m_2)/2$ is the average of the quark and
antiquark masses $am_1$ and $am_2$ in the meson and $am_s$ is the bare strange
quark mass, all in lattice units. We
neglect for the moment the chiral logarithms discussed in
\rsec{subsec-QCL} and perform a fit to the linear form
\beq
\label{eq-mpfit}
(aM_P)^2={\cal A}+{\cal B}(am)\;,
\eeq
where $aM_P$ is the pseudoscalar mass. 
If we fit all data with $am\le 0.1$ we obtain (see \refig{fig-mqmp2})
\beq
\label{eq-fpp}
{\cal A}=0.0058(15)\;,\qquad
{\cal B}=1.376(15)\qquad
\eeq
for the $PP$ correlator and 
\beq
\label{eq-fps}
{\cal A}=0.0059(16)\;,\qquad
{\cal B}=1.380(17)\qquad
\eeq
for the $PP-SS$ correlator.
(The corresponding values in
\cite{Giusti:2001pk} were ${\cal A}=0.006(4)$, ${\cal B}=1.39(3)$ and 
${\cal A}=-0.0005(68)$, ${\cal B}=1.43(7)$ for
the $PP$ and $PP-SS$ correlators, respectively.)
These results exhibit the good chiral 
behavior of the overlap formulation.  The non-zero value of the 
intercept $\cal A$ in \reqs{eq-fpp}{eq-fps}, albeit small, 
is however statistically significant.  Since our results rule out the 
possibility that this is due to zero modes, the small deviation from 
chiral behavior can originate either from finite volume effects 
or chiral logarithms.  In a later section we will argue against finite volume
effects and show that it is indeed compatible with chiral logs.

\bfig[tp]
\includegraphics*[width=8.3cm,clip]{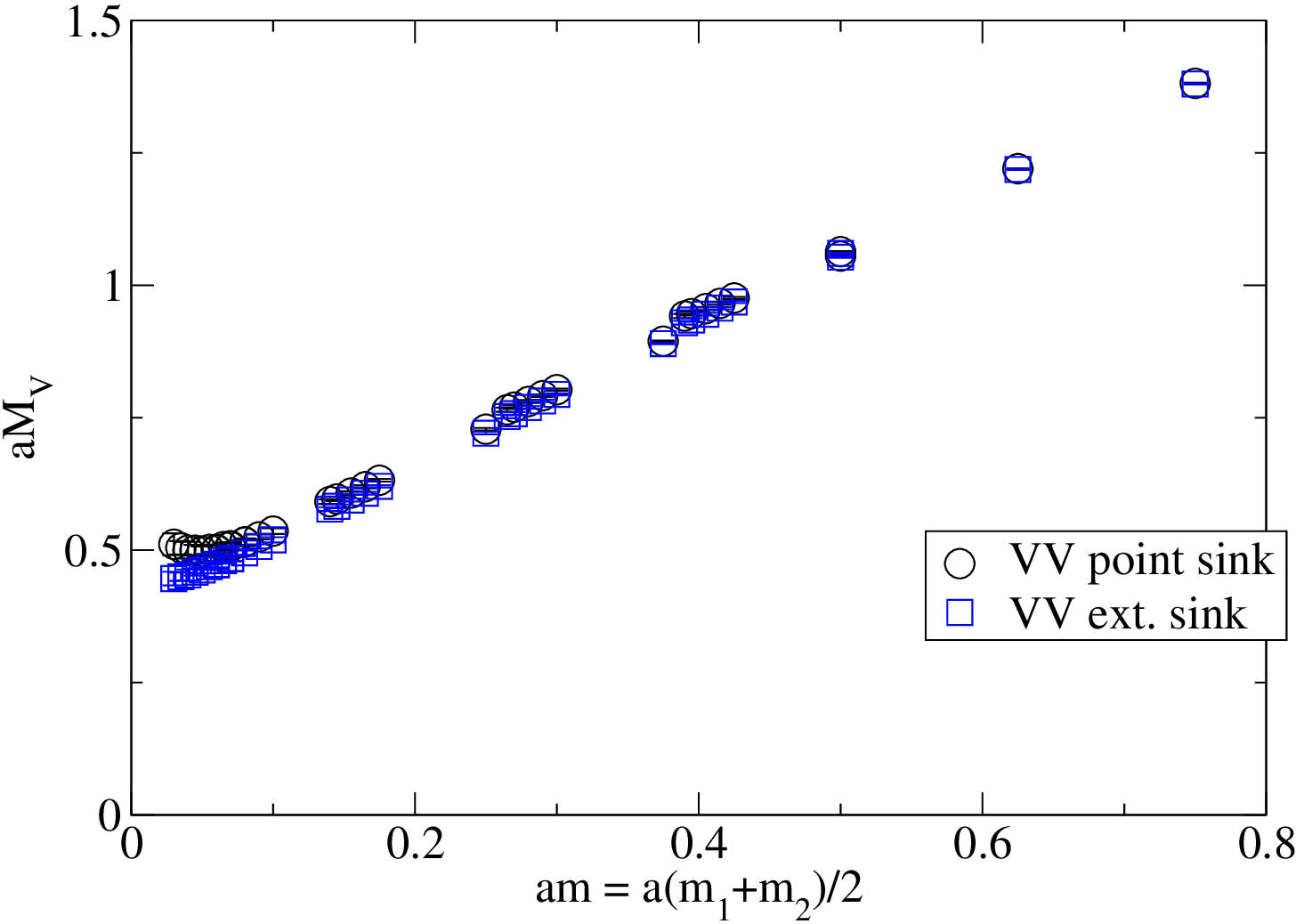}
\caption{Vector meson spectrum for point and extended sinks.}
\label{fig-VVall}
\efig

Using a larger lattice than~\cite{Giusti:2001pk} allowed for
observation of vector meson states, as shown in \refig{fig-VVall},
with $aM_V$ the vector meson mass.  
The observation of vector meson states was enhanced by use of 
extended sink operators, as discussed in \rsec{subsec-ExtSinkOps}.
Our results for the meson spectrum are reproduced in the tables
presented in Appendix~\ref{sec-Tables}.

\subsection{Axial Ward identity and $Z_A$}
\label{subsec-axialward}

Exact chiral symmetry implies a conserved axial current, and the
associated axial Ward identity (AWI) predicts a constant value for the ratio
\beq
\label{eq-rho}
\rho(t)=\fr{G_{\nabla_0 A_0 P}(t)}{G_{P P}(t)}\;.
\eeq
The conserved axial current is a local, but not ultralocal, operator.
The ultralocal axial current 
\beq
\label{eq-axialc}
A_0=\ybar_1(x)\g_0\g_5\bracks{(1-\fr{a}{2\r}D)\y_2}(x)
\eeq
differs from the exactly conserved axial current by a finite renormalization
factor $Z_A$ and possible corrections $\cO(a^2)$.  We calculated
the correlator in \req{eq-rho} with the current of \req{eq-axialc}
and using the lattice central difference for $\nabla_0$, corrected
so as to take into account the $\rm sinh$ behavior of the correlator.
\refigure{fig-rhot} shows the observation of plateaus for all of
our quark masses in the range $8 \le t/a \le 56$.

\bfig[tp]
\includegraphics*[width=8.3cm,clip]{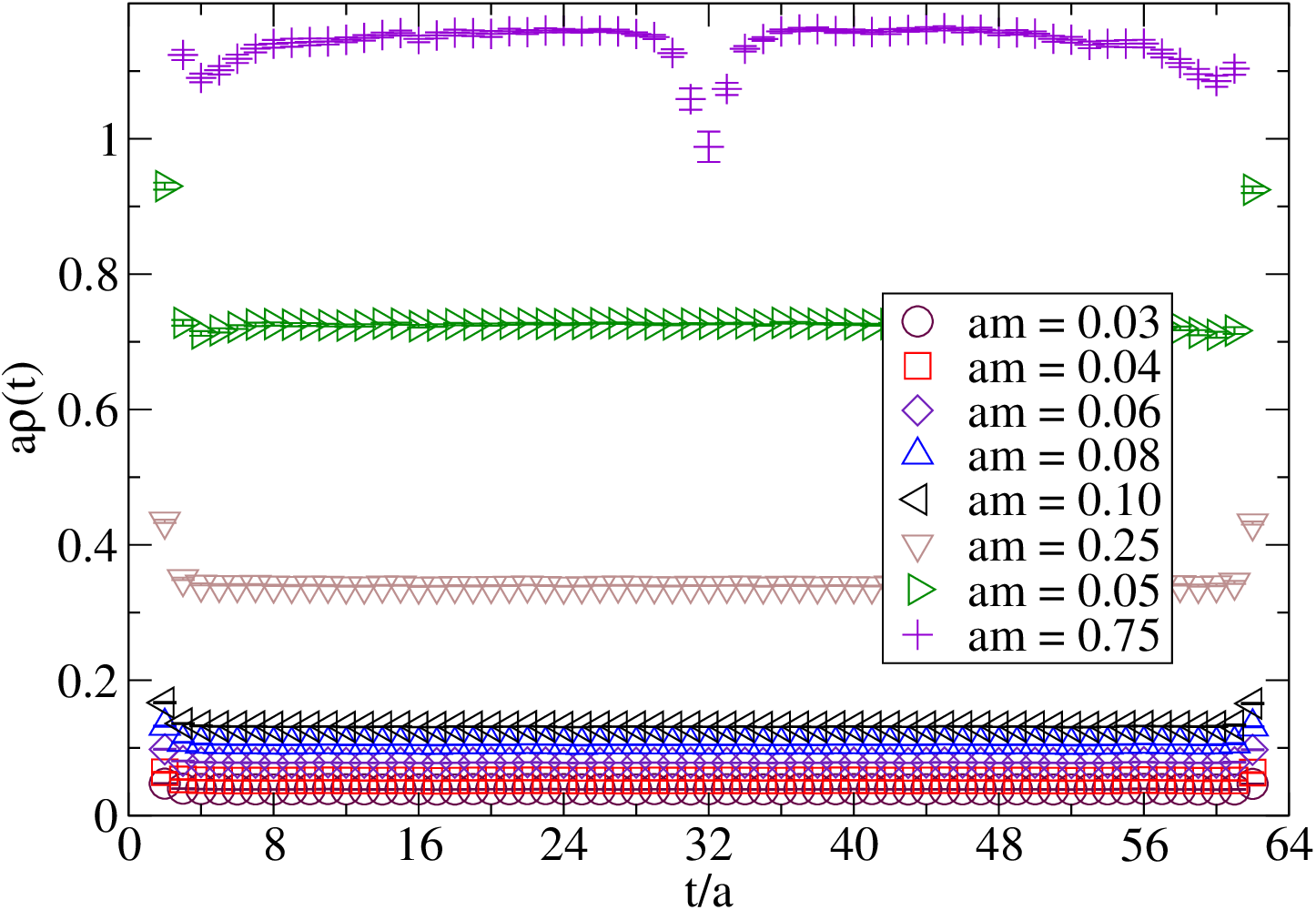}
\caption{AWI ratio as a function of time for all degenerate quark mass
combinations.}
\label{fig-rhot}
\efig
\bfig[tp]
\includegraphics*[width=8.3cm,clip]{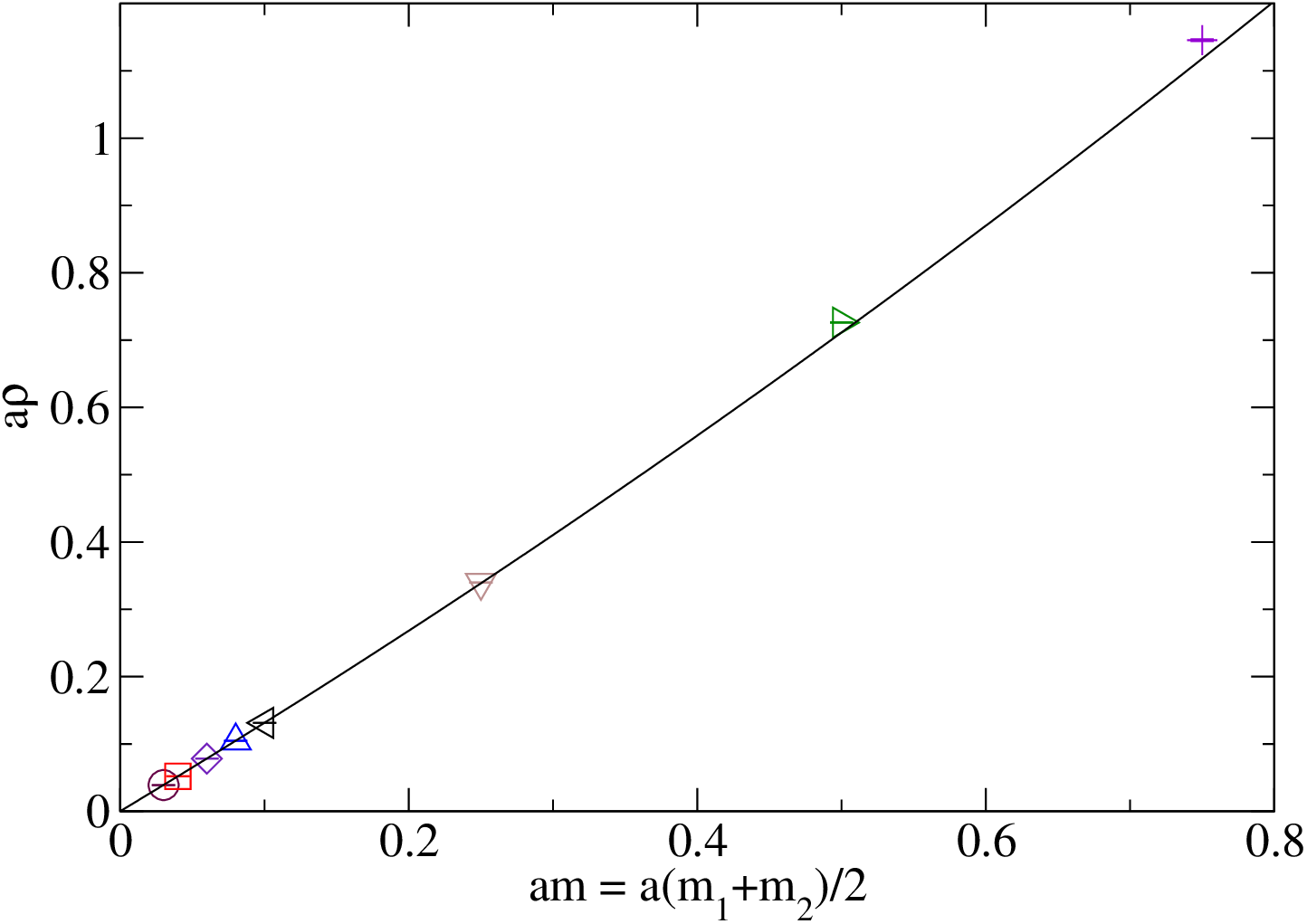}
\caption{Axial Ward identity fit.}
\label{fig-rho}
\efig

The fit shown in \refig{fig-rho} to
\beq
a\rho={\cal A}+2(am)/Z_A+ {\cal C}(am)^2
\label{eq-AWIfit}
\eeq
gives
\beq
{\cal A} = 0.00002(10)\;,\ \ 
Z_A = 1.5555(47)\;,\ \ 
{\cal C} = 0.273(32)\;.
\eeq
(In \cite{Giusti:2001pk}, the results were
${\cal A}=-0.00002(7)$, $Z_A=1.555(4)$ and ${\cal C}=0.277(12)$.)
The fact that the value of $\cA$ is consistent with zero is an
excellent indication of the good chiral behavior of overlap 
fermions (compared to the residual mass found in domain-wall fermion
calculations).  Also, $\cC$ is rather small, a possible indication
that discretization errors might be smaller than expected on the basis
of purely dimensional arguments.

\subsection{Extended sink operators}
\label{subsec-ExtSinkOps}

For some correlators, e.g.~the vector $VV$ correlators, 
the signal for the ground state is small when point sources and point
sinks are used.  In particular, the use of point sources and sinks
causes coupling to excited states that do not decay until relatively
large values of $t_{min}$.  The signal can be improved by using non-local 
extended operators instead of local point operators in the 
representation of the meson~\cite{Gottlieb:1985xq}.  Since, as
mentioned in \rsec{sec-Intro}, our 
quark propagators were calculated using point sources, we were 
limited only to the case of extended sinks.  Here we elaborate on 
our use of extended sink operators.

We consider the correlator
\beqa
G(r,t)&=&\sum_{\v{x},\v{y}} \left\langle \Tra\left[S^{f_2}(0;\v{x},t)\,
\G_A\g_5\,(S^{f_1}(0;\v{y}, t))^\dagger\,\g_5\G_B\right]\right. \nn\\
& & \qquad \x\;\left.\delta(\vert\v{x} - \v{y} \vert -r)\right\rangle\;,
\label{ext-corr}
\eeqa
where the $\delta$-function is approximated on the lattice with
a Kronecker $\delta$, taking into account the multiplicity of the sites.
The quantity $r$ is the separation of the quark and antiquark at the sink in
lattice units.  No gauge transport factor is needed to make the non-local correlator
in \req{ext-corr} well defined because we calculated the quark
propagators in Landau gauge.

\bfig[tp]
\includegraphics*[width=8.3cm,clip]{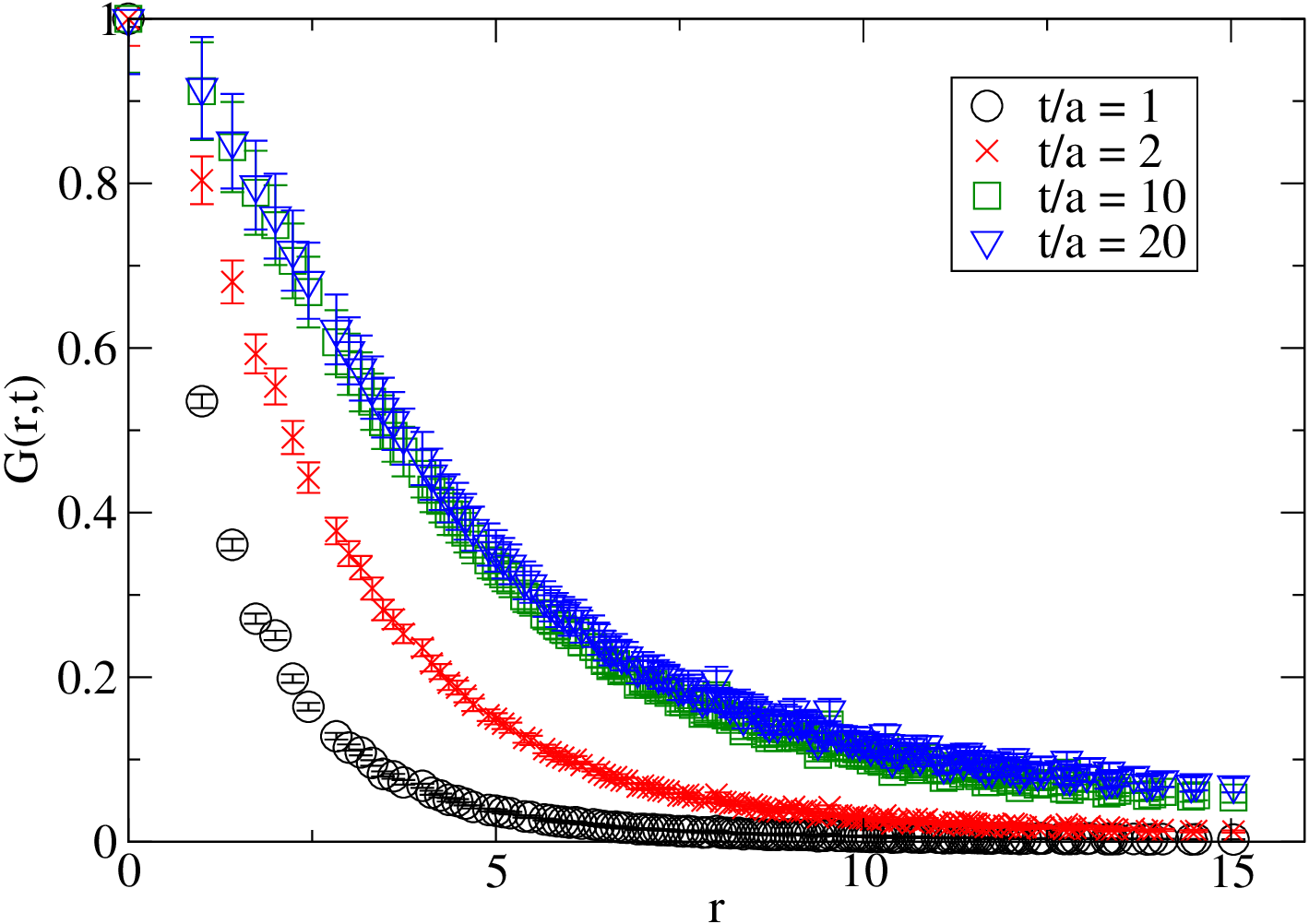}
\caption{$PP$ extended sink correlators $G(r,t)$ for various $t/a$
and $am_1=am_2=0.03$.}
\label{fig-PPGrt0303}
\efig

We calculated $G(r,t)$ for $PP$ and $VV$ correlators.
A fast Fourier transform (FFT) was used in order to speed up the calculation; we first
Fourier transformed the quark propagators and then used the convolution
theorem.  That reduced the double summation over spatial lattice sites
to a single sum, decreasing the computational time by almost
three orders of magnitude~\cite{Hauswirth:2002eq}.
\refigure{fig-PPGrt0303} illustrates the average value of the $G(r,t)$
correlators for pseudoscalar mesons with quark mass $a m =0.03$, 
each normalized to unity at $r=0$.  A clear ground state 
``wave function'' is apparent after about $t/a=8$.  

\bfig[tp]
\includegraphics*[width=8.3cm,clip]{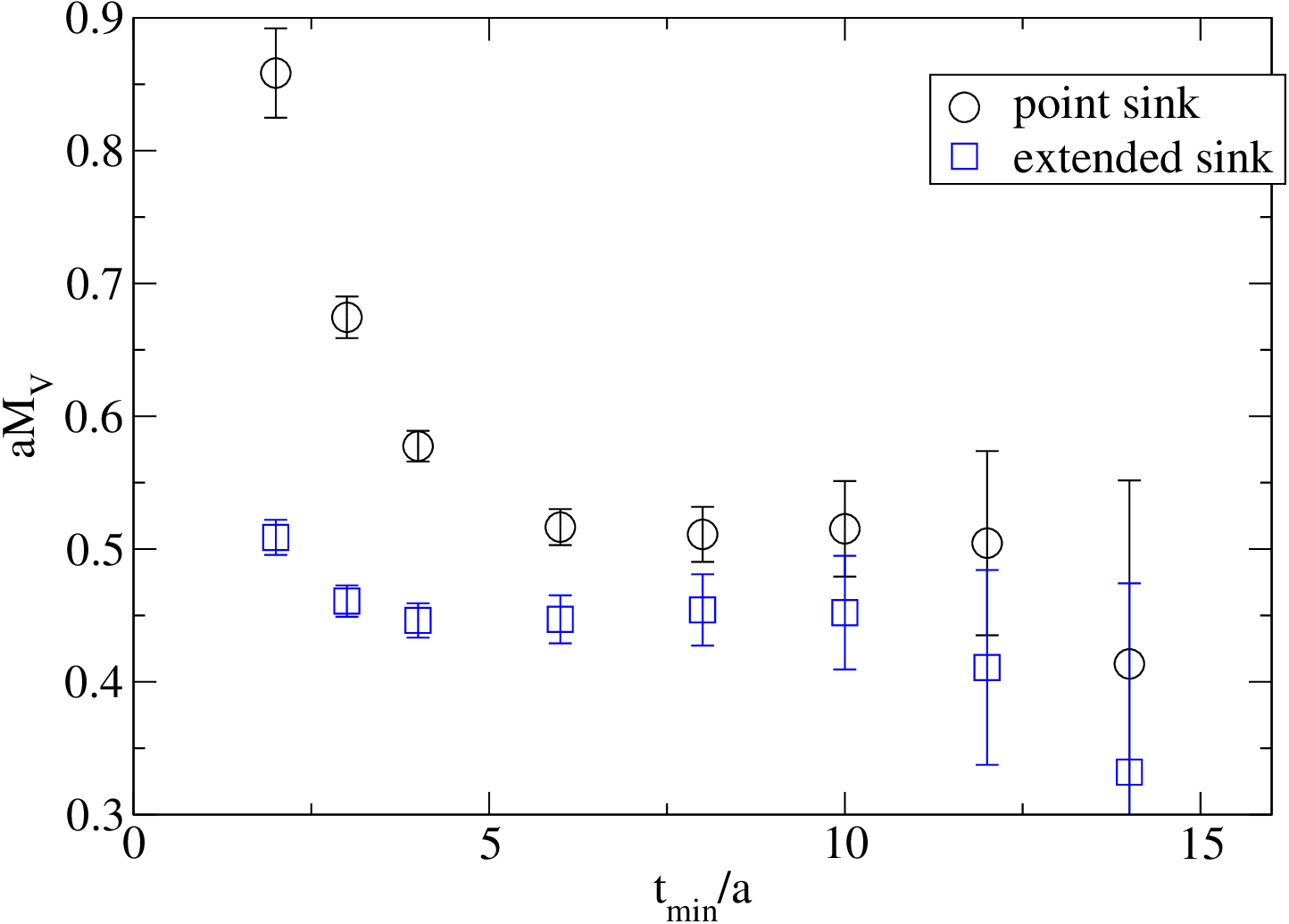}
\caption{Comparison of $t_{min}$ scans for $VV$ point and extended 
sinks with $am_1=am_2=0.03$.}
\label{fig-tminVVvsVVsm}
\efig

\bfig[tp]
\includegraphics*[width=8.3cm,clip]{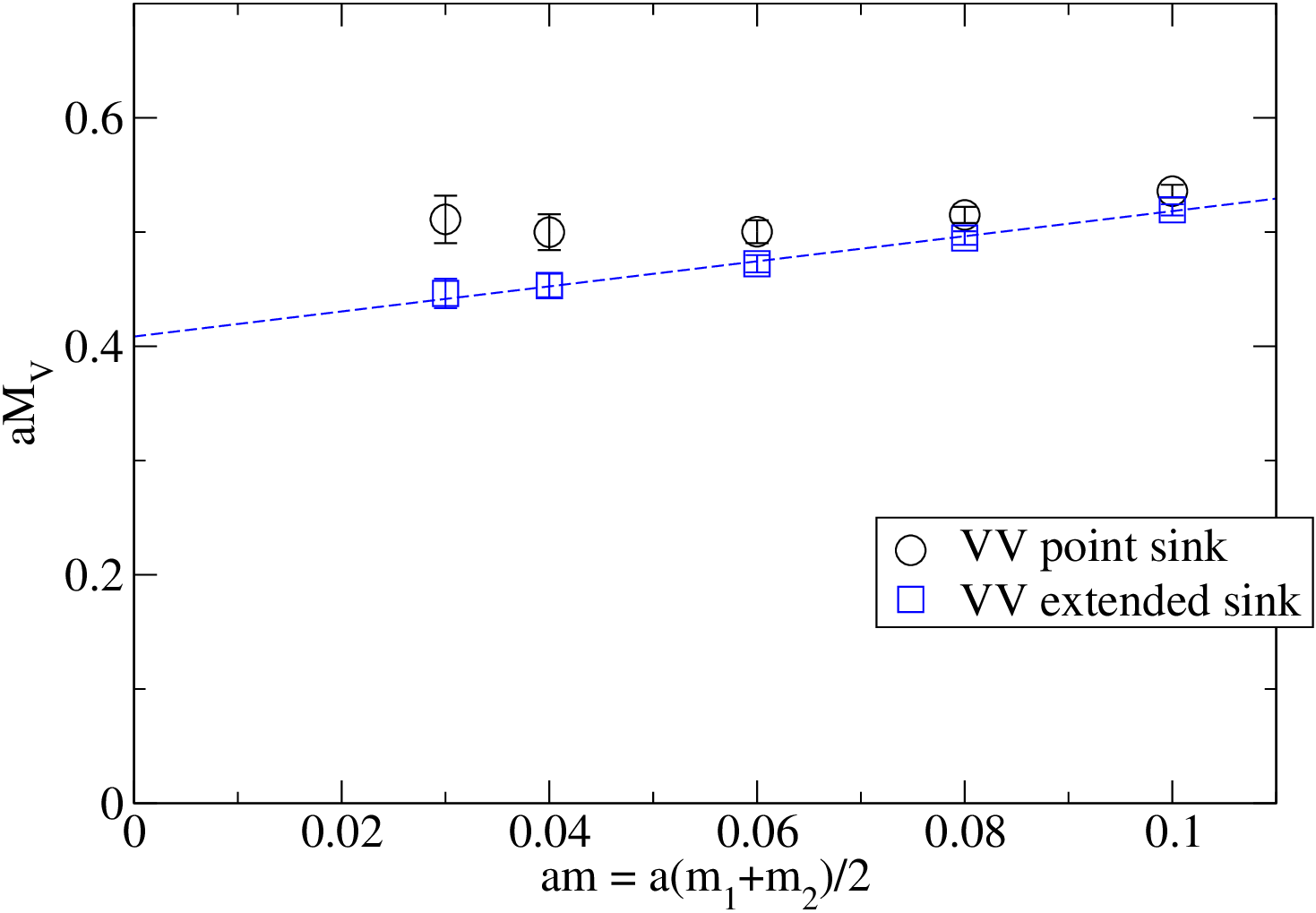}
\caption{Vector meson spectrum for $am\le 0.1$.}
\label{fig-VVlowmq}
\efig

The observation that the wave function settles into a definite 
profile representing the contribution of the ground state
allows us to use this ground state wave function to
construct extended sink correlators.  For both the $PP$
and $VV$ correlators, we used the corresponding function
$\ph(r) \equiv G(r,8a)$ (of course, with different $\ph(r)$ for the
pseudoscalar and vector states) to define an extended sink correlator
\beq
\label{ext-sink-corr}
G_{\rm ext}(t)=\sum_r \ph(r) G(r,t) \;,
\eeq
from which we then extracted the meson mass.

The use of extended sinks was most valuable in the calculation of the
vector meson spectrum, due to the resulting increase in the size of the fitting
window (see~\refig{fig-tminVVvsVVsm}).  The point sink and extended 
sink vector meson spectra are compared in \refig{fig-VVlowmq} for 
low quark mass, $am\le 0.1$.
A linear fit of the quark-mass dependence of the vector meson mass obtained
from the extended sink data gives
\beq
aM_V=0.409(15)+1.10(13)\,(am)\;.
\label{eq-vecmass}
\eeq
We note that the chiral limit value of 0.409(15) is larger than the
value 0.366 obtained using the Sommer scale value for the lattice
spacing and the experimental $\r$
mass.  As we will discuss in a later section, the
discrepancy gives an indication of the systematic error induced by the
quenched approximation.

\subsection{Quenched chiral logarithms}
\label{subsec-QCL}

A linear fit to the pseudoscalar masses for quark mass
$am \le 0.1$ (see~\refig{fig-mqmp2}) produces a line with 
an intercept which is very small, but statistically different 
from zero, indicating a deviation from the 
chiral behavior $(aM_P)^2 \propto am$.
This deviation is put in better evidence by considering the ratio
$(aM_P)^2/am$, shown in \refig{fig-msq1-PP},
which exhibits a sharp rise at low $am$.

\bfig[tp]
\includegraphics*[width=8.3cm,clip]{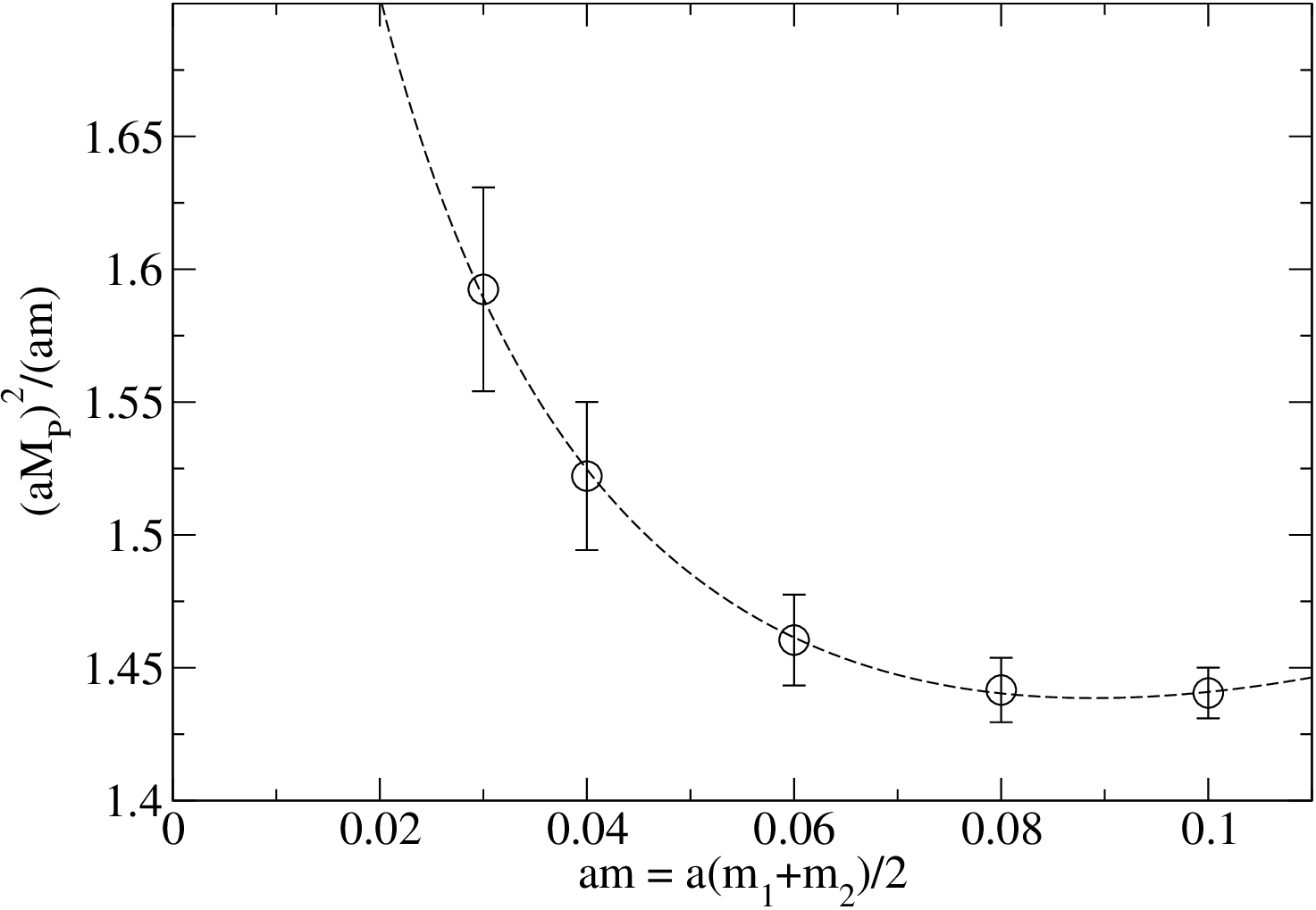}
\caption{Evidence for quenched chiral logs.}
\label{fig-msq1-PP}
\efig

Having already ruled out that the non-zero intercept may be due to
the effect of fermion zero modes (the $PP$ and $PP-SS$ correlators
produce statistically indistinguishable results for the intercept),
the small deviation from linear behavior could be due to finite
volume effects or to chiral logs.  Insofar as finite volume
effects are concerned, not having the resources needed to repeat 
the calculation on a larger lattice, we can only observe that
the Compton wavelength for our lightest pseudoscalar,
$M_P^{-1} \approx 4.5 a$, is much smaller than the size of
our lattice, $L=18a$ ($M_PL=4$).  In the rest of this section, we compare
our results for small quark masses with the predictions
from quenched chiral perturbation theory~\cite{Sharpe:1992ft,Bernard:1992mk}.

We fit the degenerate quark mass results for the pseudoscalar
masses to the expression~\cite{Sharpe:1992ft}
\beq
(aM_P)^2=A(am)^{1/(1+\d)}+B(am)^2\;,
\label{eq-MaSqQcl}
\eeq
where the leading quenched logarithms proportional to
$\d=2m_0^2/(N_c(4\pi f_\chi)^2)$ have been resummed into a power
behavior and where the term proportional to $B$ parameterizes possible
higher-order corrections in the mass expansion.  Here $m_0$ is the
singlet contribution to the $\eta'$ mass, $N_c=3$ the number of
colors, and $f_\chi$ the value of the pion decay constant in the
chiral limit.
The fit is shown in \refig{fig-msq1-PP}, and the results for the 
parameters are
\beq
A=0.680(68),\ \ B=2.98(31),\ \ \d=0.29(5)\;,
\label{chiraldelta}
\eeq
consistent with values of $\d$ presented elsewhere in the
literature~\cite{Kim:1999ur,Blum:2000kn,DeGrand:2002gm,Aoki:2002fd,Draper:2002ep,
Gattringer:2003qx,Chiu:2005ue}.
Note that a fit to $(aM_P)^2$ in \req{eq-MaSqQcl} containing an additional
constant term $C$ does not change the central values of $A$, $B$, or
$\d$ and produces a value for $C$ that is very small and consistent
with zero.

\bfig[tp]
\includegraphics*[width=8.3cm,clip]{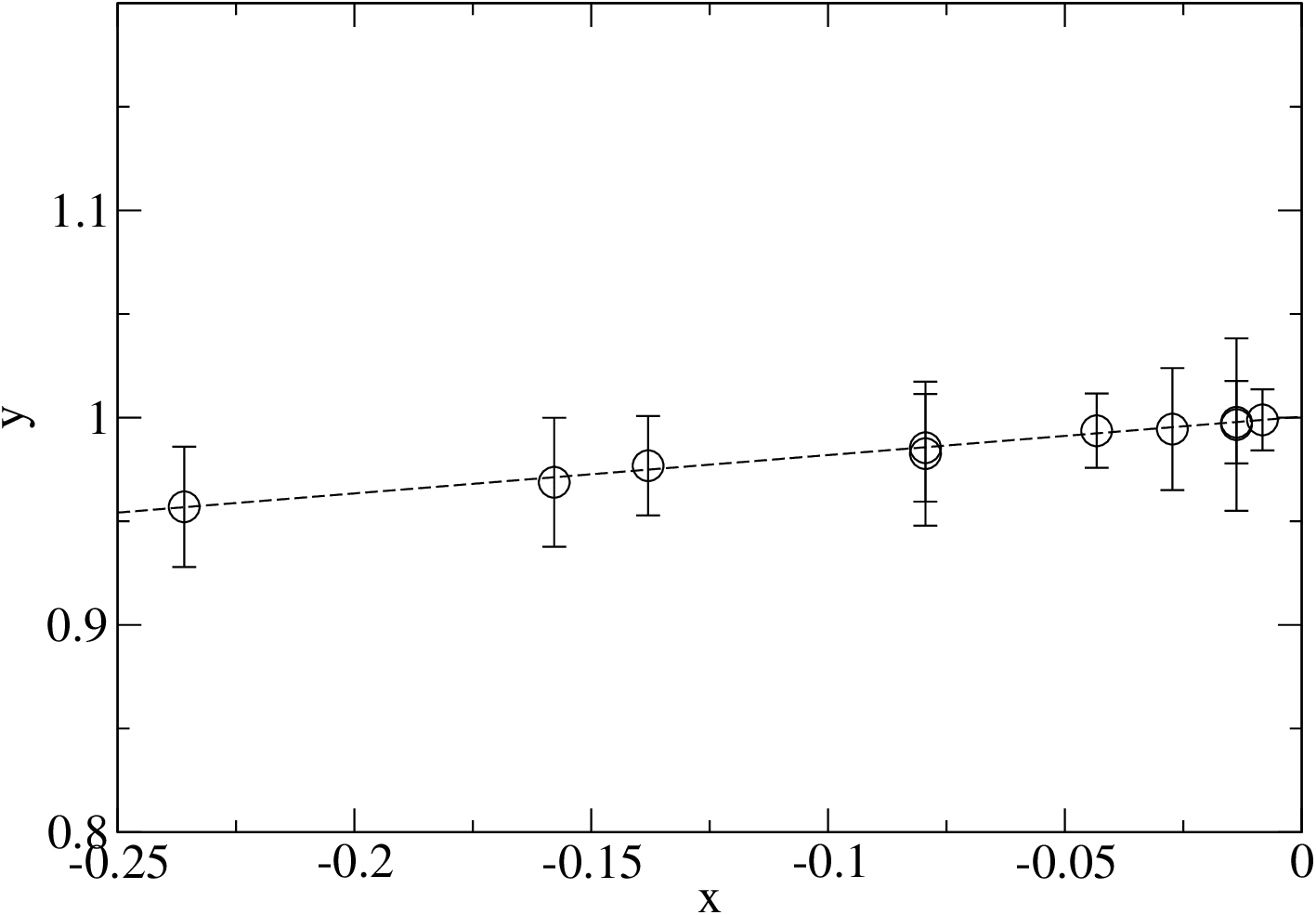}
\caption{Calculating the quenched chiral log parameter $\delta$ with
non-degenerate quark masses.}
\label{fig-PP-yx}
\efig

For the non-degenerate quark case, a value for the quenched chiral 
log parameter $\d$ was obtained via a fit to the expression
$y=1+\d\,x$~\cite{Bernard:1992mk,Aoki:2002fd}, where
\beq
\label{eq-qclndy}
y=\fr{2m_1}{m_1+m_2}\fr{M_{12}^2}{M_{11}^2}
\fr{2m_2}{m_1+m_2}\fr{M_{12}^2}{M_{22}^2}\;,
\eeq
\beq
\label{eq-qclndx}
x=2+\fr{m_1+m_2}{m_1-m_2}\,\ln\left(\fr{m_2}{m_1}\right)\;,
\eeq
and $M_{ij}$ is the mass of a pseudoscalar meson composed of quarks with
masses $m_i$ and $m_j$.  In the derivation of this functional form, the
contributions of the kinetic term of the singlet Lagrangian with coupling
$\a$, which is subleading in a $1/N_c$ counting, were neglected, as they
were in \req{eq-MaSqQcl}.  However, here the quenched chiral log
proportional to $\d$ is considered a correction and dealt with at
linear order.  In addition,
all higher order terms in the chiral expansion are neglected.  The
value of $\d$ from our non-degenerate data is
\beq
\d=0.18(8)\;, 
\eeq
which is consistent, within the large statistical
errors, with the result in \req{chiraldelta}.

As shown in \rsec{subsec-MesonSpectra}, these chiral logarithms induce only a small
deviation from linear behavior in the relation of $(aM_P)^2$ vs. $(am)$ 
at the values of light quark mass used in our simulation.  Moreover, our statistical
errors are still large and our lightest pseudoscalar meson has a mass around
that of the kaon, where the quenched theory is tuned to reproduce the
unquenched theory.  Therefore, we take the slope parameter
${\cal B}$, obtained from the linear fit of \reqs{eq-mpfit}{eq-fpp}, to 
be our estimate of the coefficient of the leading term in the chiral expansion of
$(aM_P)^2$ as a function of quark mass $am$.

\subsection{Decay constants and determination of the lattice spacing}

\bfig[tp]
\includegraphics*[width=8.3cm,clip]{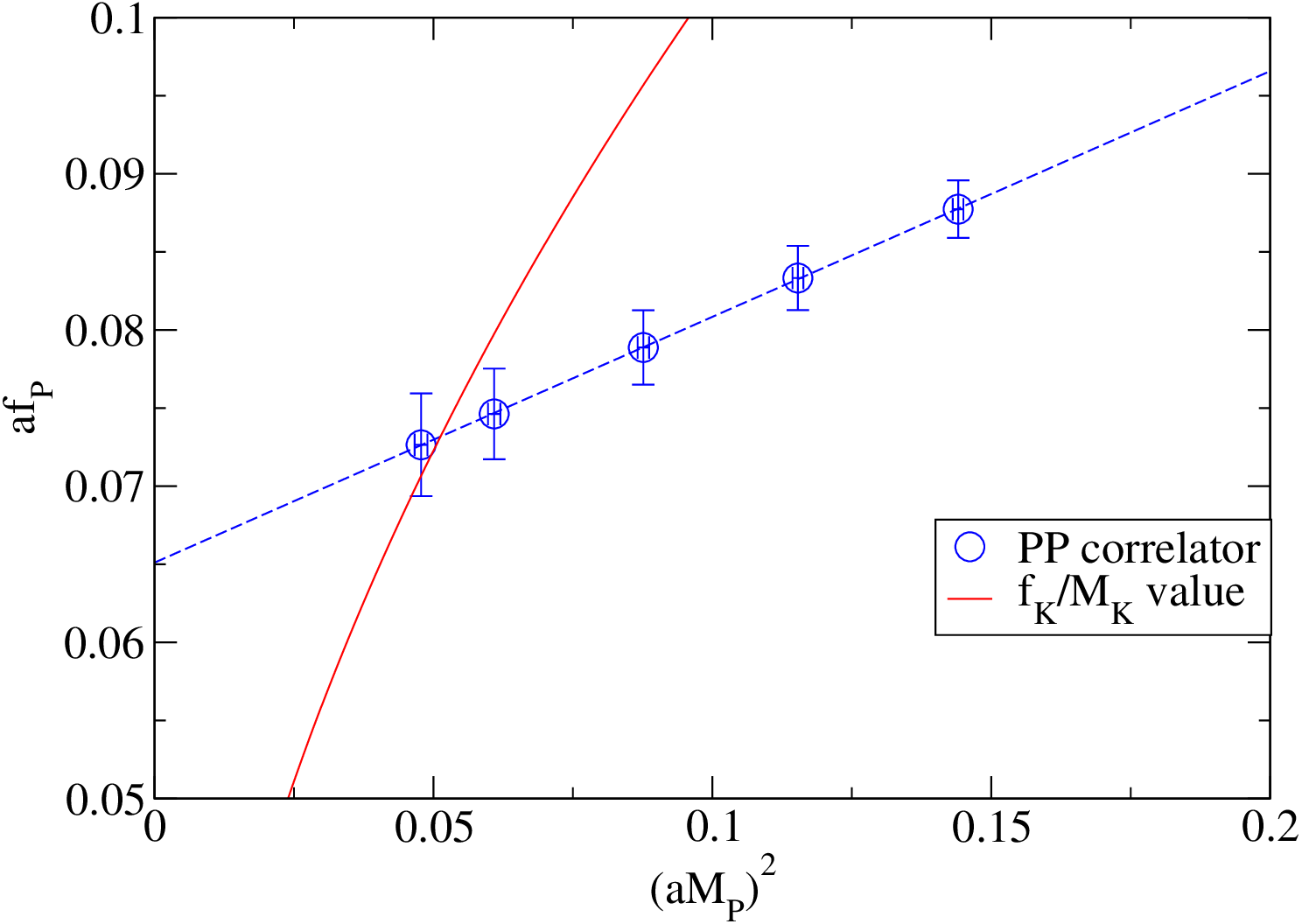}
\caption{Results for $f_P$ and determination of $a$ from the method
of physical planes.  The dashed
line corresponds to a linear fit of the data for $a f_P$ in the interval 
$am \le 0.1$; the solid curve represents the parabola $(a f_P)/(a M_P)=
(f_K/M_K)_\text{exp}$.}
\label{fig-ainv}
\efig

We extract the decay constant $f_P$ from the relation
\beq
2m|\matel{0}{P}{\pi}| = 2m\sqrt{Z} = f_PM_P^2\;,
\eeq
where the correlator matrix element $Z$ is defined in \req{eq-mesoncorrff}.
Our results for the pseudoscalar decay constant are reproduced 
in \refig{fig-ainv}, where we plot $a f_P$ as a function of $(a M_P)^2$, together with a line
representing the results of a linear fit, as predicted by NLO quenched 
chiral perturbation theory~\cite{Sharpe:1992ft,Bernard:1992mk}.
Following \cite{Allton:1996yv}, we also determine the lattice spacing
by the method of lattice physical planes. In particular, we are looking 
for the point in our lattice parameter space, at which the
dimensionless ratio of the pseudoscalar mass and decay constant 
is equal to the experimentally determined ratio
of the Kaon mass to decay constant.
The parabola in \refig{fig-ainv} shows the line, along which
\beq
\frac{af_P}{aM_P}=\Big(\frac{f_K}{m_K}\Big)_\text{exp}=0.323\;,
\eeq
where $(f_K/M_K)_\text{exp}$ stands for the experimental value 
of this ratio and we used as experimental data 
$M_K=0.495\,\text{GeV}$ and $f_K=0.16\,\text{GeV}$ \cite{PDBook}.
The intersection of the two lines gives 
\beq
\label{eq-amk}
a M_K= 0.226(6), \qquad  a f_K=0.074(2) \;,
\eeq
from which, using the input $M_K=0.495\,\text{GeV}$, one gets 
\beq
\label{eq-aipp}
a^{-1}=2.19(6)\,\GeV\;.
\eeq

Using the Sommer scale value of the lattice spacing and 
the experimental value of the pion mass, $M_{\pi} =
0.135\,\GeV$~\cite{PDBook}, yields
\beq
af_\pi=0.065(2)\;,
\eeq
and thus
\beq 
\fr{f_K}{f_\pi}=1.13(4)\;,
\label{eq-fKfpi}
\eeq
which is below the experimental value of 1.22 but is compatible with other
quenched calculations~\cite{Butler:1993zx,Becirevic:1998jp,Burkhalter:1998wu,Heitger:2000ay,Gattringer:2005ij}.

The method of lattice physical planes has the advantage
of using data in a region of quark masses accessible to the lattice
calculation, avoiding the need to perform a chiral extrapolation
to low quark mass. 

The value for $a^{-1}$ derived with the method of 
lattice physical planes should be contrasted with the one derived 
from the Sommer scale, namely $a^{-1} = 2.12\,\text{GeV}$ at $\b=6$.  
A further independent determination
of the lattice spacing can be obtained by extrapolating our results 
for the vector meson spectrum to the chiral limit.  
This gives $a M_\rho = 0.409(15)$ (see \req{eq-vecmass}), from which
one would infer
\beq
a^{-1} = 1.90(4)\,\GeV\;.
\eeq

The discrepancy among the three values of $a$ obtained above,
namely $a^{-1} = 1.90(7)\,\text{GeV}$ from the $\rho$ mass,
$a^{-1} = 2.12\,\text{GeV}$ from the Sommer scale, and 
$a^{-1}=2.19(6)\,\text{GeV}$ from the method of physical planes, 
is substantially larger than what could be due to
statistical errors alone and should be attributed for the most
part to the quenched approximation.
The other sources of error, namely those due to the
finite lattice spacing, extrapolation to small quark masses,
and finite volume effects, are substantially smaller, as can be 
inferred from the data presented in this paper or, insofar
as finite volume effects are concerned, argued from the size
of the lattice.  Taking the maximum variance in the three numbers
above, $\Delta a^{-1}=0.29 \,\text{GeV}$, as an indication of
the systematic errors in the quenched approximation, the corresponding
relative error is $\Delta a^{-1}/a^{-1} \approx 14\%$.  In the rest of this paper,
whenever we quote quantities in physical units rather than
lattice units, we will use the value of $a^{-1}$ from the Sommer scale 
for the conversion and systematic errors on such quantities, when given,
will include an estimate of the scale setting ambiguity from the
method of physical planes.

\reqn{eq-amk} together with the linear fit of \reqs{eq-mpfit}{eq-fpp},
which was justified at the end of \rsec{subsec-QCL},
gives us 
\beq
\label{eq-bqm-mlpp}
a(m_s +\hat{m}) = 0.0661(44)\;, 
\eeq
where $m_s$ stands for the bare mass of the $s$-quark and $\hat m$ 
for the average bare mass of the light $u$ and $d$ quarks.  
We note in passing that using the value of
\req{eq-bqm-mlpp}, together with our results for the vector meson
masses (from the extended sink correlators), give
\beq
\fr{M_{K^*}}{M_\r}=1.09(5)\;,
\eeq
which is compatible within errors with the experimental value of 1.15
as well as with other quenched calculations~\cite{Becirevic:1998jp}.
Use of the Sommer scale value of the lattice spacing, together 
with our data for the pseudoscalar spectrum and the experimental value 
for the kaon mass, produces a slightly larger value for $a(m_s+\hat m)$
than the one obtained from the method of physical planes, namely
\beq
\label{eq-bqm-sommer}
a(m_s +\hat{m}) = 0.0709(17) \;.
\eeq
We will use this value in the rest of the paper.  The variation that
the change in $a(m_s+\hat m)$ induces in the values already quoted
for $f_K/f_\pi$ and $M_{K^*}/M_\rho$ are minimal and well below
the statistical errors.

We calculated the vector decay constant using
\beq
af_V = aZ_V\sqrt{\fr{Z}{M_V^2}}\;,
\eeq
where $Z$ is the matrix element appearing in the $\text{cosh}$ fit
of the point source, point sink vector meson correlator
and $Z_V$ is the renormalization constant of the ultralocal
vector current. For $M_V$, we used the extended sink vector masses.
Like the axial current, the conserved vector current in the
overlap discretization is a local, but not ultralocal, operator,
and the ultralocal current used in the vector meson correlator
must be renormalized.  The chiral symmetry properties of the
overlap formulation guarantee, however, that $Z_V=Z_A$, and so we
can use the result obtained in Sect.~\ref{subsec-axialward}.

\bfig[tp]
\includegraphics*[width=8.3cm,clip]{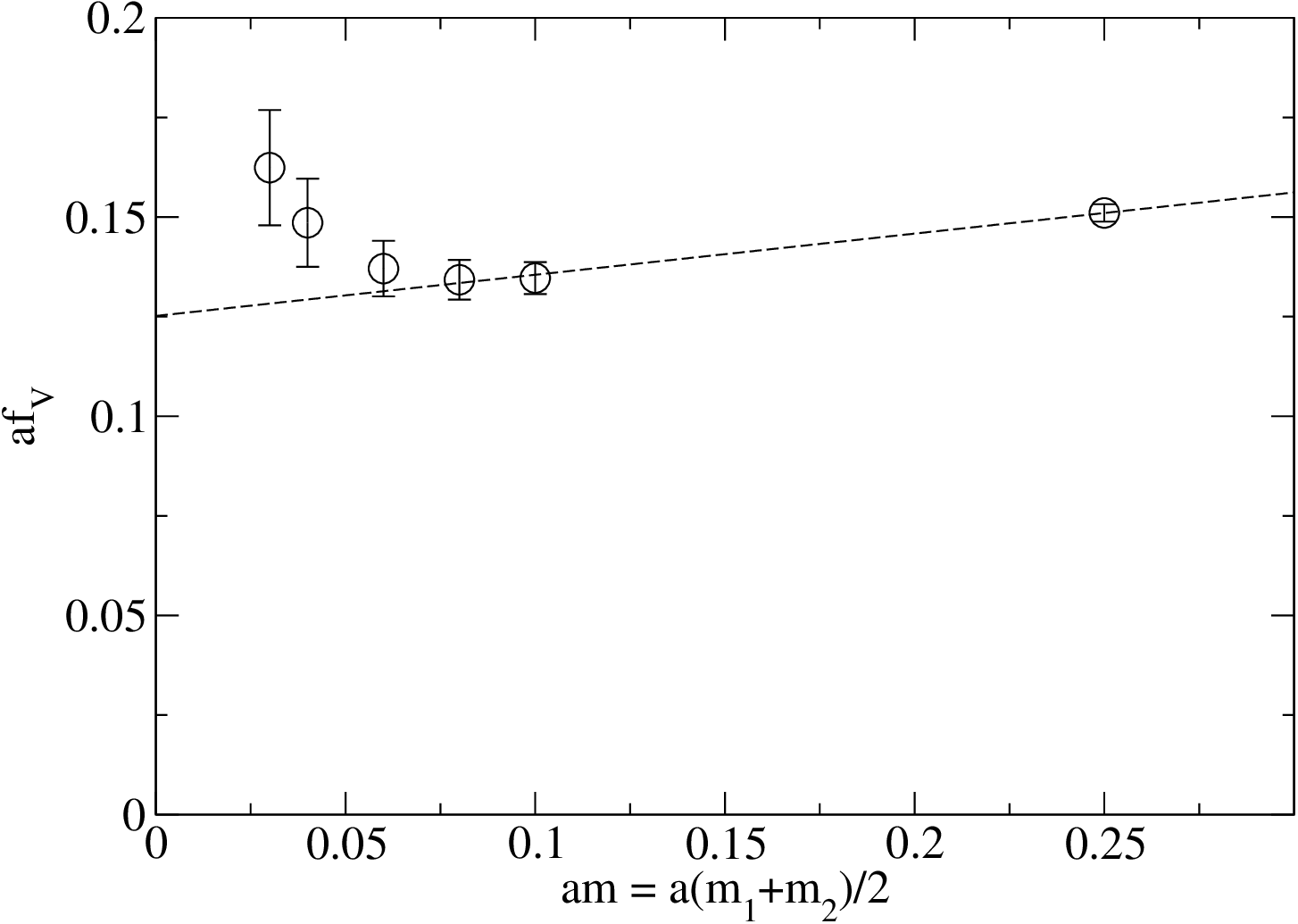}
\caption{
\label{fig-afV}
Vector decay constant vs.~bare quark mass.}
\efig

Unfortunately, as described in Sect.~\ref{subsec-ExtSinkOps}, the the
coupling of the vector meson ground state to the point sink is rather
small. Therefore it is difficult to find a valid plateau range especially for
small quark masses.  In fact, as can be seen in \refig{fig-VVlowmq}, only at
around $am>0.08$ do the masses extracted from point and extended sink
propagators coincide. We therefore also expect the $Z$ extracted from smaller
masses to be heavily affected by excited state contributions and choose to
ignore them. Consequentlly, we only use the points with quark mass $am=0.08,
0.10, 0.25$ in the fit of the $af_V$ data.  Our results for $a f_V$ are
shown in \refig{fig-afV}.  A linear fit of the data, along with the quark
mass of \req{eq-bqm-sommer}, yields
\beq
\label{eq-fKstarfrho}
af_\r=0.125(5),\ \ af_{K^*}=0.128(5),\ \ \fr{f_{K^*}}{f_\r}=1.03(6)\;.
\eeq
The ratio $f_{K^*}/f_\r$ agrees well with the experimental value
1.03(4)~\cite{Becirevic:1998jp}.

Using the Sommer scale value for the lattice spacing, the above 
results give in turn
\beq
f_\r=265(11)\,\MeV,\ \ \ f_{K^*}=272(10)\,\MeV\;.
\eeq

\subsection{Quark masses and chiral condensate}
\label{subsec-ChiralCond}

Our result for the bare quark masses, $a(m_s +\hat{m})$,
can be converted into a corresponding result
for the renormalized quark masses.  The bare quark mass
$m(a)$ is related to the renormalized quark mass $\bar m(\mu)$ by
\beq
\label{eq-massrenorm}
\bar m(\mu)=\lim_{a \to 0} Z_m(a\mu)m(a)\;.
\eeq
The mass renormalization constant $Z_m$ is in turn related
to the renormalization constant $Z_S$ for the non-singlet scalar
density by $Z_m(a \mu)=1/Z_S(a \mu)$. 
We calculated $Z_S$ in the RI-MOM scheme starting from
the identity
\beq
\label{eq-zsza}
Z_S^{\text{RI}}(a \mu)=\lim_{m \to 0}Z_A
\left.\frac{\Gamma_A(p,m)}{\Gamma_S(p,m)}
\right|_{p^2=\mu^2}\;,
\eeq
where $\Gamma_A(p,m)$ and $\Gamma_S(p,m)$ are suitably defined
Green's functions for the axial current and the scalar density
in Landau gauge and $Z_A$ is the renormalization constant 
for the axial current
calculated in Sect.~\ref{subsec-axialward}.  Details of the 
procedure will be presented in a separate publication.  
The Green's functions in \req{eq-zsza} have been calculated
non-perturbatively in a window of momenta which extends into
the perturbative QCD domain, where contact can be made with
perturbatively calculated renormalization constants.
We extracted our central value of $Z_S$ by performing a
combined fit to $am = 0.03$ and $am = 0.1$ data in a momentum
range $p^2=3-14\,\text{GeV}^2$ using four-loop
running~\cite{Chetyrkin:1999pq} and additional $(ap)^2$ terms to account
for discretization effects as well as $1/p^2$ and $1/p^4$ terms to
account for mass effects and other possible subleading terms in the
operator product expansion of the relevant correlation function.
We obtain
\beq
Z_S^{\text{RI}}(2\,\text{GeV})=1.25(2)(2)\;,
\eeq
where the first error is statistical and the second is an estimate
of the systematic error obtained by varying the fit range and
dropping additional terms when indicated.
With this result, one can use the three-loop perturbative calculation 
of the ratio $Z_S^{\overline{\text{MS}}}/Z_S^{\text{RI}}$ 
\cite{Chetyrkin:1999pq} to calculate
\beq
\label{eq-zsmsb}
Z_S^{\overline{\text{MS}}}(2\,\text{GeV})=1.44(2)(3)\;.
\eeq
Putting together \reqs{eq-bqm-sommer}{eq-zsmsb}, we obtain
\beq
(m_s+\hat{m})^{\overline{\text{MS}}}(2\,\text{GeV})=105(3)(4)\,\MeV
\eeq
for the sum of strange and light quark masses.
Using the value $m_s/\hat{m}=24.4(1.5)$ from chiral perturbation
theory~\cite{Leutwyler:1996qg}, we obtain
\beq
m_s^{\overline{\text{MS}}}(2\,\text{GeV})=101(3)(4)\,\MeV
\eeq
for the strange quark mass, which agrees well with the quenched lattice
world average~\cite{Lubicz:2000ch,Durr:2005ik,Gattringer:2005ij}.

In the unquenched theory, the bare chiral condensate with valence
overlap quarks is defined as
\beq
\c(a) \defined \lim_{m \to 0} \fr{1}{N_f}
\left\<\ybar(0)\bracks{\parens{1-\fr{a}{2\r}D}\y}(0)\right\>\;,
\eeq
where $m$ is the common mass given to the light quarks.  It
satisfies the integrated non-singlet chiral Ward identity
\beq
\fr{1}{N_f}\left\<\ybar(0)\bracks{\parens{1-\fr{a}{2\r}D}\y}(0)\right\>
=m\sum_x\<P(x)P^c(0)\>\;,
\label{eq-cwi}
\eeq
where $P$ is the pseudoscalar density composed of two mass-degenerate quarks
of different flavor and $P^c$ is the density obtained by interchanging
the quark flavors.
Inserting a complete set of states in $\<P(x)P^c(0)\>$ gives
\beq
\c(a) = -\lim_{m \to 0} \fr{m}{M_P^2}|\matel{0}{P}{P}|^2\;,
\eeq
where $M_P$ is the mass of the pseudoscalar state $\ket{P}$.  Using
\beq
2m|\matel{0}{P}{\pi}|=f_PM_P^2\;,
\eeq
where $f_P$ is the pseudoscalar decay constant, yields the familiar
Gell-Mann-Oakes-Renner (GMOR) relation
\beq
\c(a) = -\lim_{m \to 0}\fr{f_P^2M_P^2}{4m}\;.
\eeq
Due to quenched chiral logarithms, $\c(a)$ is ill-defined in quenched
QCD.  However, as argued in \rsec{subsec-QCL}, the slope parameter
$\cB$, obtained from the linear fit of \reqs{eq-mpfit}{eq-fpp}, should
be a reasonable estimate of $\lim_{m \to 0}\,M_P^2/m$.  Thus, we
take
\beq
\label{eq-gmorb}
\c(a) = -\fr{1}{4}f_\chi^2\,\cB\,a^{-1}
\eeq
as our determination of the physical quark condensate.
Using our data for $f_P$ we get
\beq
a^3 \chi(a)= -0.00144(10)
\label{eq-cond1}
\eeq
or
\beq
\chi(a)= -0.0137(10)\,\text{GeV}^3 \;.
\label{eq-cond1gev}
\eeq
Note that using the value $f_\chi = 0.123 \,\text{GeV}$~\cite{Amoros:2001cp}
would instead give $\chi(a)= -0.0110(1)\,\text{GeV}^3$.

Using $\langle \bar \psi \psi \rangle^{\overline{\text{MS}}}=
Z_S^{\overline{\text{MS}}} \chi(a)$, we finally get
\beqa
\langle \bar \psi \psi \rangle^{\overline{\text{MS}}}(2\,\text{GeV})
&=&-0.0197(14)(20) \,\text{GeV}^3 \nn\\
&=& - [270(6)(9) \,\text{MeV}]^3 \;.
\label{eq-cond1ren}
\eeqa
Our results for $m_s$ and $\langle \bar \psi \psi \rangle$ are in good
agreement with the results presented in \cite{Giusti:2001pk}. The values 
were
$m_s^{\overline{{\text MS}}}(2\,\GeV)=102(6)(18)\,\MeV$ and $\langle
\ybar \y \rangle^{\overline{\text{MS}}}(2\,\text{GeV})= - [267(5)(15)
\,\text{MeV}]^3$.  They also agree with the 
determinations of the condensate using finite-size scaling
techniques~\cite{Hernandez:1999cu,Hernandez:2001yn}, as well as with
the recent continuum-limit calculation of~\cite{Wennekers:2005wa}, all
of which were obtained using overlap fermions.

\bfig[tp]
\includegraphics*[width=8.3cm,clip]{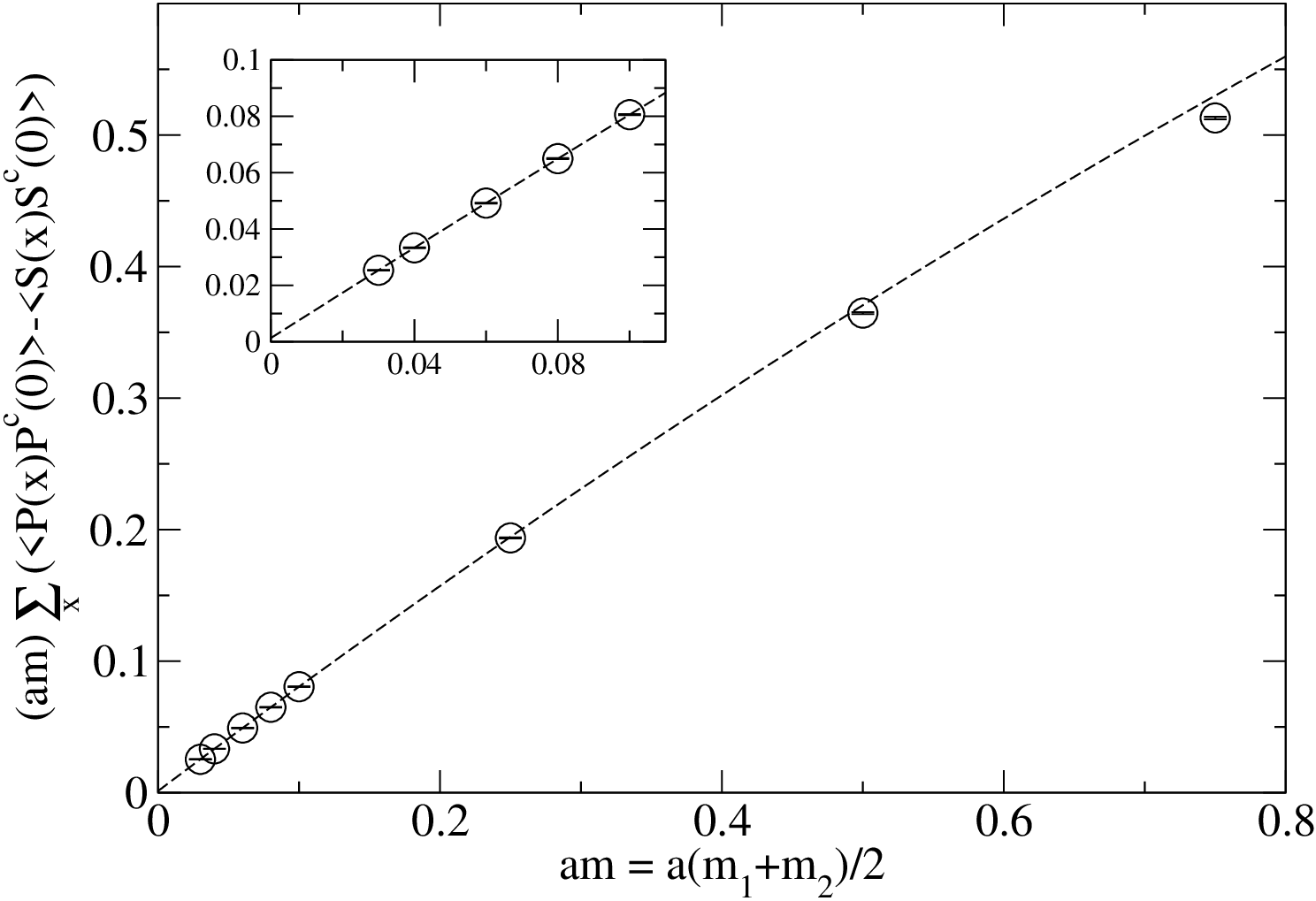}
\caption{Direct determination of the chiral
condensate.  The curve corresponds to a quadratic fit of the data
according to \req{eq-ccd} in the interval $am \le 0.1$.}
\label{fig-ccdirect}
\efig

One could also attempt a determination of $\chi$ directly from a fit
to the mass dependence of the quantity
\beqa
-a^3\hat{\chi}(m)&=&am\sum_x\langle P(x)P^c(0)\rangle \nn\\ 
&=& \left\<\ybar(0)\bracks{\parens{1-\fr{a}{2\r}D}\y}(0)\right\>\;.
\label{hatcc}
\eeqa
This is, however, made difficult by the very steep dependence of $\hat 
\chi(m)$ on $m$ and also by possible infrared divergent contributions 
from zero modes.  Contrary to the case of the determination of the 
pseudoscalar spectrum, where we found the effect of zero modes to 
be suppressed because of the large size of our lattice, zero modes 
are likely to contribute to the expression in \req{hatcc}, because it 
involves the short distance behavior of the quark propagator.  
The contribution from zero modes can be eliminated by instead considering 
the subtracted expression
\beq
-a^3\tilde{\chi}(m)=am\sum_x(\langle P(x)P^c(0)\rangle-
\langle S(x)S^c(0)\rangle)\;,
\eeq
which has the same $m \to 0$ limit as $am\sum_x(\<P(x)P^c(0)\>$
and where the contributions from chiral zero modes cancel. Of course,
in the quenched theory, that quantity diverges in the chiral limit
due to quenched chiral logarithms. However, here again the effect of
the quenched chiral logarithms appear to be small for the light quark
masses reached in our simulation, and we assume that a polynomial
extrapolation of our quenched results to the chiral limit gives a
reliable estimate of the physical condensate.
Our results for $-a^3\tilde{\chi}(m)$ are shown in 
\refig{fig-ccdirect}.
A quadratic fit  
\beq
\label{eq-ccd}
-a^3\tilde{\chi}(m)=-a^3\chi+{\cal B}(am)+{\cal C}(am)^2
\eeq
gives the result
\beq
-a^3\chi=0.00131(8)\;,\ \ 
{\cal B}=0.806(3)\;,\ \ 
{\cal C}=-0.14(1)\;.
\eeq
The value we obtain in this way for $ a^3 \chi$ is compatible
with the value in \req{eq-cond1}. The renormalized value is
\beqa
\langle \bar \psi \psi \rangle^{\overline{\text{MS}}}(2\,\text{GeV})
&=&-0.0179(11)(18) \,\text{GeV}^3 \nn\\
&=& - [262(5)(9) \,\text{MeV}]^3 \;.
\eeqa

\section{Meson Scaling Analysis}
\label{sec-Scaling}
Here we present our results for the coarser $14^3 \x 48$ lattice, with
$\b=5.85$, as well as comparisons between the two lattices.

\subsection{Meson spectra}

We extract meson masses from the correlators $G_{PP}(t)$ and
$G_{PP-SS}(t)$. The effective mass plateau is plotted in \refig{fig-em_106-1448}
for quark mass $am=0.106$. The data suggest use of a symmetrized fit
range $10\le t/a \le 24$ in order to extract the pseudoscalar mass.

\bfig[tp]
\includegraphics*[width=8.3cm,clip]{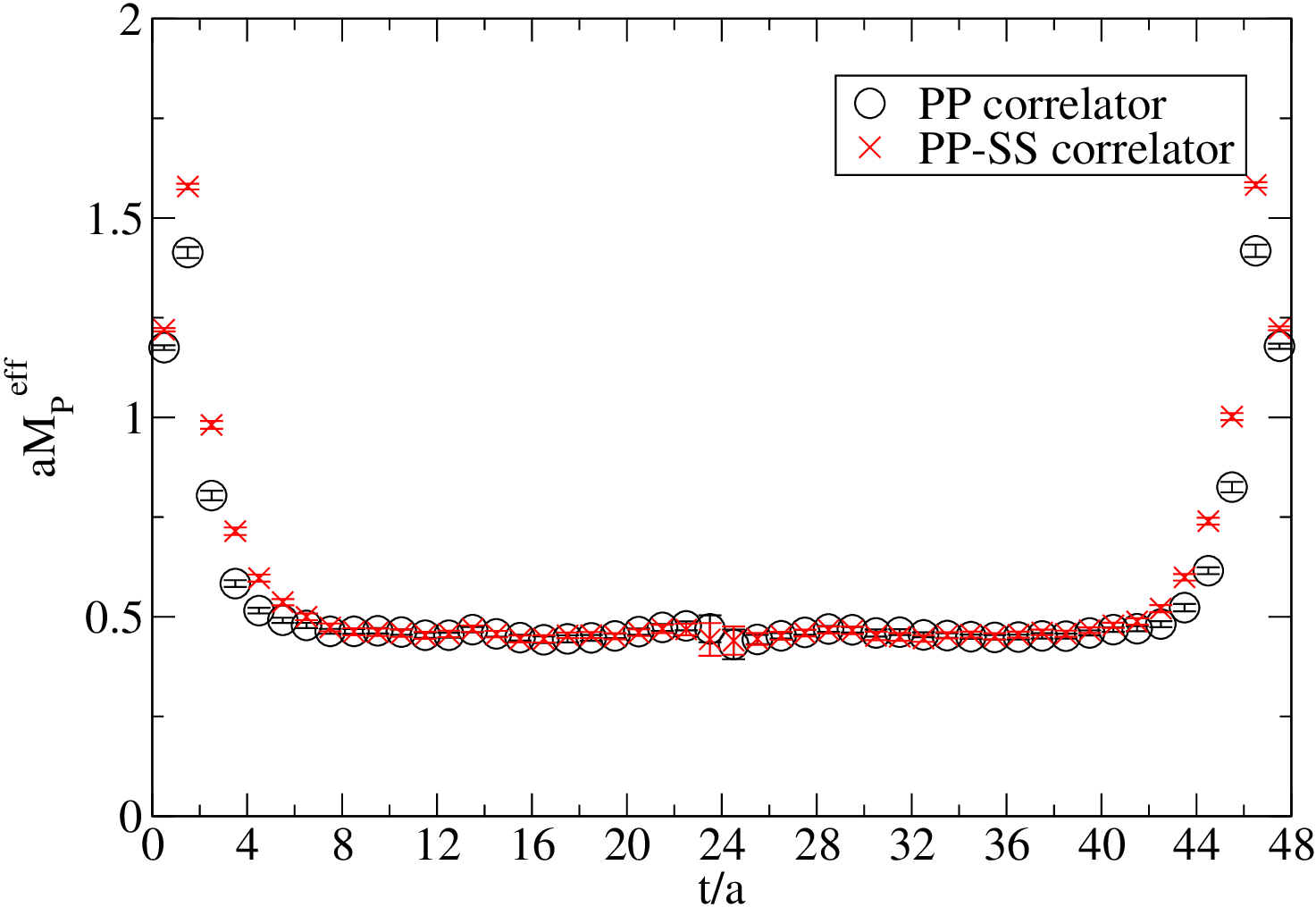}
\caption{Pseudoscalar effective mass plateau for different channels at $am=0.106$.}
\label{fig-em_106-1448}
\efig

The extracted meson masses and matrix elements using this fit range
are reported in \refts{tab-PPpoint-1448}{tab-PP-SSpoint-1448} in Appendix B.
We neglect for the moment chiral logarithms and perform a fit to
\beq
\label{eq-mpfit-1448}
(aM_P)^2={\cal A}+{\cal B}(am)\;.
\eeq
Fitting all data with $am\le 0.132$, we obtain (see \refig{fig-mqmp2-1448})
\beq
{\cal A}=0.0045(15)\;,\qquad
{\cal B}=1.923(16)
\label{eq-fpp-1448}
\eeq
in the $PP$ channel and
\beq
\label{eq-fps-1448}
{\cal A}=0.0007(29)\;,\qquad
{\cal B}=1.959(23)
\eeq
in the $PP-SS$ channel. In general, as for the $18^3 \x 64$ lattice, 
the $PP$ and $PP-SS$ channel results 
were compatible on the $14^3 \x 48$ lattice, so all further
pseudoscalar data, unless otherwise specified, are from the $PP$ channel.

\bfig[tp]
\includegraphics*[width=8.3cm,clip]{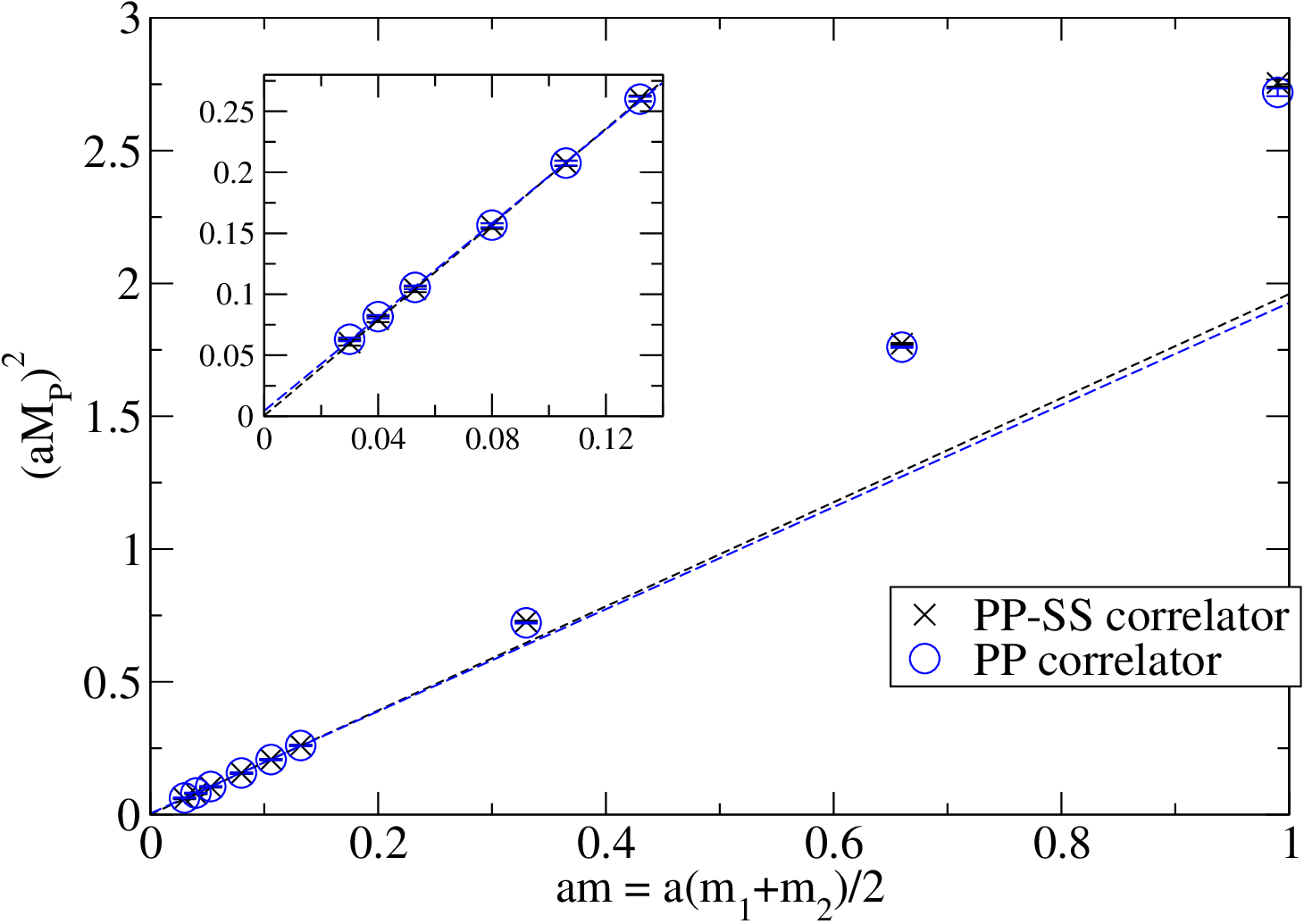}
\caption{Chiral behavior of meson masses. The lines correspond
  to the best fit to \req{eq-mpfit-1448} in the interval $am \le 0.132$.}
\label{fig-mqmp2-1448}
\efig

\bfig[tp]
\includegraphics*[width=8.3cm,clip]{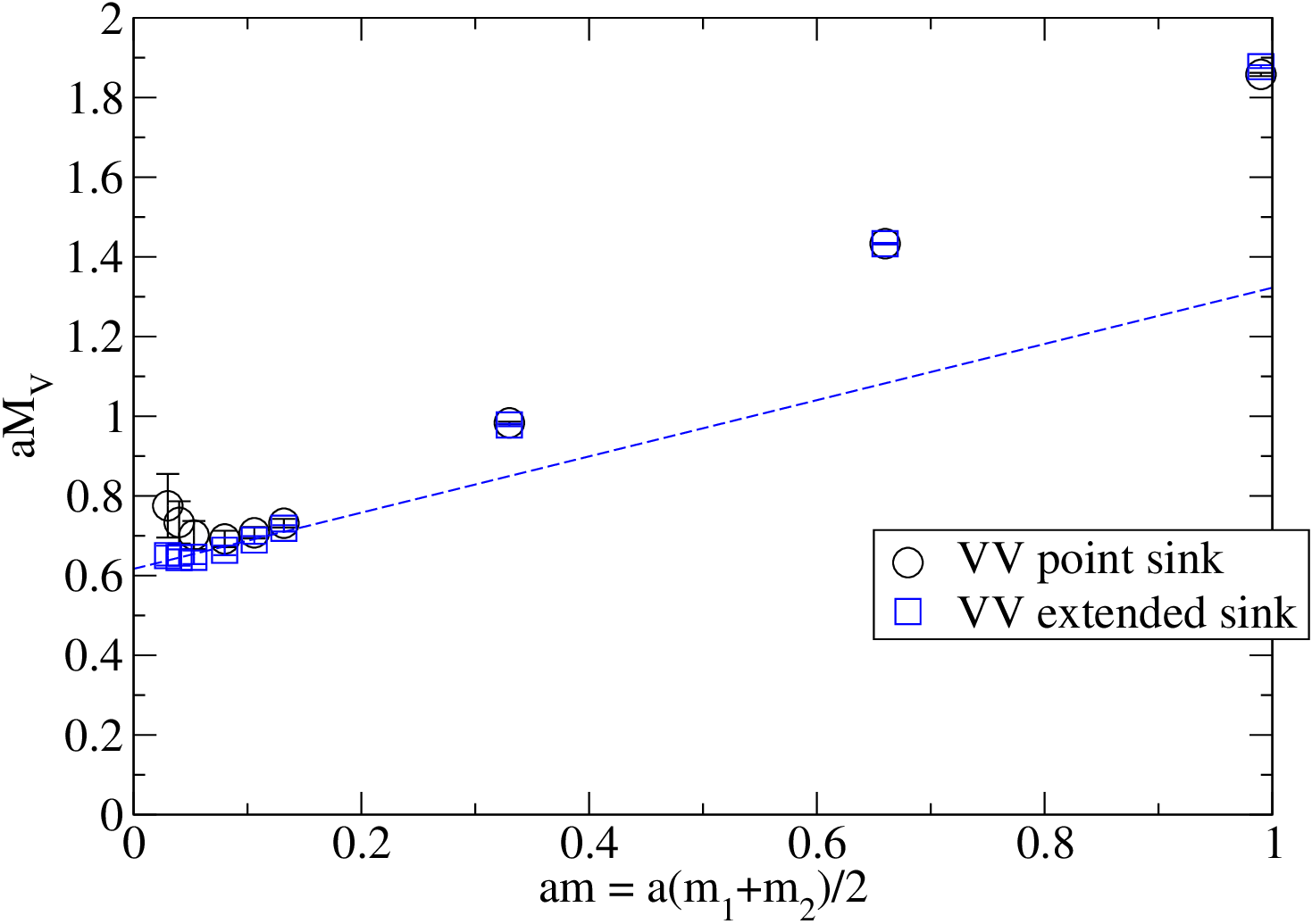}
\caption{Vector meson masses on the coarser lattice. The line
  corresponds to the best fit of the extended sink data to
  \req{eq-vecmass-1448} in the interval $am \le
  0.132$.}
\label{fig-mv-1448}
\efig

A fit of the quark-mass dependence of the vector meson mass obtained
from the extended sink data, shown in \refig{fig-mv-1448}, gives
\beq
\label{eq-vecmass-1448}
aM_V=0.605(29)+0.84(22)\,(am)\;.
\eeq
We note that the chiral limit value of 0.605(29) is larger than the
value 0.483 obtained using the Sommer scale value of the lattice
spacing and the experimental $\r$ mass.  
As for the finer lattice, we expect that this is a result
of quenching errors.

\subsection{Axial Ward identity and $Z_A$}

\refigure{fig-rhot-1448} shows the ratio 
$\rho(t)=G_{\nabla_0 A_0P}(t)/G_{PP}(t)$ for all of our bare quark
masses. We observe a
nice plateau for all quark masses, except the largest bare quark
mass $am = 0.99$, in a range $8\le t\le 40$. The lack of a plateau for
$am = 0.99$ we believe is due to discretization effects, since that
quark mass is large and comparable to the inverse lattice spacing.

\bfig[tp]
\includegraphics*[width=8.3cm,clip]{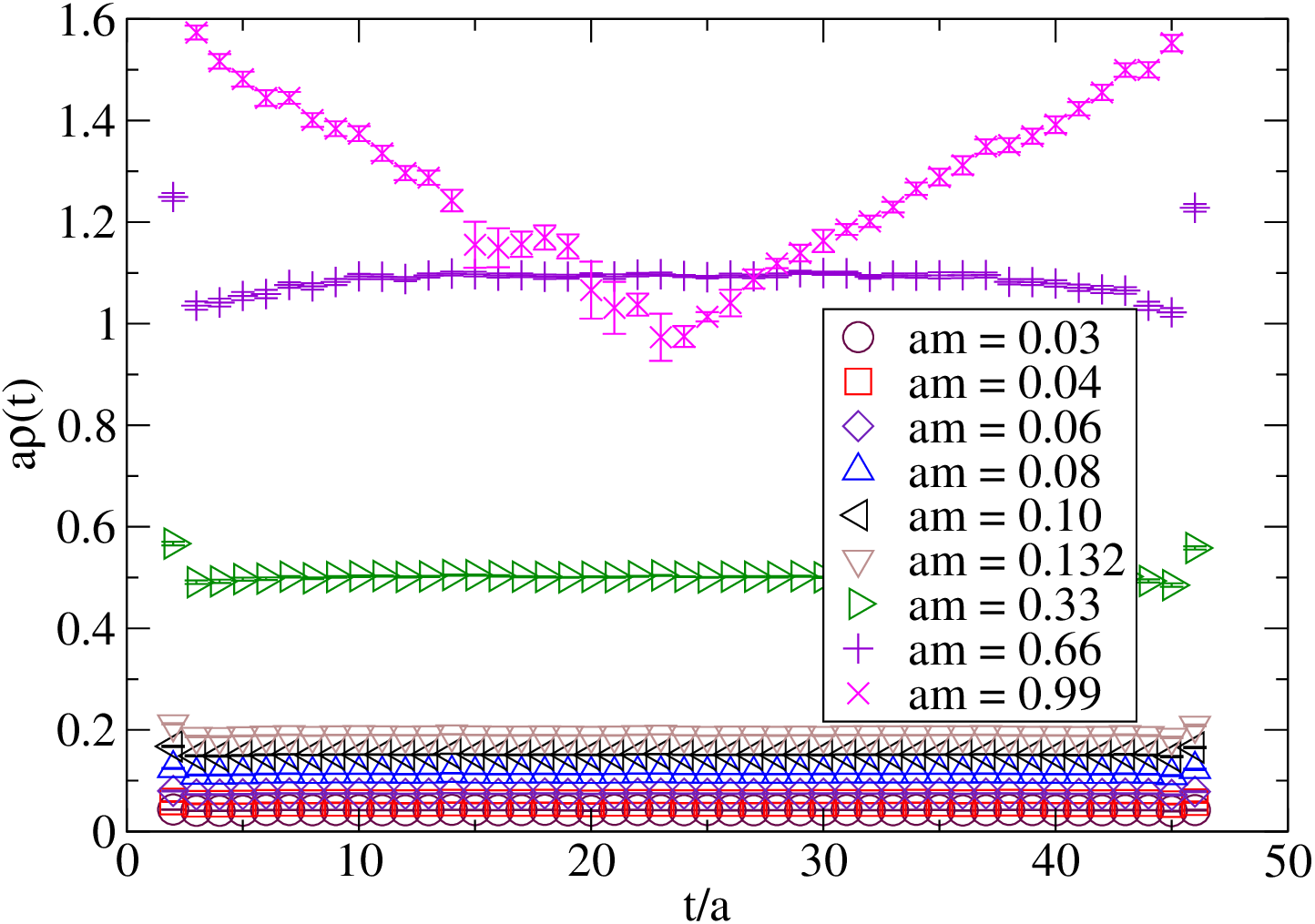}
\caption{The plateau of the AWI ratio $a\rho(t)$ for different bare quark masses.}
\label{fig-rhot-1448}
\efig

\bfig[tp]
\includegraphics*[width=8.3cm,clip]{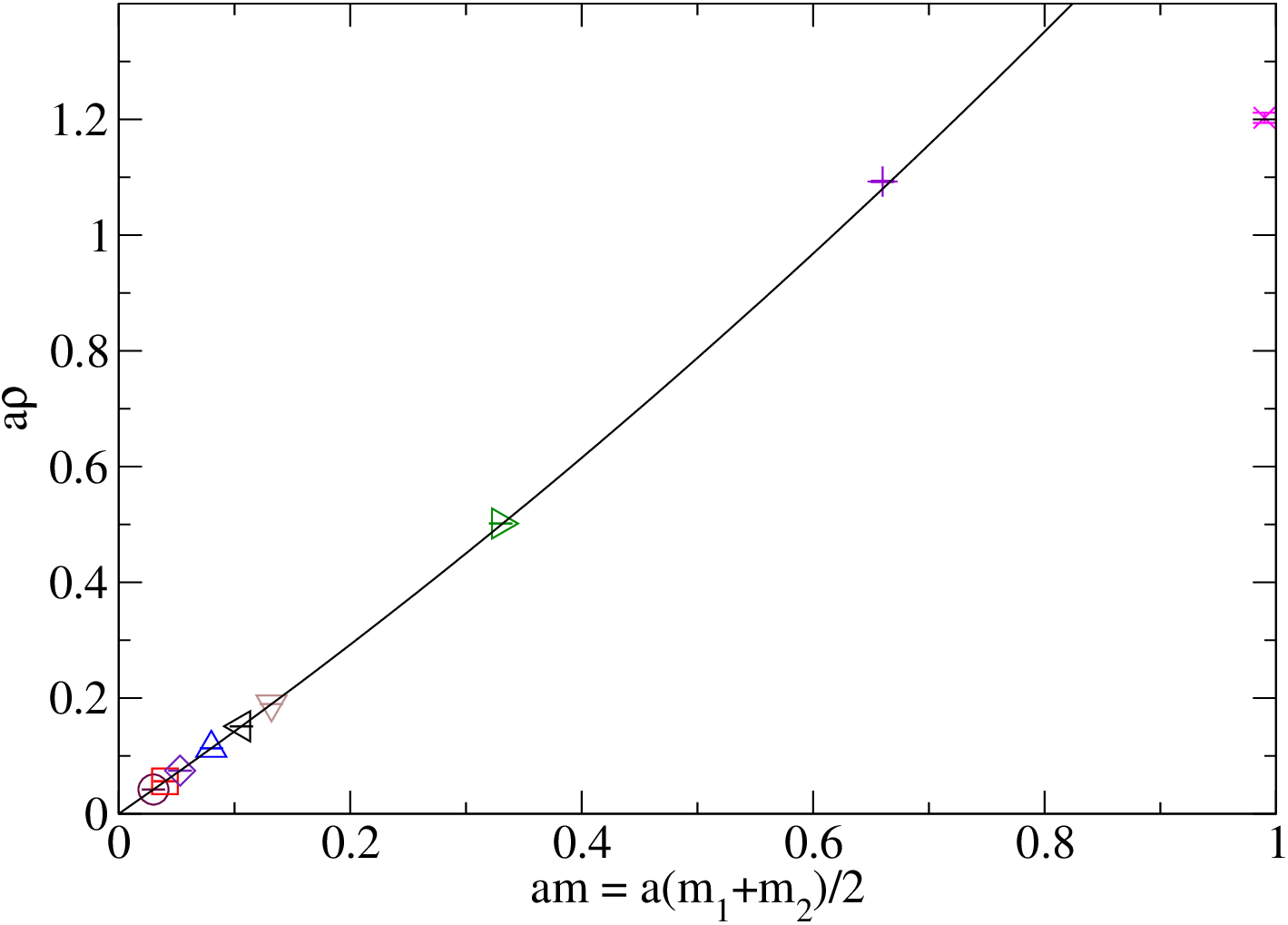}
\caption{The plateau value of $a\rho$ vs. bare quark
  mass. The curve corresponds to the best fit to~\req{eq-AWIfit} in the
  interval $am \le 0.132$.}
\label{fig-rho-1448}
\efig

Taking the aforementioned plateau range and performing a fit to~\req{eq-AWIfit}
with all data for $am \le 0.132$ (\refig{fig-rho-1448}), we obtain 
\beq
{\cal A}=0.00004(5)\;,\ \ 
Z_A=1.4434(18)\;,\ \ 
{\cal C}=0.381(8)\;.
\eeq
Our results are compatible with chiral symmetry for all bare quark
masses except $am=0.99$ and the resulting $Z_A$ is in agreement with
\cite{Bietenholz:2004wv,Fukaya:2005yg}.

\subsection{Quenched chiral logarithms}
\label{subsec-QCL-1448}

\bfig[tp]
\includegraphics*[width=8.3cm,clip]{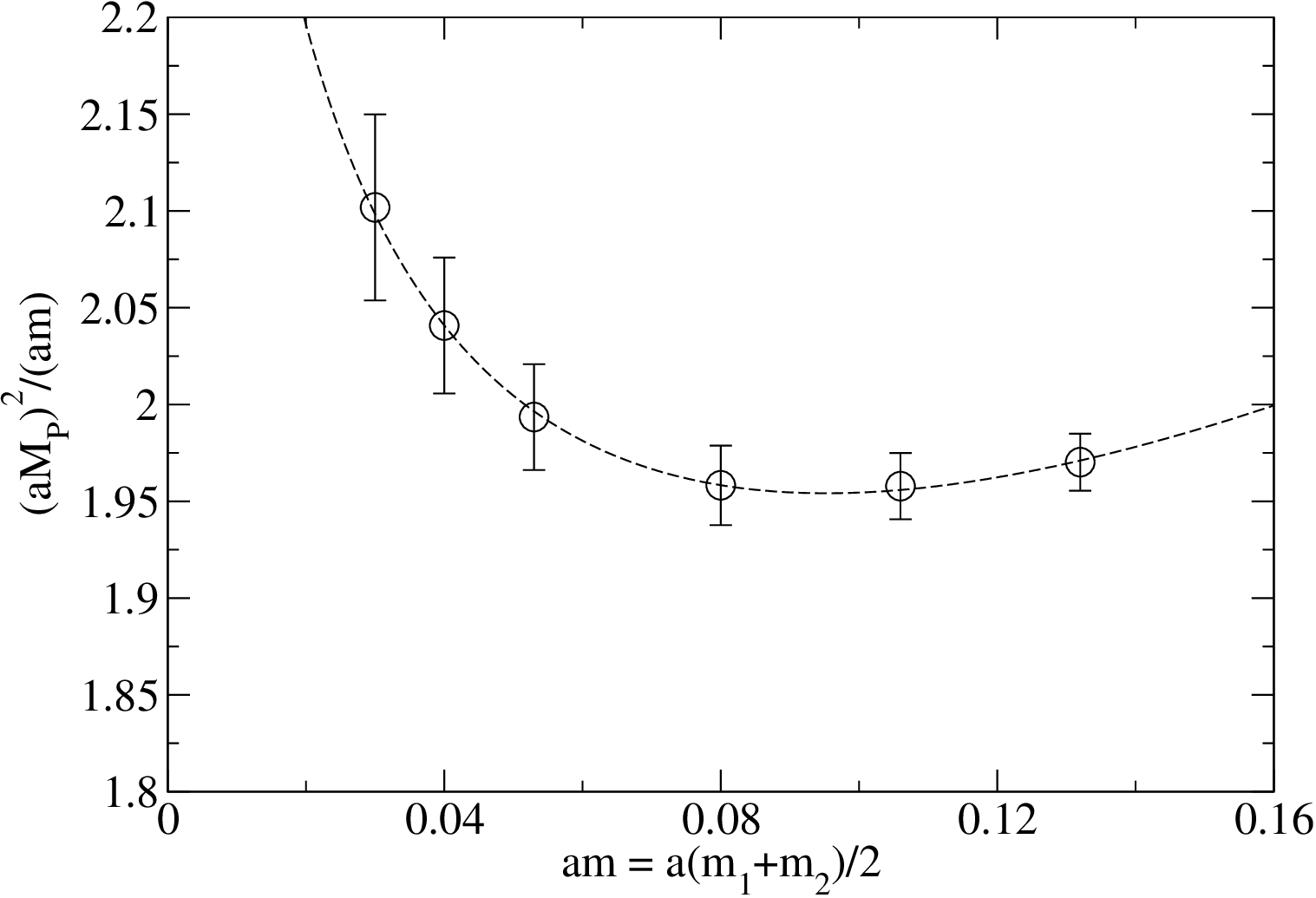}
\caption{Evidence for quenched chiral logs.}
\label{fig-msq1-PP-1448}
\efig

A fit of the degenerate quark mass $PP$ channel results to the expression 
$(aM_P)^2=A(am)^{1/(1+\d)}+B(am)^2$ of \req{eq-MaSqQcl} (see \refig{fig-msq1-PP-1448})
gives 
\beq
A=1.20(11)\;,\ \ B=2.66(41)\;,\ \ \d=0.17(4)\;.
\eeq

Considering the case of unequal quark masses, a fit to the expressions
of \reqs{eq-qclndy}{eq-qclndx} (see \refig{fig-PP-yx-1448}) gives $\d=0.22(4)$.

\bfig[tp]
\includegraphics*[width=8.3cm,clip]{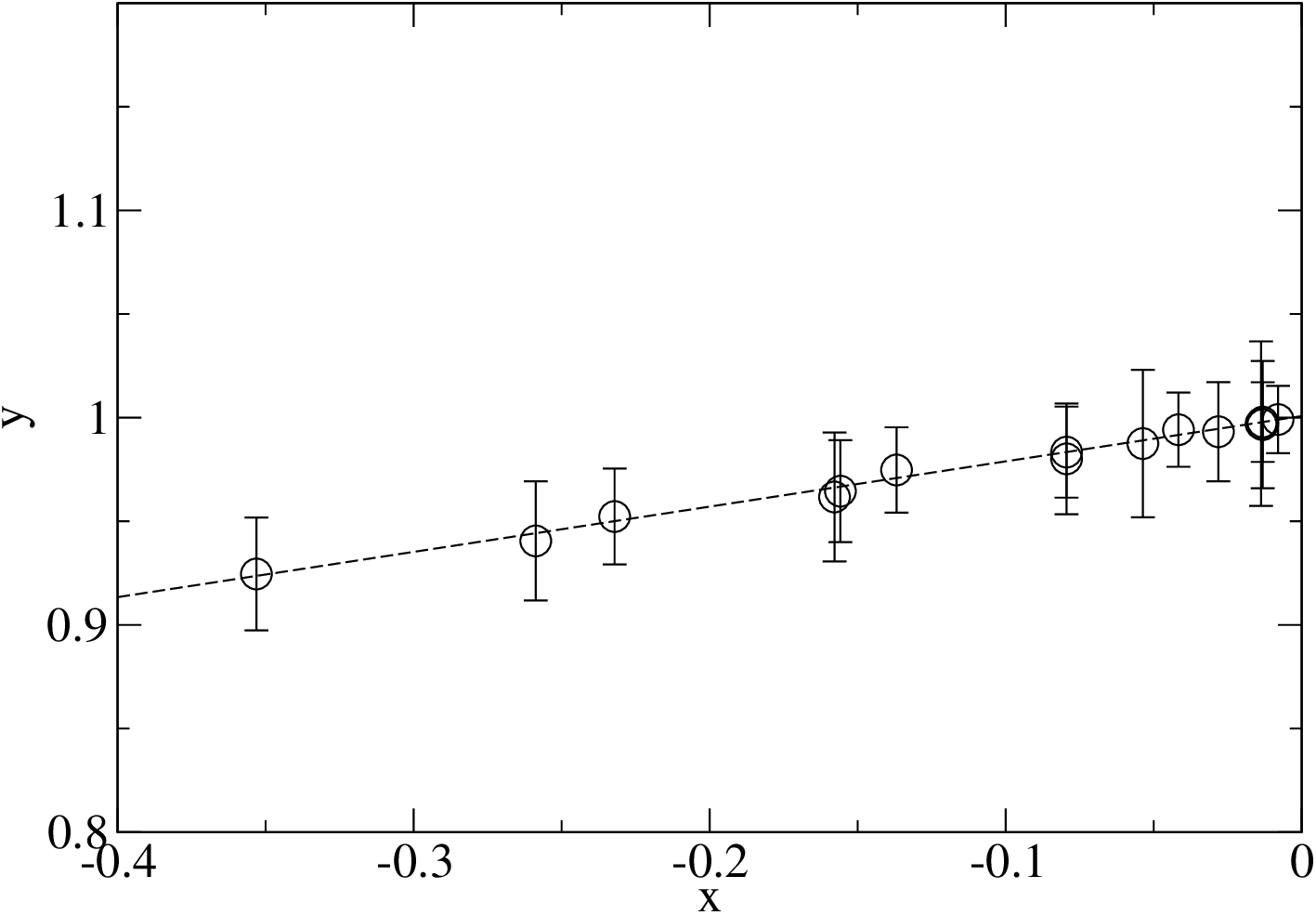}
\caption{Calculating the quenched chiral log parameter $\d$.}
\label{fig-PP-yx-1448}
\efig

As discussed in \rsec{subsec-QCL} for the results obtained on the
finer lattice, we will take the slope parameter $\cB$, obtained from
the linear fit of \reqs{eq-mpfit-1448}{eq-fpp-1448}, to be our
estimate of the coefficient of the leading term in the chiral
expansion of $(aM_P)^2$ as a function of quark mass $am$.

\subsection{Decay constants and determination of the lattice spacing}

As for the $18^3 \x 64$ lattice, we plot in \refig{fig-ainv-1448} the lattice
decay constant $af_P$ versus meson masses $(aM_P)^2$ as obtained by the
simulation. The continuous curve represents the physical ratio of these
quantities, $(f_K/m_K)_{exp} = 0.323$. It turns out that the mesons at our bare quark mass
$am =0.053$ correspond most closely to the physical
kaons.

\bfig[tp]
\includegraphics*[width=8.3cm,clip]{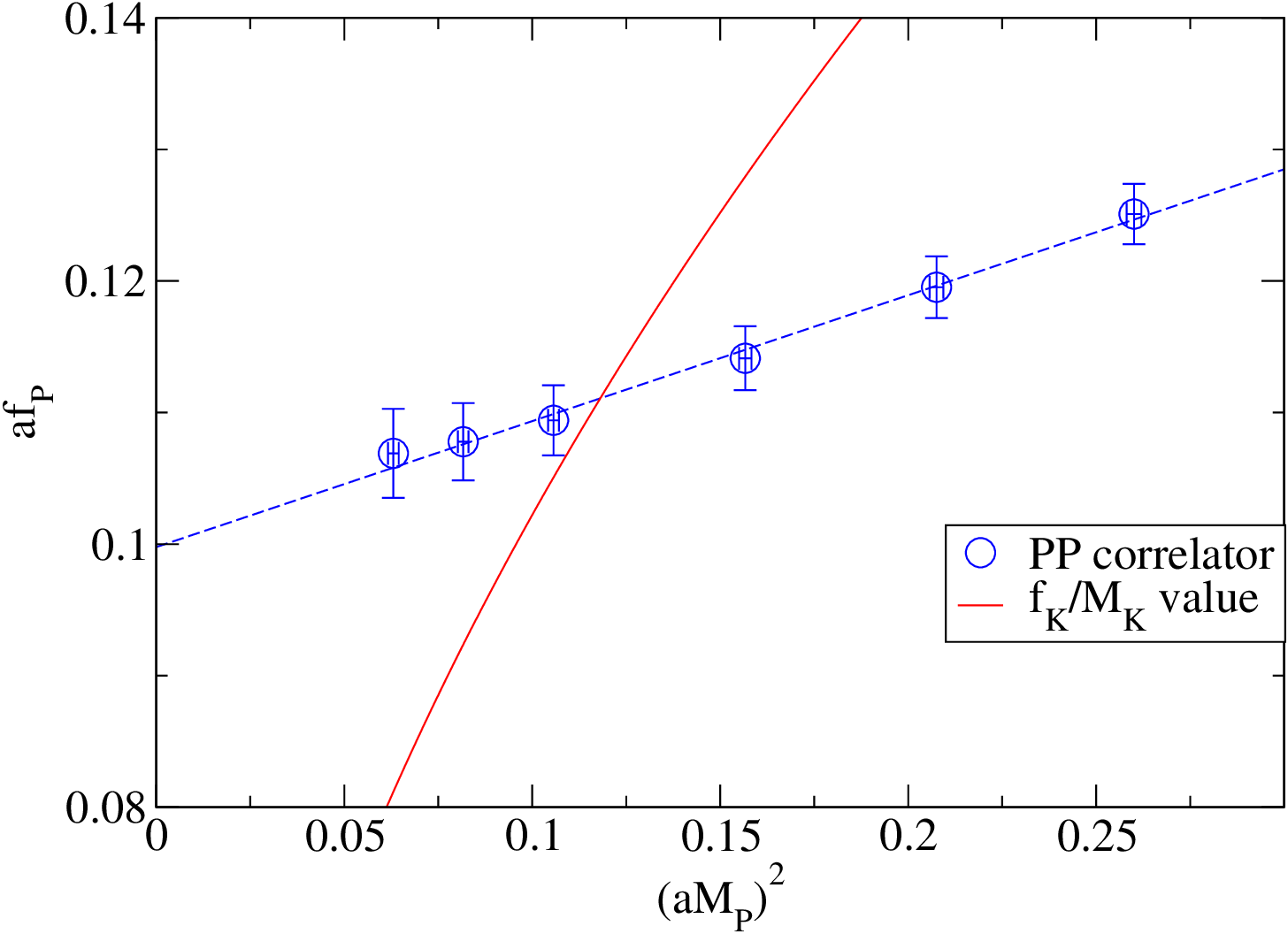}
\caption{Results for $f_P$ and determination of $a$ from the method
of physical planes. The dashed
line corresponds to a linear fit of the data for $a f_P$ in the interval 
$am \le 0.132$, the solid curve represents the parabola $(a f_P)/(a M_P)=
(f_K/M_K)_\text{exp}$.}
\label{fig-ainv-1448}
\efig

The intersection of the two curves gives
\beq
\label{eq-amk-1448}
a M_K= 0.343(9)\;, \qquad  a f_K=0.109(2)\;,
\eeq
from which, using the input $M_K=0.495\,\text{GeV}$, one gets 
\beq
\label{eq-aipp-1448}
a^{-1}=1.44(4)\,\text{GeV}\;.
\eeq

Use of the Sommer scale value of the lattice spacing and the experimental pion mass yields
\beq
af_\pi=0.100(3)
\eeq
and thus
\beq 
\fr{f_K}{f_\pi}=1.09(4)\;,
\eeq
which agrees with~\req{eq-fKfpi}.

As for the $18^3 \x 64$ lattice, the value for $a^{-1}$ derived with the method of 
lattice physical planes should be contrasted with the one derived 
from the Sommer scale, namely $a^{-1} = 1.61\,\text{GeV}$
at $\beta=5.85$. 
The lattice spacing obtained by extrapolating our results 
for the vector meson spectrum to the chiral limit, $a M_\rho =
0.605(29)$ (see \req{eq-vecmass-1448}), is
\beq
a^{-1} = 1.28(6)\,\text{GeV} \;.
\eeq

As for the finer lattice, the discrepancy between the three values for the
lattice spacing can be attributed to the quenched approximation, with
a relative error of $\Delta a^{-1}/a^{-1} \approx 20\%$.

\reqn{eq-amk-1448}, together with the linear fit of
\reqs{eq-mpfit-1448}{eq-fpp-1448}, which was argued for at the end of
\rsec{subsec-QCL-1448}, gives
\beq
\label{eq-bqm-mlpp-1448}
a(m_s +\hat{m}) = 0.1176(59) \;.
\eeq
As for the finer lattice, we note in passing that the 
value of \req{eq-bqm-mlpp-1448}, together
with our results for meson masses, give
\beq
\fr{M_{K^*}}{M_\r}=1.07(6)\;.
\eeq
Using the Sommer scale value of the lattice spacing and our pseudoscalar
spectrum data yields a slightly smaller value for $a(m_s+\hat{m})$ of
\beq
\label{eq-bqm-sommer-1448}
a(m_s+\hat{m}) = 0.0943(15)\;.
\eeq
We will use this value for the rest of the scaling analysis.

We again calculated the vector decay constant using
$af_V = aZ_V\sqrt{Z/M_V^2}$. In the fit of the $af_V$ data, we used only the points with
quark mass $am=0.106, 0.132, 0.33$, since the point-point correlators 
give poor results for $Z$ for lower quark masses.  Our results for
$a f_V$ are shown in \refig{fig-afV-1448}.
A linear fit of the data, along with the quark mass of \req{eq-bqm-sommer-1448}, yields
\beq
af_\r=0.175(13)\;,\ \ af_{K^*}=0.179(12)\;,\ \ \fr{f_{K^*}}{f_\r}=1.02(10)\;.
\eeq
The ratio $f_{K^*}/f_\r$ agrees well with both the value on the finer
lattice, \req{eq-fKstarfrho}, and the experimental value 
1.03(4)~\cite{Becirevic:1998jp}.

Using the Sommer scale value for the lattice spacing, the above 
results give in turn
\beq
f_\r=281(22)\,\MeV\;,\ \ \ f_{K^*}=287(19)\,\MeV\;.
\eeq

\bfig[tp]
\includegraphics*[width=8.3cm,clip]{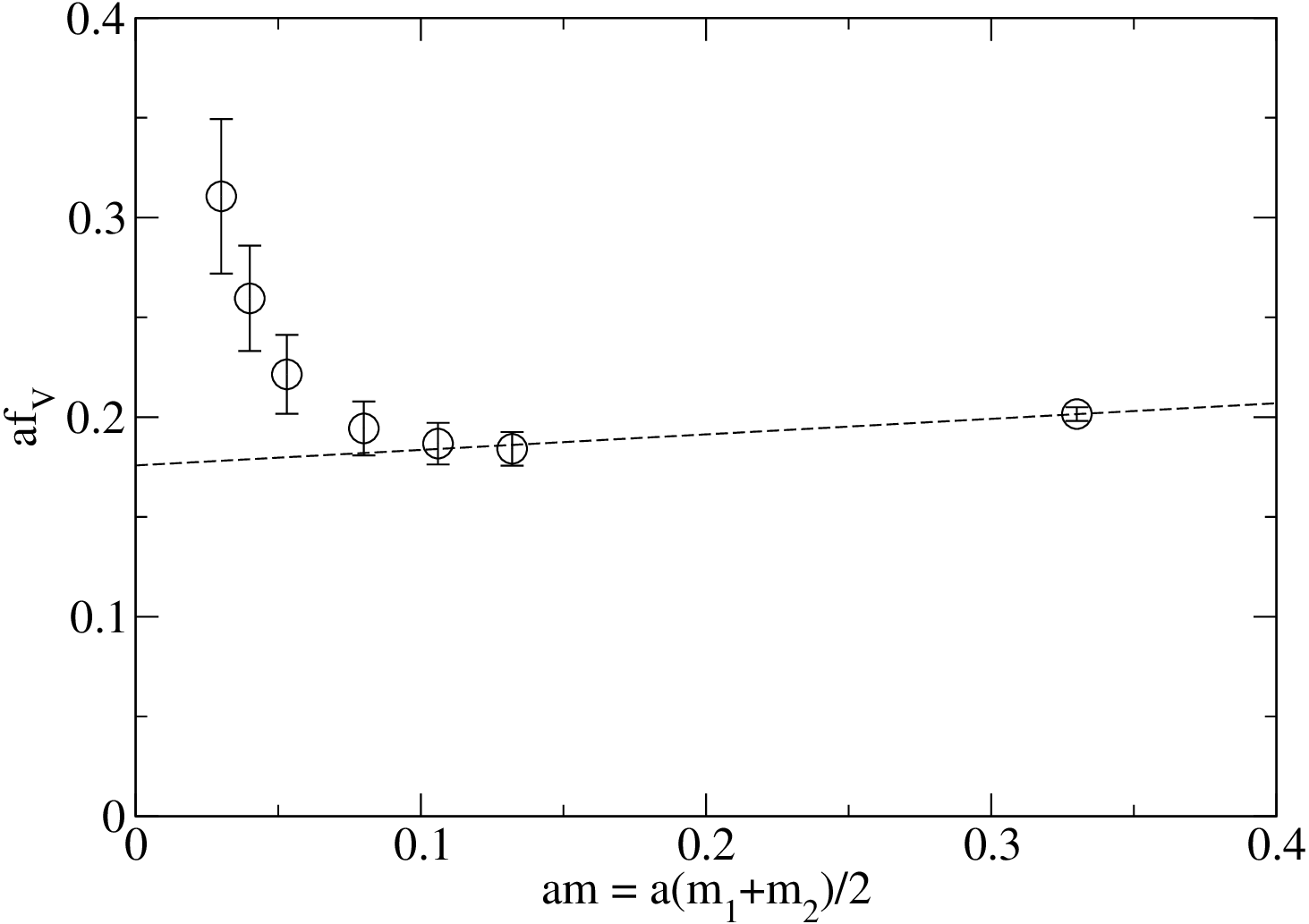}
\caption{
\label{fig-afV-1448}
Vector decay constant vs. bare quark mass.}
\efig

\subsection{Quark masses and chiral condensate}

As in \rsec{subsec-ChiralCond}, we calculate $Z_S$ in the RI-MOM scheme from
\beq
\label{eq-zsza-1448}
Z_S^{\text{RI}}(a\mu)=\lim_{m \to 0}Z_A\left.\frac{\Gamma_A(p,m)}{\Gamma_S(p,m)}\right|_{p^2=\mu^2}\;.
\eeq
Only data with $am \le 0.132$ were used for this purpose.
In order to obtain our final result, we used a combined fit to 
$am = 0.04$ and $am=0.132$ in the range $p^2=2-8\,\text{GeV}^2$ 
for central values and obtained an estimate of the systematic
error by varying the fit range and dropping additional terms
when indicated. We fit $Z_S$ to the same functional form 
as for $\beta=6.0$, yielding
\beq
Z_S^{\text{RI}}(2\,\text{GeV})=1.29(3)(14)
\eeq
and
\beq
\label{eq-zsmsb-1448}
Z_S^{\overline{\text{MS}}}(2\,\text{GeV})=1.48(3)(16)\;.
\eeq

Putting together \reqs{eq-bqm-sommer-1448}{eq-zsmsb-1448}, we obtain
\beq
(m_s+\hat{m})^{\overline{\text{MS}}}(2\,\text{GeV})=102(3)(16)\,\text{MeV}
\eeq
for the sum of the strange and light quark masses.
As for the finer lattice, using the value $m_s/\hat{m}=24.4(1.5)$ from chiral perturbation
theory~\cite{Leutwyler:1996qg}, we obtain
\beq
m_s^{\overline{\text{MS}}}(2\,\text{GeV})=98(3)(15)\,\MeV
\eeq
for the strange quark mass, which agrees well with the quenched lattice
world average~\cite{Lubicz:2000ch,Durr:2005ik,Gattringer:2005ij}.

Using the GMOR inspired relation of \req{eq-gmorb} and our data for $f_P$ we get
\beq
a^3 \chi(a)= -0.00473(28)
\label{eq-cond1-1448}
\eeq
or
\beq
\chi(a)= -0.0195(12)\,\text{GeV}^3 \;.
\label{eq-cond1gev-1448}
\eeq
Note that using the value $f_\chi = 0.123 \,\text{GeV}$~\cite{Amoros:2001cp}
would instead give $\chi(a)= -0.0117(1)\,\text{GeV}^3$.

Using $\langle \bar \psi \psi \rangle^{\overline{\text{MS}}}=
Z_S^{\overline{\text{MS}}} \chi(a)$, we finally get
\beqa
\langle \bar \psi \psi \rangle^{\overline{\text{MS}}}(2\,\text{GeV})
&=&-0.0292(18)(98) \,\text{GeV}^3 \nn\\
&=& - [308(6)(34) \,\text{MeV}]^3 \;.
\label{eq-cond1ren-1448}
\eeqa

\bfig[tp]
\includegraphics*[width=8.3cm,clip]{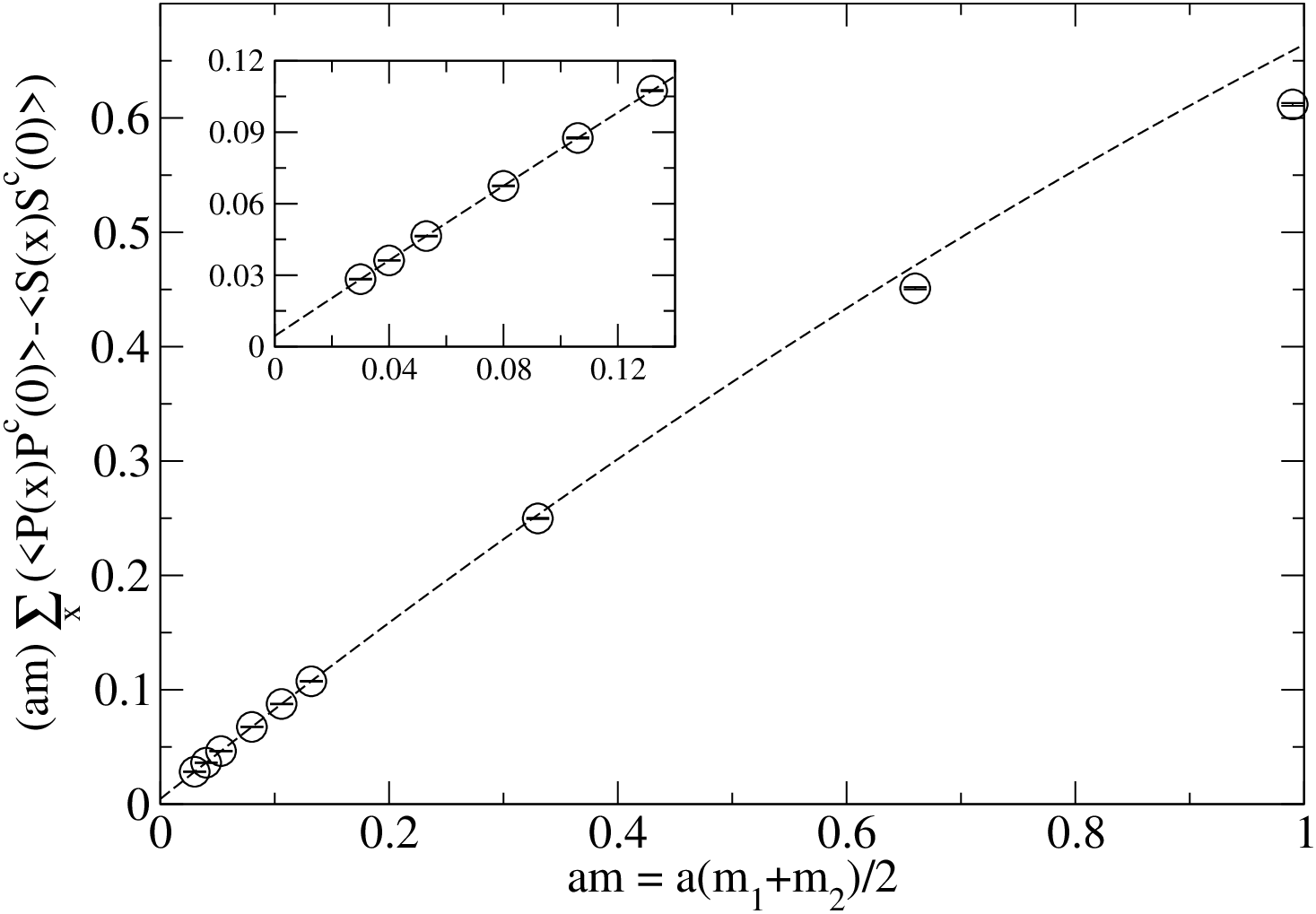}
\caption{Direct determination of the chiral
condensate. The curve corresponds to a quadratic fit of the data
according to \req{eq-ccd-1448} in the interval $am \le 0.132$.}
\label{fig-ccdirect-1448}
\efig

Our results for the direct calculation of $-a^3\tilde{\chi}(m)$ using
\beq
-a^3\tilde{\chi}(m)=am\sum_x(\langle P(x)P^c(0)\rangle-
\langle S(x)S^c(0)\rangle)
\eeq
are shown in \refig{fig-ccdirect-1448}.
The quadratic fit discussed in \rsec{subsec-ChiralCond},
\beq
\label{eq-ccd-1448}
-a^3\tilde{\chi}(m)=-a^3\chi+{\cal B}(am)+{\cal C}(am)^2\;,
\eeq
gives the result
\beq
-a^3\chi=0.00448(22)\;,\ \ 
{\cal B}=0.798(4)\;,\ \ 
{\cal C}=-0.14(2)\;.
\eeq
The value we obtain in this way for $ a^3 \chi$ is compatible
with the value in \req{eq-cond1-1448}. The renormalized value is
\beqa
\langle \bar \psi \psi \rangle^{\overline{\text{MS}}}(2\,\text{GeV})
&=&-0.0277(15)(93) \,\text{GeV}^3 \nn\\
&=& - [302(5)(34) \,\text{MeV}]^3 \;.
\eeqa

\subsection{Direct comparison of the two lattices}
\label{subsec-mesoncomp}

We compare in \refigsdash{fig-MPSq_scaling_PP}{fig-MV_scaling} our
results for the pseudoscalar and vector spectra for the finer and
coarser lattices, using the Sommer scale value for the lattice spacing
to express masses in physical units. We neglect logarithmic effects in
the lattice spacing and plot the mass spectra as a function of bare
quark mass.  It is interesting to observe that our results for the mass spectra on
the two different lattices are very similar. A
qualitatively equivalent conclusion would be reached with the
renormalized quark mass.  Our results suggest that the scaling
violations for the quantities that we consider may be quite small.
However, one has to keep in mind, that we only have data for
two values of the lattice spacing and that the mass parameter $\rho$
is not held fixed (for a detailed discussion of this point see Appendix~\ref{sec-Overlap}).

\reft{tab-scaling} in Appendix~\ref{sec-Tables} shows a direct comparison of data from the
two lattices.

\bfig[tp]
\includegraphics*[width=8.3cm,clip]{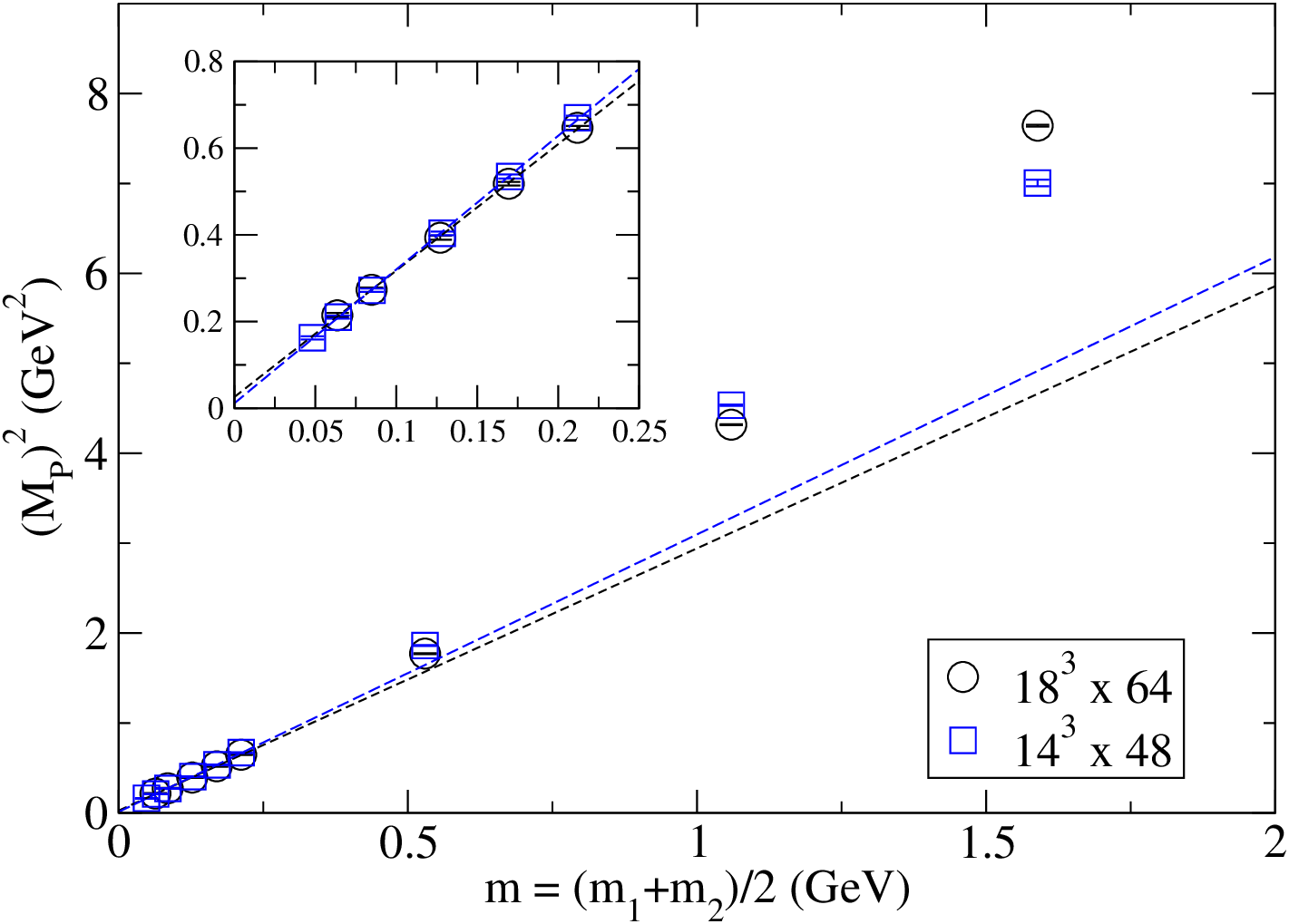}
\caption{Pseudoscalar $PP$ spectrum comparison.}
\label{fig-MPSq_scaling_PP}
\efig

\bfig[tp]
\includegraphics*[width=8.3cm,clip]{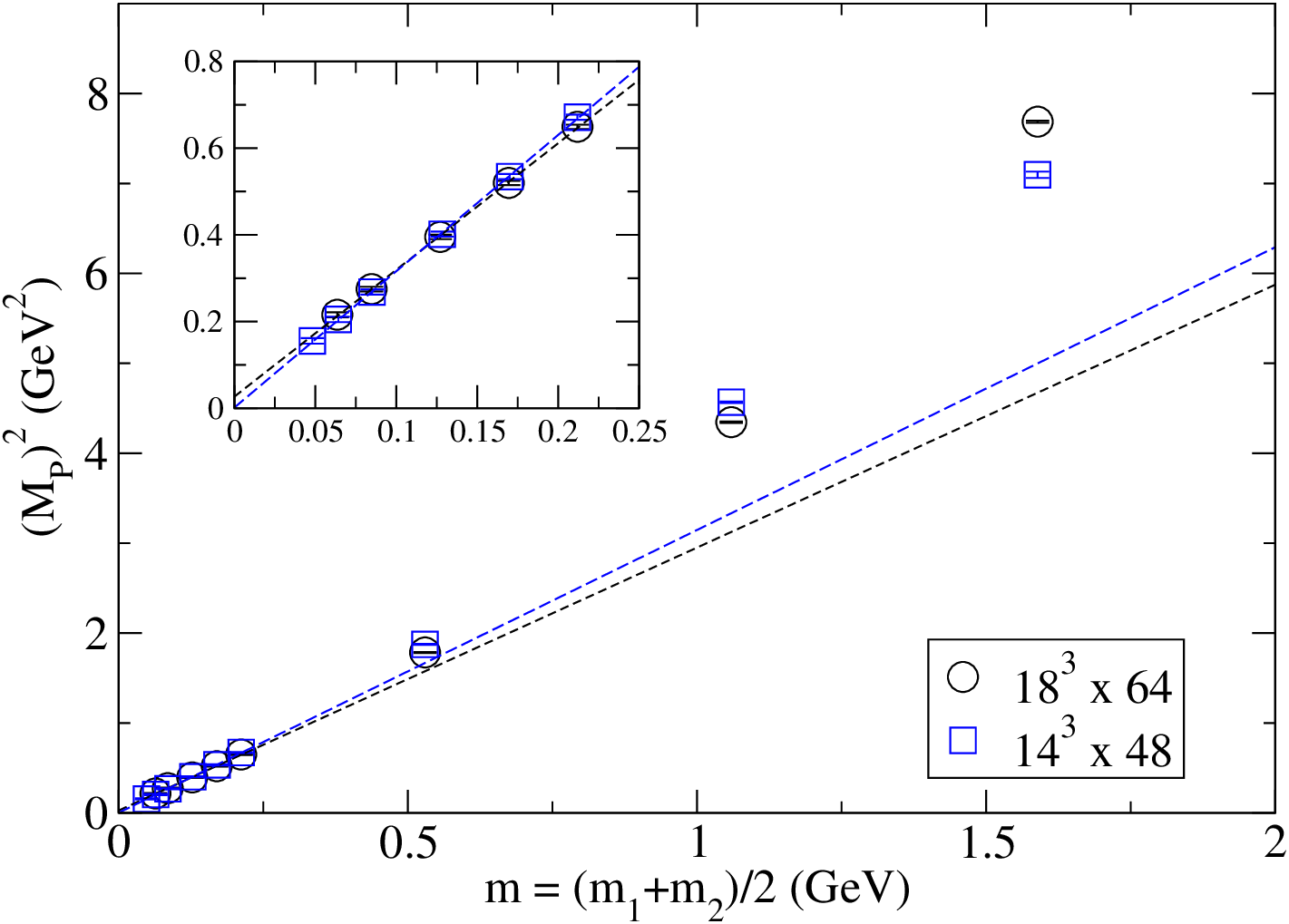}
\caption{Pseudoscalar $PP-SS$ spectrum comparison.}
\label{fig-MPSq_scaling_PP-SS}
\efig

\bfig[tp]
\includegraphics*[width=8.3cm,clip]{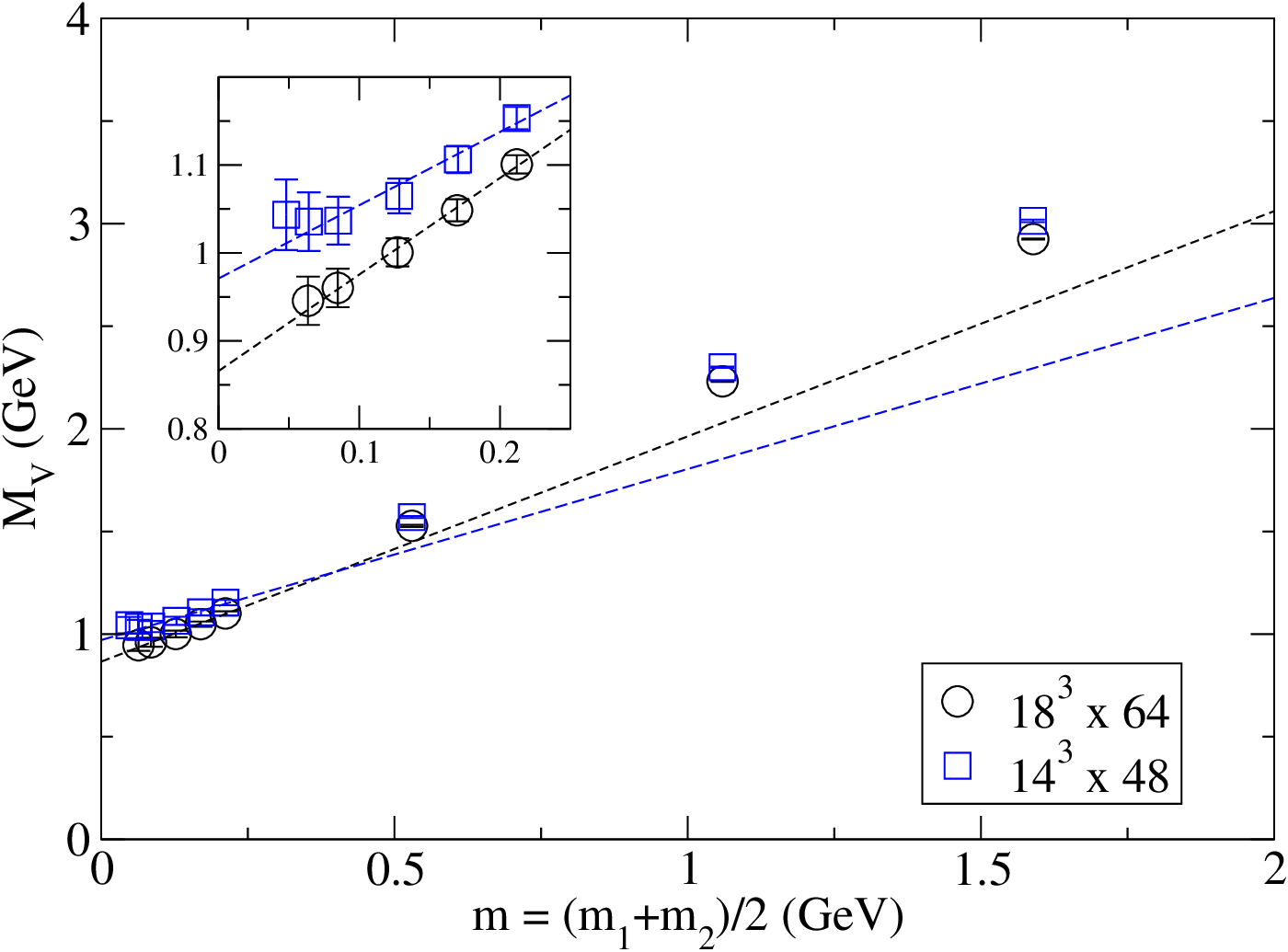}
\caption{Vector (extended sink) spectrum comparison.}
\label{fig-MV_scaling}
\efig

\section{Light Baryon Spectra}
\label{sec-LightBaryons}
\subsection{Baryon spectra}

The increased temporal extent of our lattice with respect to that in 
\cite{Giusti:2001pk} has made possible the calculation of the light baryon
spectrum.  Despite the challenges of limited statistics and the use
of point sources and sinks, the lightest octet and decuplet masses were
measured, together with those of the corresponding negative-parity states.
For other recent determinations of the baryon spectrum with chiral fermions,
see \cite{Mathur:2003zf,Dong:2001kv,Lee:2002gn,Galletly:2003vf,Chiu:2005zc,Aoki:2004ht,Sasaki:2001tg,Sasaki:2001nf}.

Baryon correlation functions were constructed with the following interpolating
operators.  For the octet, we use
\beq
O^{f_1 f_2 f_3}_\alpha=\epsilon_{abc}(\psi^{f_1 T}_a C\gamma_5 \psi^{f_2}_b)
                       \psi^{f_3}_{c,\alpha}\;,
\eeq
where $C=\gamma_2\gamma_4$ is the charge conjugation matrix, $\alpha$ is a
Dirac index, $f_i$ denotes flavor, and $a,b,c$ denote color.  For the decuplet
states, we use a notation similar to that in \cite{Aoki:2002fd} and define
\beq
\Gamma_\pm=(\gamma_2\mp i\gamma_1)/2\;, \qquad \Gamma_0=i\gamma_3\;.
\eeq
We work in a representation of the Dirac matrices where 
$\gamma_4=\mathrm{diag}(1,1,-1,-1)$.
Labeling the channels by $J_z$, the decuplet operators are then given by
\beqa
D_{3/2}^{f_1 f_2 f_3}&=&\epsilon_{abc}(\psi^{f_1 T}_a C\,\Gamma_+ \psi^{f_2}_b)
                        \psi^{f_3}_{c,\alpha=1}\;,\\
D_{-3/2}^{f_1 f_2 f_3}&=&\epsilon_{abc}(\psi^{f_1 T}_a C\,\Gamma_- \psi^{f_2}_b)
                        \psi^{f_3}_{c,2}\;,         \\
D_{1/2}^{f_1 f_2 f_3}&=&\epsilon_{abc}[(\psi^{f_1 T}_a C\,\Gamma_0 \psi^{f_2}_b)
                        \psi^{f_3}_{c,1} \nn\\
           & & \qquad +(\psi^{f_1 T}_a C\,\Gamma_+ \psi^{f_2}_b)\psi^{f_3}_{c,2}]/3\;,   \\
D_{-1/2}^{f_1 f_2 f_3}&=&\epsilon_{abc}[(\psi^{f_1 T}_a C\,\Gamma_0 \psi^{f_2}_b)
                        \psi^{f_3}_{c,2} \nn\\
           & & \qquad + (\psi^{f_1\,T}_a C\,\Gamma_- \psi^{f_2}_b) \psi^{f_3}_{c,1}]/3\;.
\eeqa

Without loss of generality, we may assume that the three quark flavors are
distinct and for simplicity call the flavors $u,d,s$.  For the results
that follow, two quarks are always taken to have the same mass, and assigning
them identical flavor would merely change the normalization of the
correlator.  The two possible octet states may be identified with the
$\Sigma^0$ and $\Lambda^0$ and are given by
\beqa
\Sigma_\alpha&=&O_\alpha^{dsu}+O_\alpha^{usd}\;, \\
\Lambda_\alpha&=&O_\alpha^{dsu}-O_\alpha^{usd}-2O_\alpha^{uds}\;.
\eeqa
These give identical correlators only when all three quark masses are
degenerate.  When propagating forward, the $\alpha=1,2$ components of these
operators have positive-parity, while the $\alpha=3,4$ components have 
negative-parity.  To extract a mass, we must project out states of definite parity
by defining
\beq
G_{\pm}(t)=\sum_\mathbf{x} \langle 0|\Sigma_\alpha(\mathbf{x},t)
\big(\frac{1\pm\gamma_4}{2}\big)_{\alpha\beta} 
\bar\Sigma_\beta(\mathbf{0},0)|0\rangle\;.
\eeq
The mass $m_+$ of the positive-parity state is then extracted by fitting
$A e^{-m_+ t} = G_+(t)+G_-(T-t)$ for an appropriate range of times $t$
(where we can neglect the backward-propagating negative-parity contribution to
$G_+(t)$ and vice versa).  Here $T$ is the total extent of the lattice in the
time direction.  Similarly, $A e^{-m_- t} = G_-(t)+G_+(T-t)$ yields the mass
of the negative-parity state.

For the decuplet we use
\beq
(\Sigma^*)_{J_z}=D_{J_z}^{uds}+D_{J_z}^{sud}+D_{J_z}^{dsu}
\eeq
and combine correlators for the four spin states with the corresponding
time-reversed correlators of opposite parity.  The negative-parity operators
are defined by replacing the Dirac index of $\psi^{f_3}$ in
$D_{J_z}^{f_1 f_2 f_3}$ according to $1 \rightarrow 3$, $2 \rightarrow 4$.

Here we have followed convention by defining our operators in covariant form.
We note, however, that one may also work directly with components and
construct the spin wave functions explicitly.  For example, using the notation
$\ket{\alpha\beta\gamma}=\epsilon_{abc}\psi^u_{a,\alpha} \psi^d_{b,\beta}
\psi^s_{c,\gamma}$, alternate operators for the two spin-up octet states are
\beqa
\Lambda &=& (\ket{121}-\ket{211})/\sqrt{2}\;, \\
\Sigma  &=& (\ket{121}+\ket{211}-2\ket{112})/\sqrt{6}\;.
\eeqa
By construction, these have only terms with upper components, whereas the
operators given in covariant form also include terms involving lower components
(e.g.~$\ket{341}$).  Contributions from these additional terms are suppressed
since the lower components vanish in the non-relativistic limit, and the two
types of operators were in fact found to give compatible results for both
the octet and decuplet.  For the results presented here, only the covariant
forms were used.

Baryon masses were calculated at $\beta=6$ with two degenerate quarks having
each of the five lightest available masses
($am_1=am_2=0.03,0.04,0.06,0.08,0.1$) and for all available masses 
of the third quark (including 0.25, 0.5, and 0.75).  Errors were estimated by
a bootstrap procedure where correlators for the four or eight spin channels are
grouped by configuration prior to sampling.  Fitting windows were chosen based
on plots of the effective mass $M_{eff}=\mathrm{ln}[C(t-a)/C(t)]$ for
$am_1=am_2=am_3=0.03$.  The preferred window (used below for chiral fits) was
determined by choosing $t_{min}$ such that $M_{eff}(t_{min})$ and
$M_{eff}(t_{min}+a)$ are compatible to within $1\,\sigma$ and $t_{max}$ such
that the error in $M_{eff}(t_{max})$ (determined by the bootstrap method) does not exceed
30 percent.  The latter criterion was made more stringent for the
positive-parity decuplet state for reasons described below.  The windows
chosen on the basis of the lightest quark mass were then used for all other
quark masses.

\begin{figure}[tp]
\includegraphics*[width=8.3cm,clip]{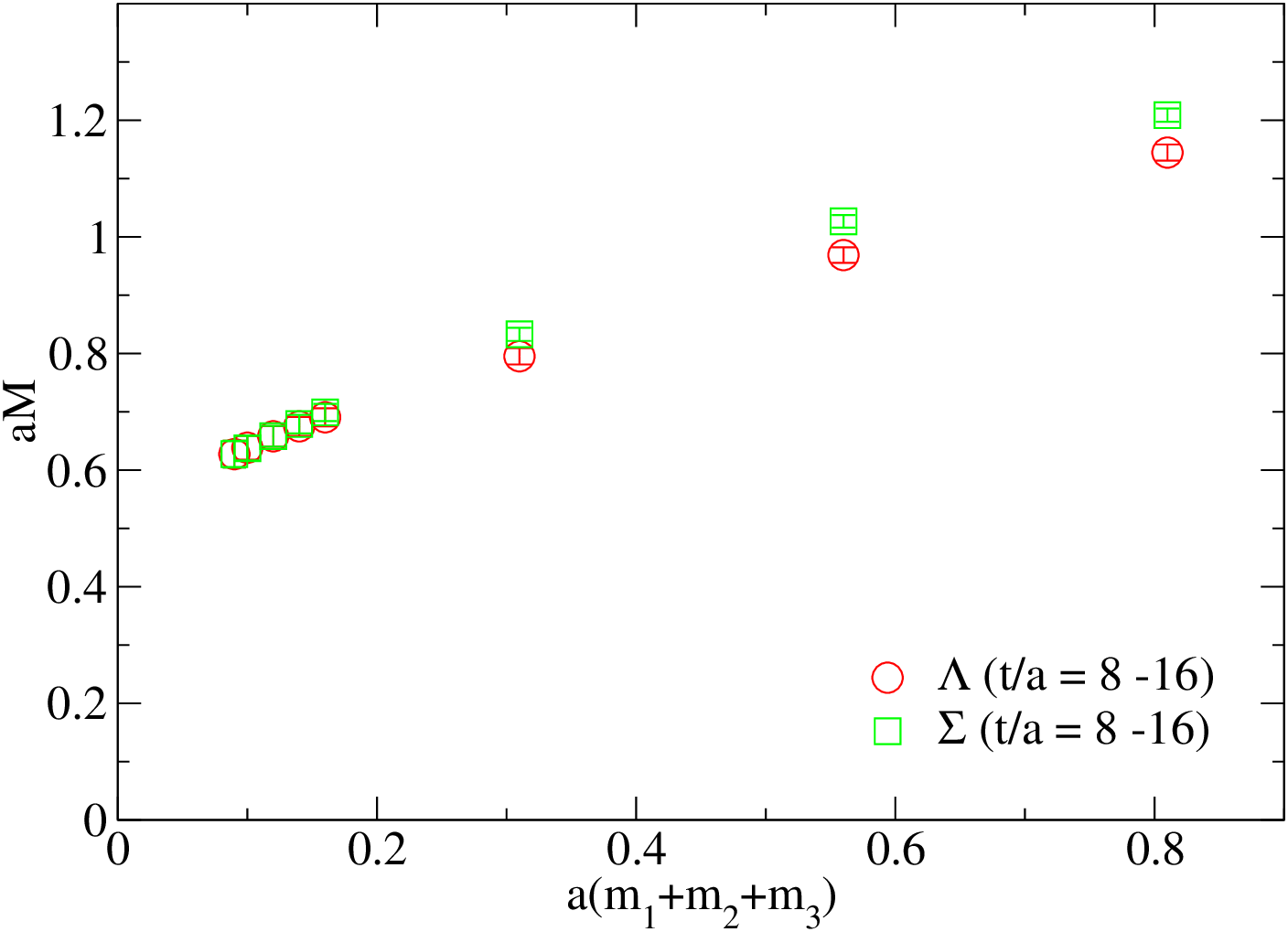}
\caption{Masses of the two octet states with quark masses $am_1=am_2=0.03$}
\label{fig-octets}
\end{figure}

For the octet states, the windows were $8 \le t/a \le 16$ for $J^P=\frac{1}{2}^+$ and
$6 \le t/a \le 8$ for $J^P=\frac{1}{2}^-$.
Data for a range of windows are provided in \reft{tab-octet} in Appendix~\ref{sec-Tables}.
In \refig{fig-octeta}, we plot the $\Lambda$-like octet masses as a
function of total quark mass.  Measurements for two values of $t_{min}$ are
shown in order to give some indication of the dependence on fitting window.
\refigure{fig-octets} shows the splitting between the two octet states as the
mass of the third quark is increased away from $am_1=am_2=0.03$.

\begin{figure}[tp]
\includegraphics*[width=8.3cm,clip]{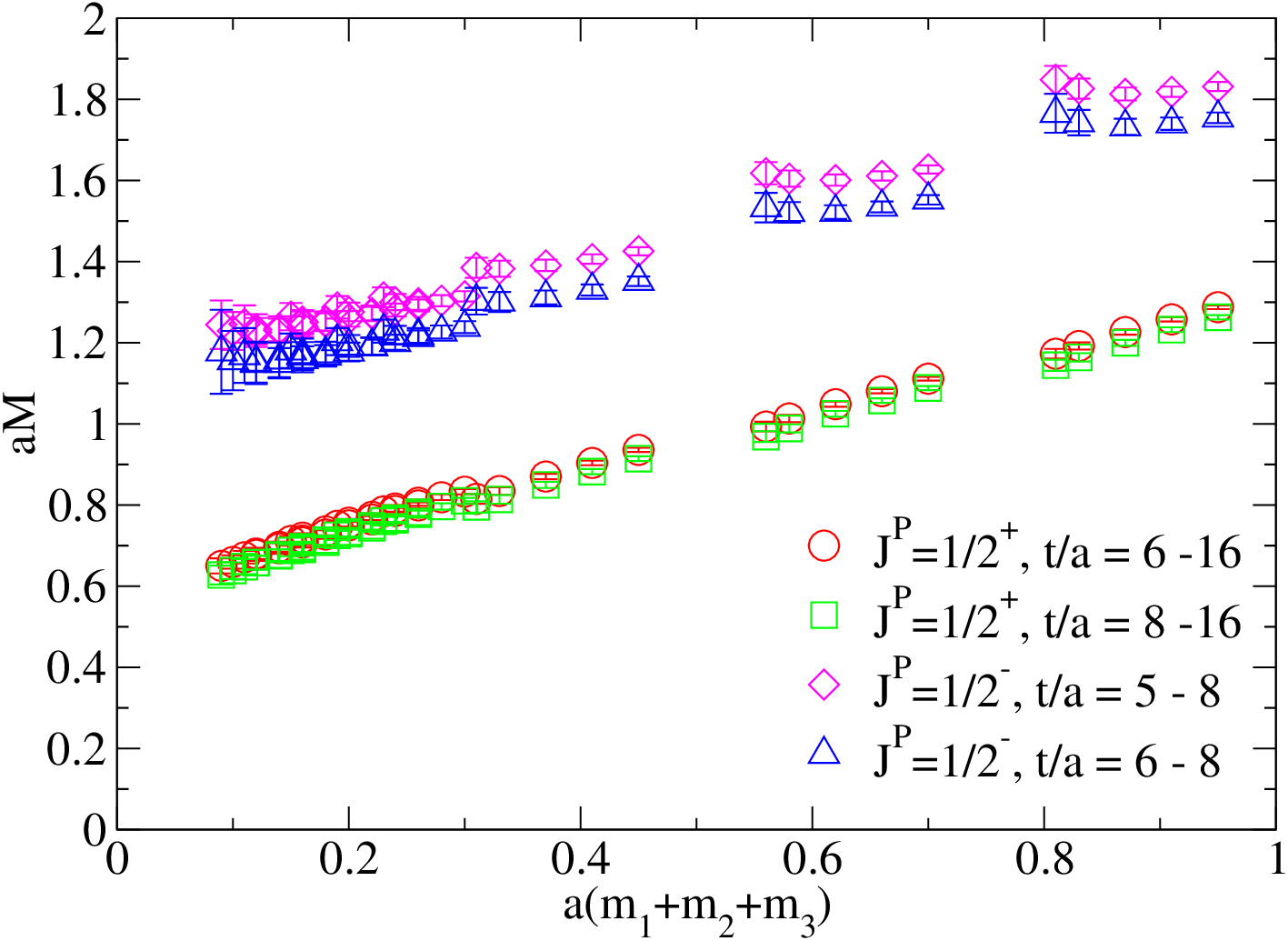}
\caption{Positive- and negative-parity octet masses for two
fitting windows and quark masses $m_1=m_2$ degenerate.}
\label{fig-octeta}
\end{figure}

\begin{figure}[tp]
\includegraphics*[width=8.3cm,clip]{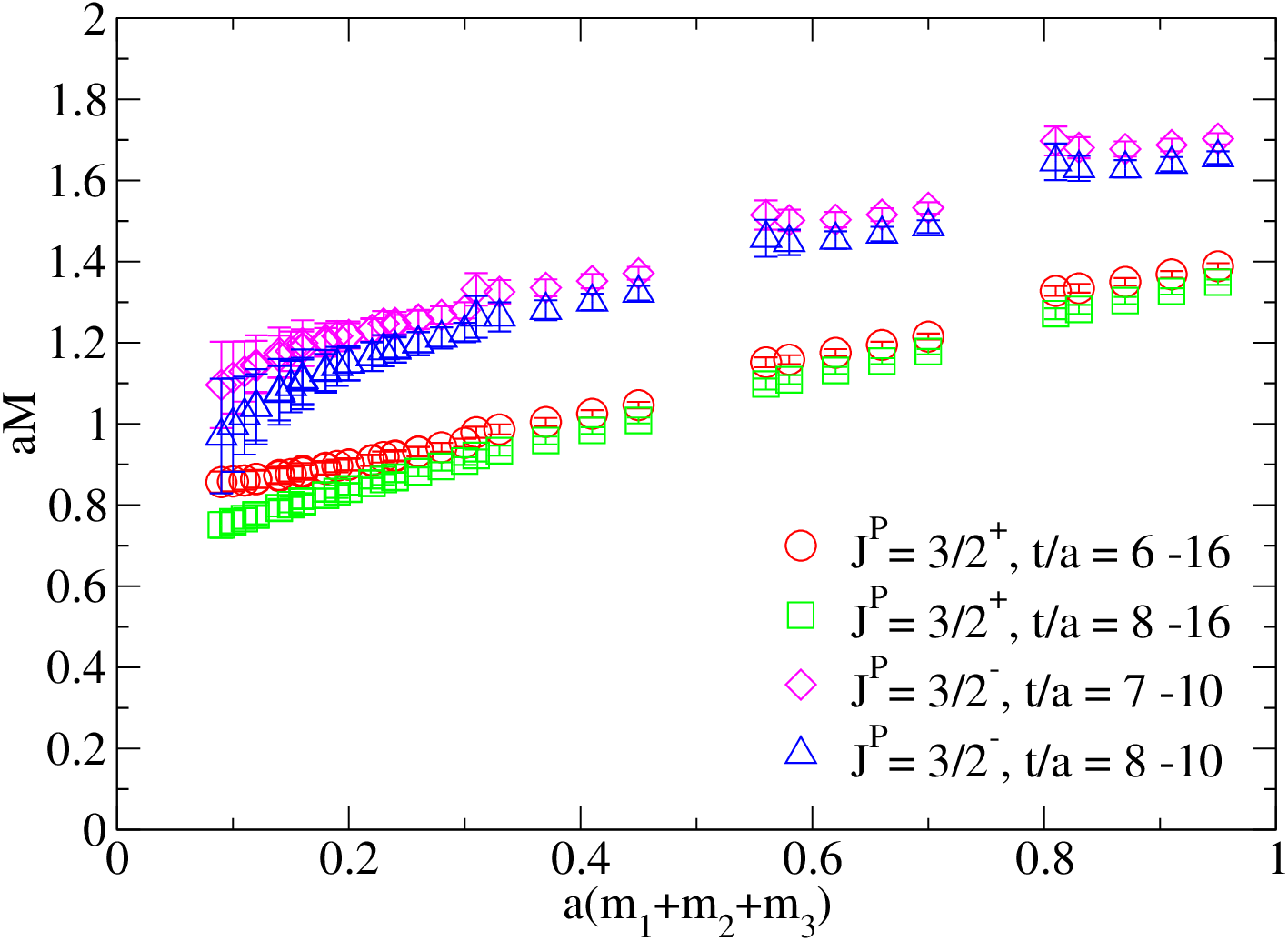}
\caption{Positive- and negative-parity decuplet masses for two fitting
windows and quark masses $m_1=m_2$ degenerate.}
\label{fig-decuplet}
\end{figure}

Results for the decuplet states are shown in \refig{fig-decuplet}.
The data are reproduced for a range of fitting windows in \reft{tab-decuplet}.
We note that the masses of the heavier states
($J^P=\frac{1}{2}^-, \frac{3}{2}^\pm$) exhibit a substantial dependence on the
fitting window.  In particular, the effective mass of the positive-parity
decuplet state is seen to first plateau and then, at large times, continue
downward toward the octet mass.  This behavior may
be explained by considering what gives rise to the more rapid fall-off
of correlators for excited states.  Correlators for the octet and the
decuplet are constructed from the same ingredients, the same quark propagators.
The more rapid fall-off of the decuplet arises because of cancellations
between terms.  A fit of the correlator gives a reliable determination of
the mass only insofar as these cancellations are not overwhelmed by
fluctuations.  At large times, the remaining fall-off is due primarily to
the constituent masses of the quarks rather than the baryon mass itself.
For this reason, we have constrained $t_{max}$ to the value used for the
octet, giving the fitting window $8 \le t/a \le 16$.  The uncertainty associated with
this choice of window is relatively large and of the same order as the
statistical error.  The window for the $J^P=\frac{3}{2}^-$ state is $8
\le t/a \le 10$.
We also note that the windows for both negative-parity states are rather 
small due to the early onset of fluctuations.  For all of these reasons, our
results for the negative-parity octet and decuplet states should be considered
to have indicative value only.

\begin{figure}[tp]
\includegraphics*[width=8.3cm,clip]{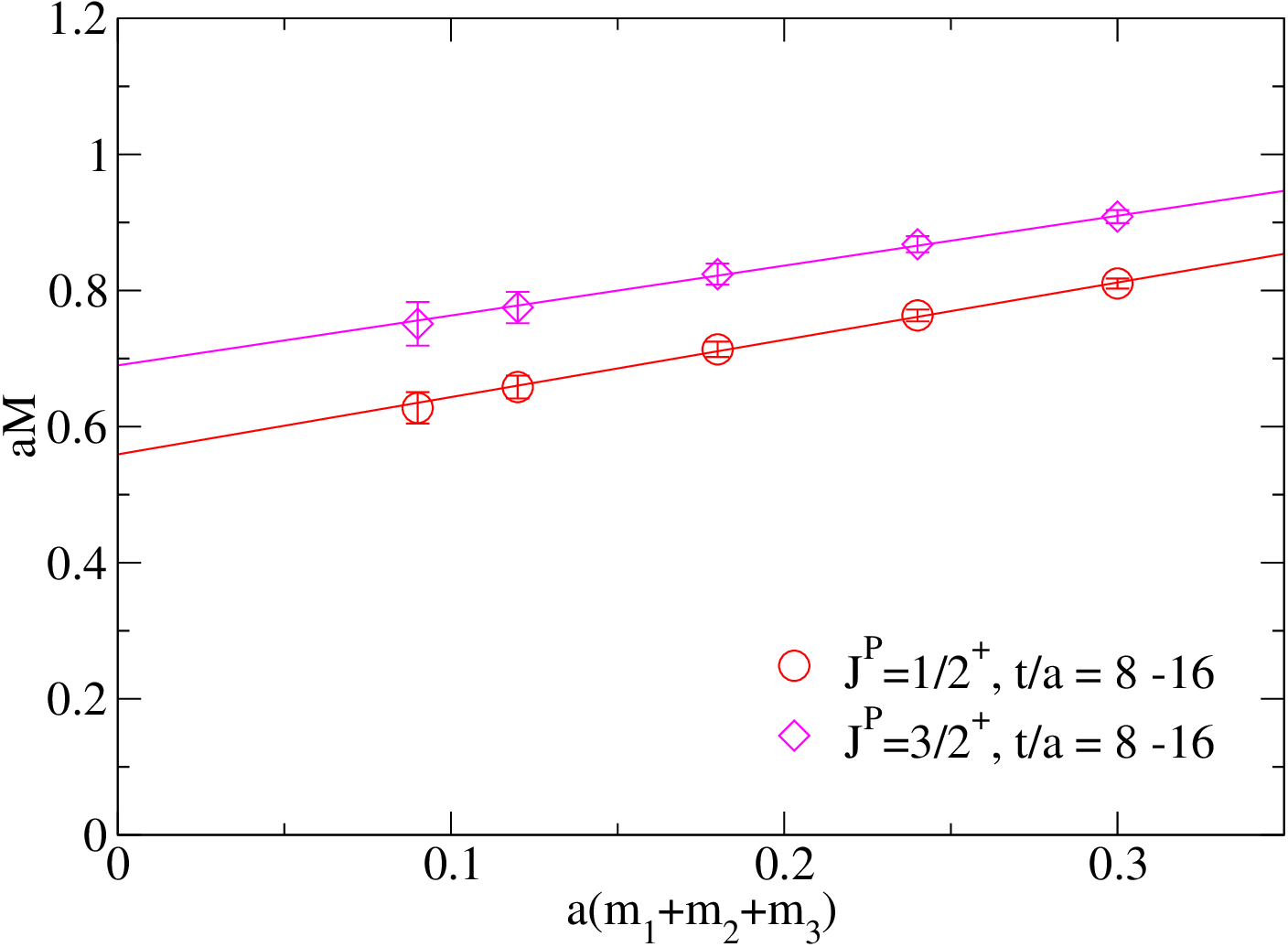}
\caption{Baryon masses with three light degenerate quarks were extrapolated
to the chiral limit with assumed linear dependence on quark mass.}
\label{fig-barfit}
\end{figure}

Finally, in \refig{fig-barfit} we plot the masses of the positive-parity
states for light degenerate quark masses
$am_1=am_2=am_3=0.03,0.04,0.06,0.08,0.1$.  A linear extrapolation to the chiral
limit gives $aM_8=0.559(24)$ and $aM_{10}=0.690(32)$.  In this limit,
using the value of $aM_\rho$ from~\req{eq-vecmass}, we find
\beqa
M_8/M_\rho &=& 1.37(8)\;, \\
M_{10}/M_8 &=& 1.23(8)\;.
\eeqa
The limited statistics of our data do not warrant a more complicated
fitting form. Experimentally, $M_N/M_\rho=1.21$ and $M_\Delta/M_N=1.31$~\cite{PDBook}.

\subsection{Baryon scaling}

\begin{figure}[tp]
\includegraphics*[width=8.3cm,clip]{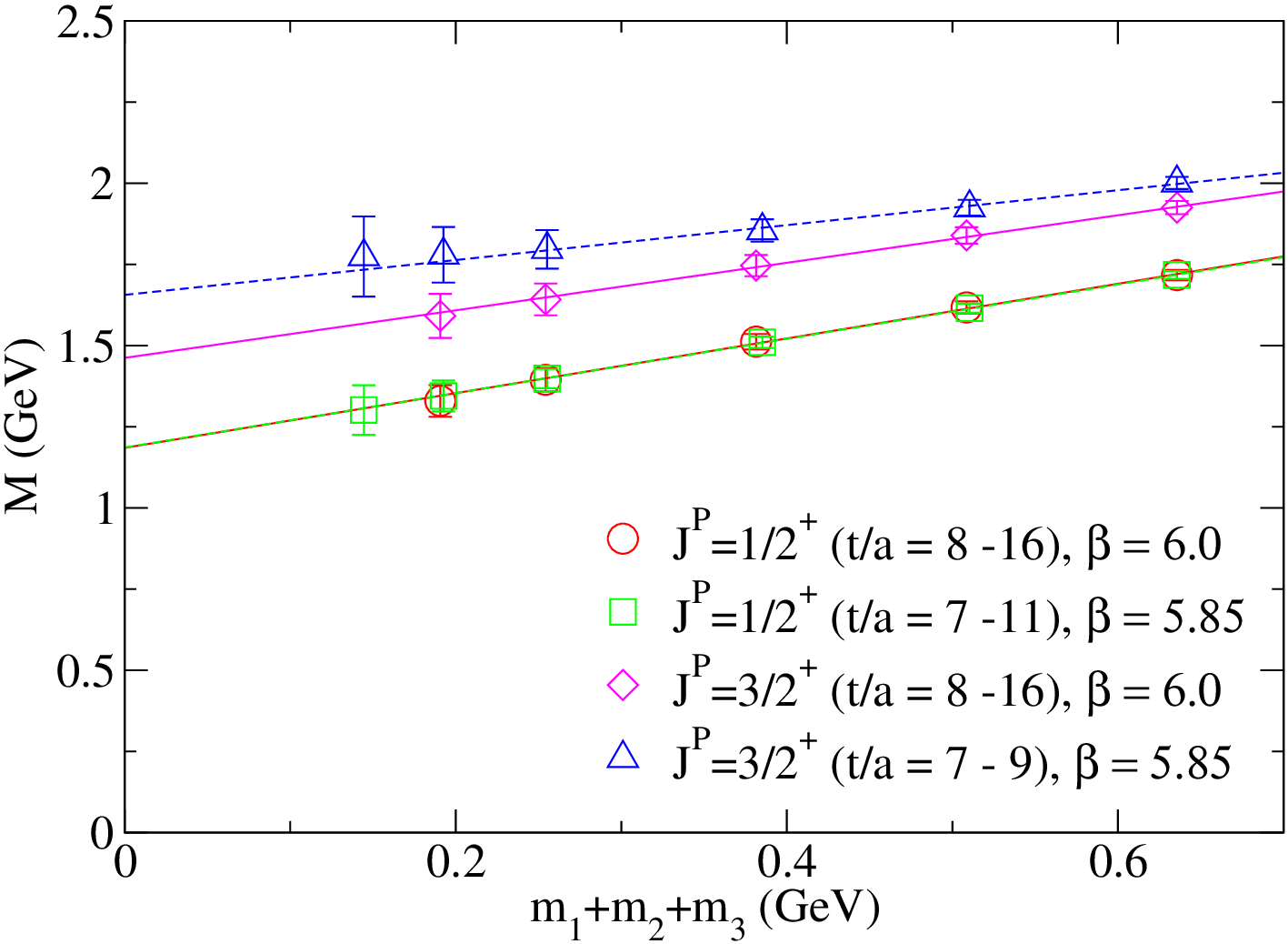}
\caption{Baryon masses with light degenerate quarks at two values of $\beta$,
with lattice spacing set by the Sommer scale.}
\label{fig-barscal}
\end{figure}

To investigate scaling, masses of the positive-parity states were also
calculated at $\beta=5.85$ on the $14^3 \times 48$ lattice.  In light of the
small fitting windows that had been required on the finer lattice, fits for the
corresponding negative-parity masses were not attempted.  Here we find
$aM_8=0.741(28)$ and $aM_{10}=1.033(55)$ in the chiral limit, yielding
\beqa
M_8/M_\rho &=& 1.23(8)\;, \\
M_{10}/M_8 &=& 1.39(9)\;.
\eeqa
As in the meson analysis, we make use of the Sommer scale,
which gives $a^{-1}=2.12$~GeV at $\beta=6$ and $a^{-1}=1.61$~GeV
at $\beta=5.85$. 
\refigure{fig-barscal} shows the $J^P=\frac{1}{2}^+$
and $J^P=\frac{3}{2}^+$ states for light degenerate quarks on both lattices.
Bare masses are rescaled by the corresponding values of $a^{-1}$.  
We see that the octet spectrum exhibits
good scaling while the decuplet shows some indication of scaling violation.
Comments in \rsec{subsec-mesoncomp} regarding
renormalized quark masses, however, apply here as well. Also, as we have already
observed, the decuplet masses suffer from
uncertainty in the choice of fitting window and so the apparent lack of
scaling should not be considered to have much significance.

\section{Diquark Correlations}
\label{sec-DiquarkCorr}
\subsection{Diquark correlators}

In the past two years, experiments have produced indications of bound quark systems beyond
the usual quark-antiquark mesons and three quark baryons
\cite{Nakano:2003qx,Barmin:2003vv,Stepanyan:2003qr,Barth:2003es,Hicks:2004vd}. The most
prominent example is the five particle $\Theta^+(1540)$
pentaquark. The reality of such states is in question, since a number
of attempts at verifying the pentaquark observations have failed
\cite{Bai:2004gk,Aubert:2004bm,Abe:2004wf,Armstrong:2004gp,
Abt:2004tz,Antipov:2004jz,Longo:2004gd,Litvintsev:2004yw}.

Nevertheless, lattice calculations should provide a definitive check
on whether QCD predicts such states. In certain models
\cite{Jaffe:2003sg} of the
$\Theta^+(1540)$, which consists of the bound valence quarks
$uudd\bar{s}$, the four quarks bind into two pairs of diquarks,
and the two diquarks bind with the
remaining $\bar{s}$.

Information on possible diquark states can be obtained by measuring
diquark correlations on the lattice \cite{Csikor:2003ng}.
A study of correlations inside
baryons, using a method similar to the one we used to investigate the
$q\bar{q}$ wave functions in \rsec{subsec-ExtSinkOps}, is in
progress.

The fact that our propagators were calculated in Landau gauge
allows us also to measure quark-quark correlations directly and to
fit their decay in Euclidean time in terms of an effective ``diquark
mass'' \cite{Hess:1998sd}. Of course, because of the gauge fixing, one should not consider
such a mass parameter to be the mass of a physical state. Nevertheless,
it can produce an indication of the relative strength of quark
bindings inside diquark states. We consider correlations for the
diquark operators
\beq
\cO_{c}^{s_1s_2}(x)=\e_{cc_1c_2}\y_{c_1}^{s_1}(x)\,\y_{c_2}^{s_2}(x)\;,
\eeq
and
\beq
\cO_{c_1c_2}^{s_1s_2}(x)=\fr{1}{\sr{2}}
(\y_{c_1}^{s_1}(x)\,\y_{c_2}^{s_2}(x)+\y_{c_2}^{s_1}(x)\,\y_{c_1}^{s_2}(x))\;,
\eeq
which are a $\bo{\bar{3}}$ and $\bo{6}$ of color, respectively. Using
these operators, we form four types of diquark states: (i) color
$\bo{\bar{3}}$, spin-0, flavor $\bo{\bar{3}}$, (ii) color
$\bo{\bar{3}}$, spin-1, flavor $\bo{6}$, (iii) color $\bo{6}$, spin-0,
flavor $\bo{6}$, and (iv) color $\bo{6}$, spin-1, flavor $\bo{\bar{3}}$.

Diquark correlation functions for $am_1=0.03$ and $am_2=0.03$, the
lightest quark mass combination, are displayed in
\refigsdash{fig-3b_12+34}{fig-6_13+24}. Since $\g_4$ was diagonal in
the $\g$-matrix basis we used, upper components were combined with
time-reversed lower components to form positive-parity states, while
mixed upper and lower components were combined to form negative-parity states.
\refigure{fig-qp-all} shows a plot of the quark correlator for
input quark mass 0.03.

\bfig[tp]
\includegraphics*[width=8.3cm,clip]{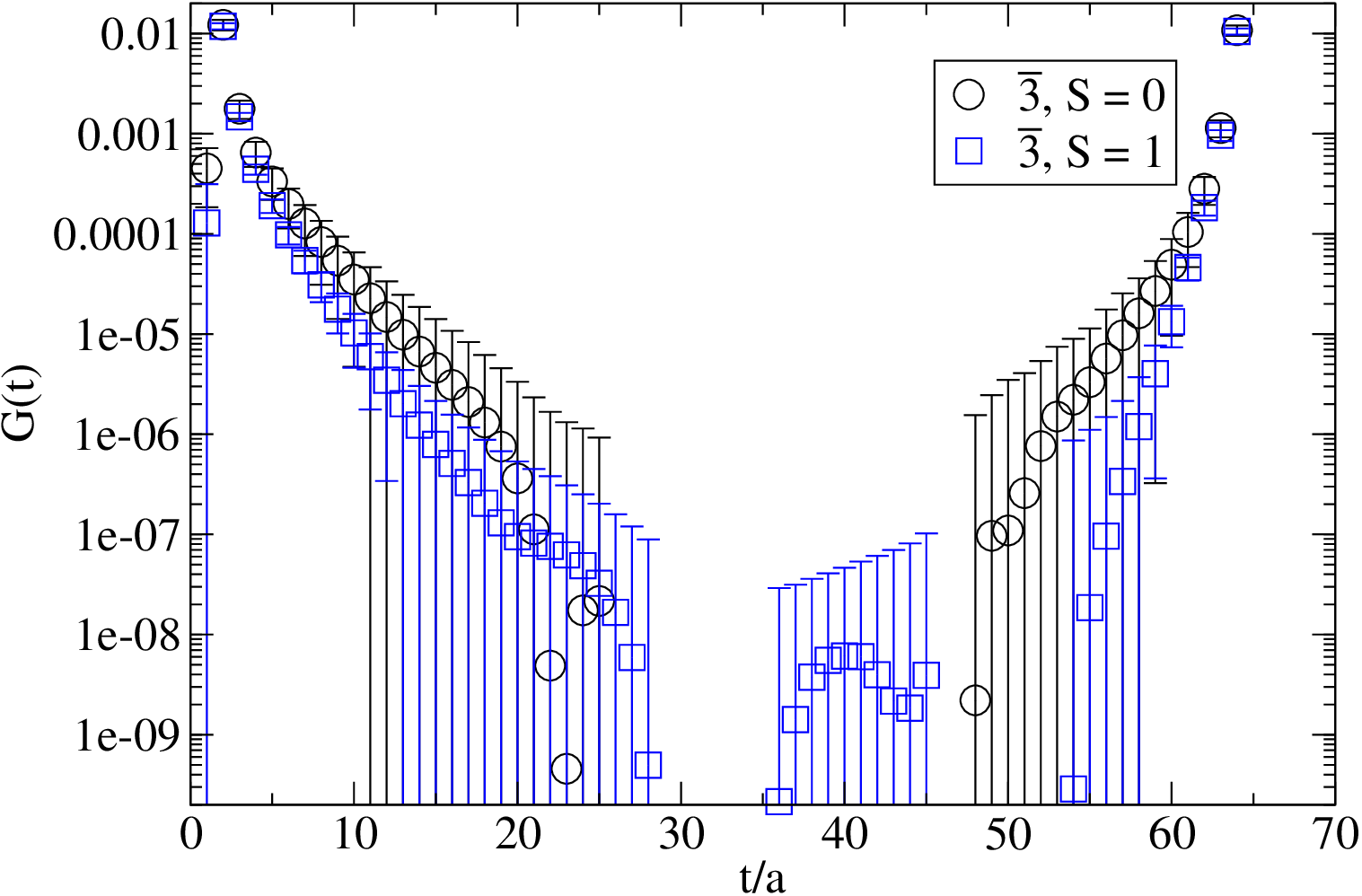}
\caption{$\bo{\bar{3}}$ positive-parity diquark correlation
function; $am_1=0.03$, $am_2=0.03$.}
\label{fig-3b_12+34}
\efig

\bfig[tp]
\includegraphics*[width=8.3cm,clip]{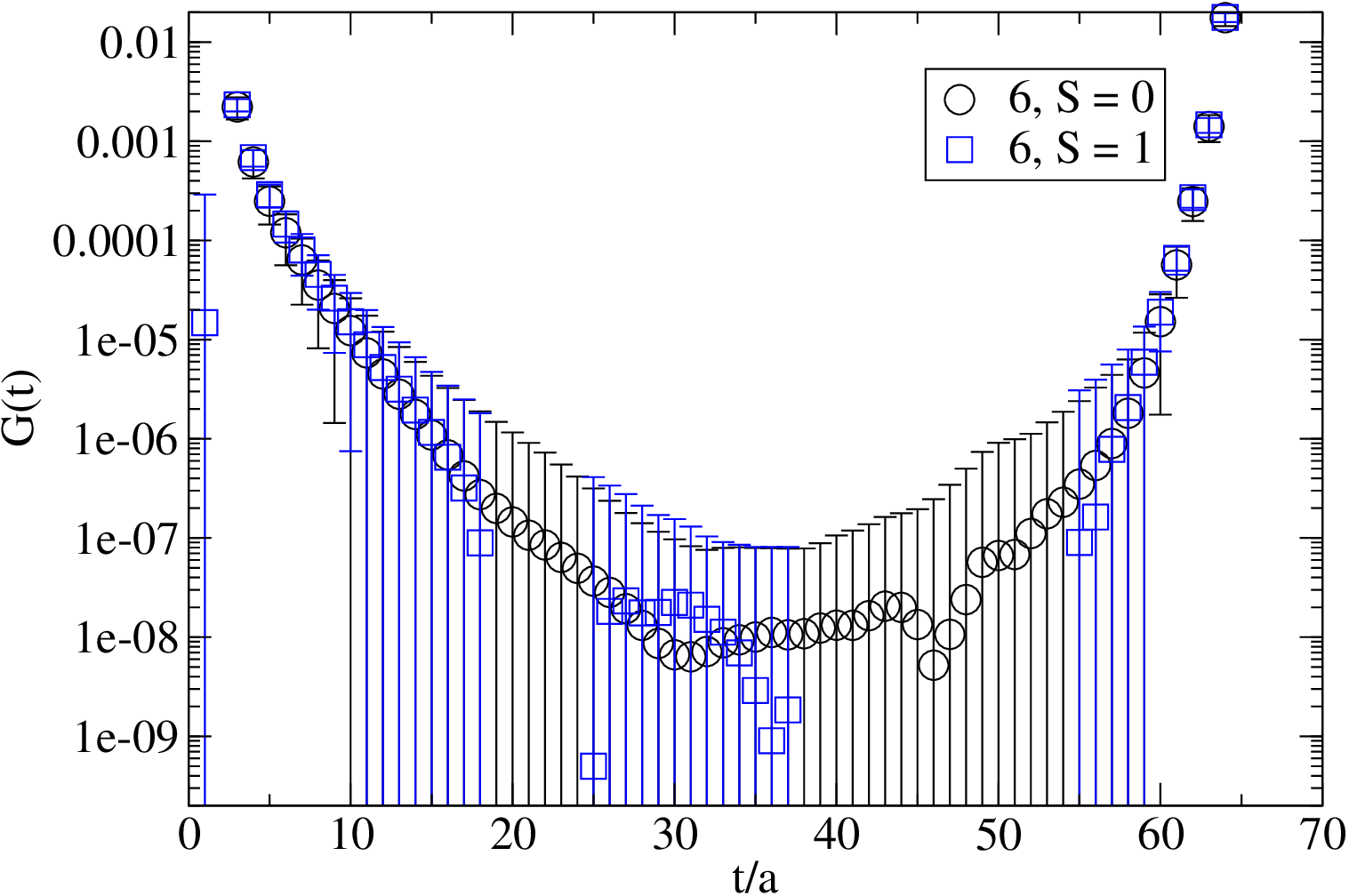}
\caption{$\bo{6}$ positive-parity diquark correlation
function; $am_1=0.03$, $am_2=0.03$.}
\label{fig-6_12+34}
\efig

\bfig[tp]
\includegraphics*[width=8.3cm,clip]{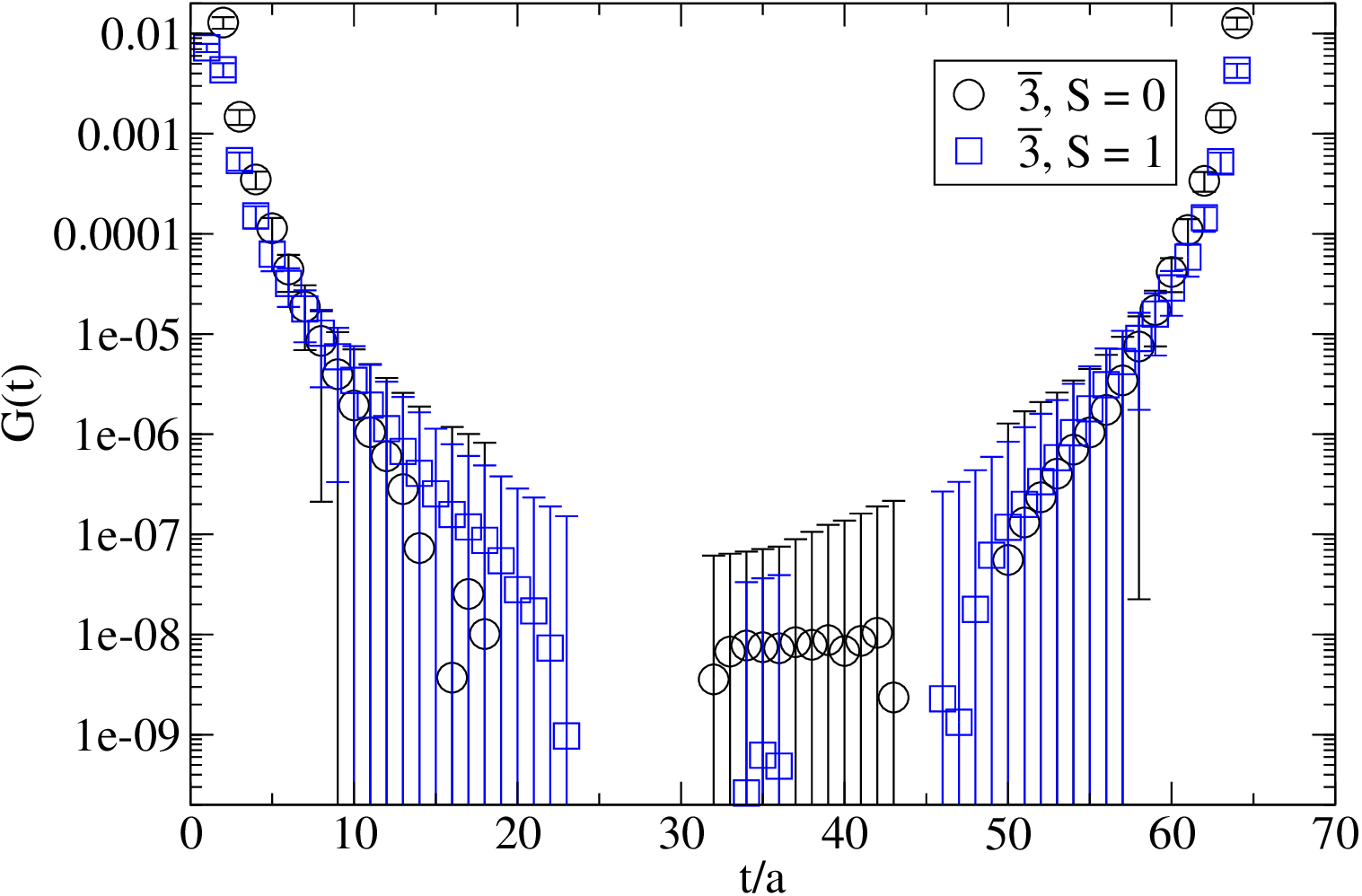}
\caption{$\bo{\bar{3}}$ negative-parity diquark correlation
function; $am_1=0.03$, $am_2=0.03$.}
\label{fig-3b_13+24}
\efig

\bfig[tp]
\includegraphics*[width=8.3cm,clip]{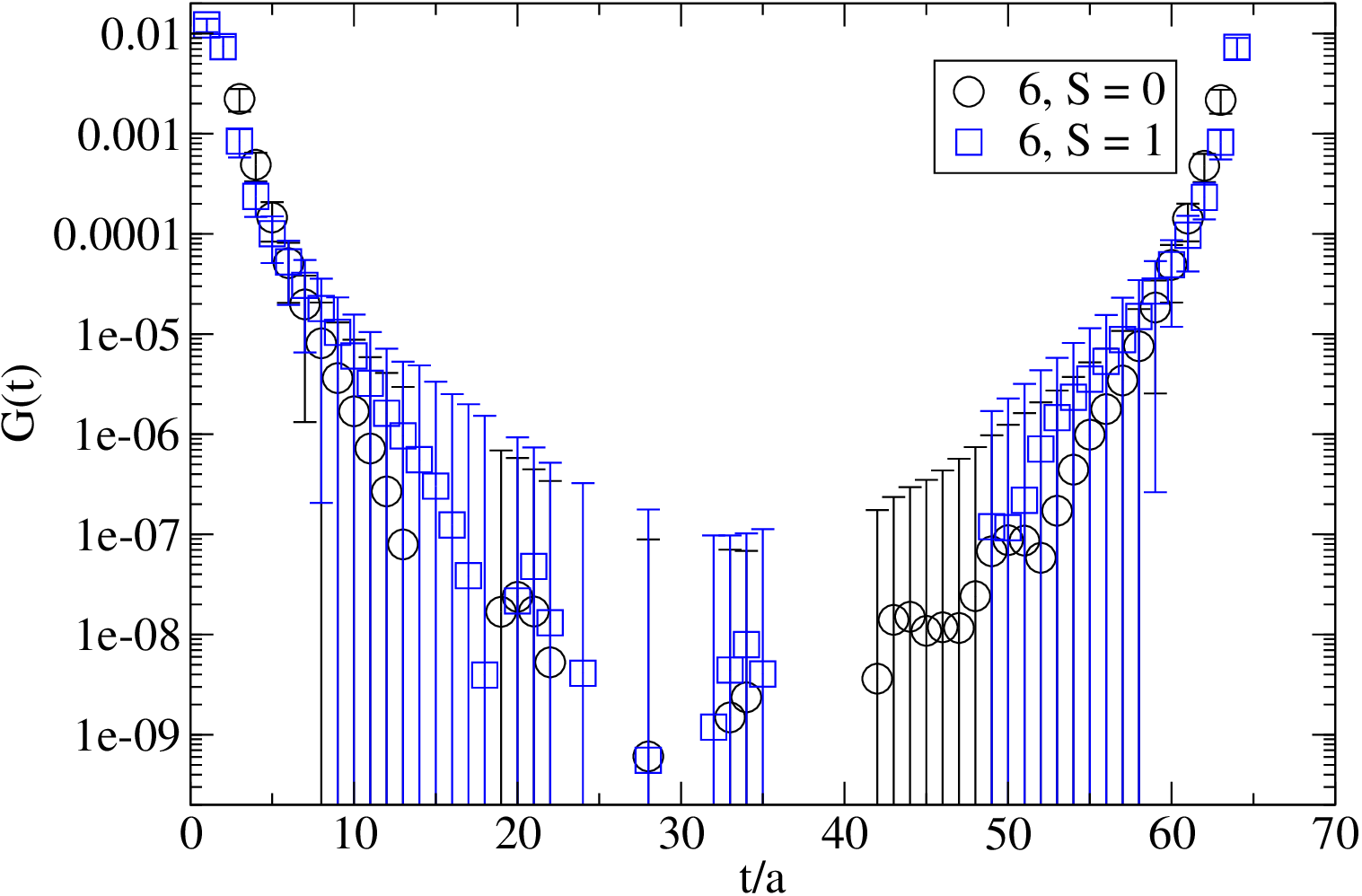}
\caption{$\bo{6}$ negative-parity diquark correlation
function; $am_1=0.03$, $am_2=0.03$.}
\label{fig-6_13+24}
\efig

\bfig[tp]
\includegraphics*[width=8.3cm,clip]{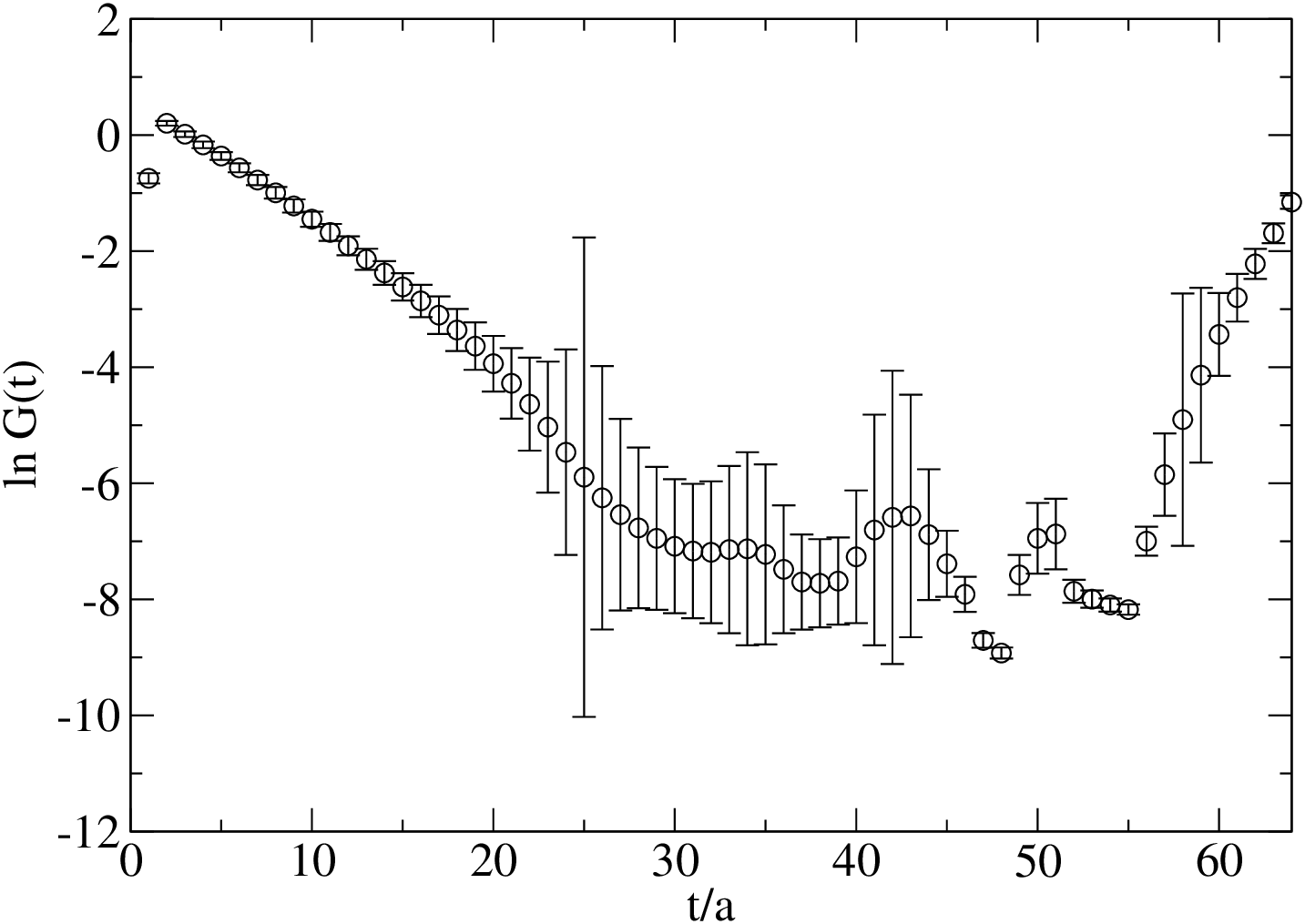}
\caption{Quark correlation function for input mass
  $am=0.03$.}
\label{fig-qp-all}
\efig

\subsection{Diquark spectra}

\refigure{fig-diquarks-12+34} shows the (positive-parity) diquark spectrum and the constituent
quark masses as a function of input quark mass. The fitting window used
was $5 \le t/a \le 15$. That figure also
includes a plot of twice the constituent quark mass and extrapolations
to zero quark mass for the ``twice quark mass'' and $\bo{\bar{3}}$ spin-0
results. \refigure{fig-diquarks-13+24} shows the same for the 
negative-parity states, with a fit for the lowest energy diquark state,
the $\bo{\bar{3}}$ spin-1.

It is interesting to observe that the $\bo{\bar{3}}$ diquark spin-0 extrapolation in
\refig{fig-diquarks-12+34} is below twice the quark mass extrapolation
and that the $\bo{\bar{3}}$ diquark spin-0 state
is significantly more strongly bound than the $\bo{\bar{3}}$
diquark spin-1 state. Such a result is consistent with the predictions
of diquark models \cite{Anselmino:1992vg,Jaffe:2003sg}. 
However, a much more detailed analysis must be done, particularly on
diquarks within baryon states, in order for rigorous conclusions to be reached.

The diquark data is shown in
\refts{tab-dqspectrum1}{tab-dqspectrum2} in Appendix~\ref{sec-Tables}.
Constituent quark masses, calculated from fits ($5 \le t/a \le 15$) 
to the quark correlators, are shown in
\reft{tab-quark} in Appendix~\ref{sec-Tables}.

\bfig[tp]
\includegraphics*[width=8.3cm,clip]{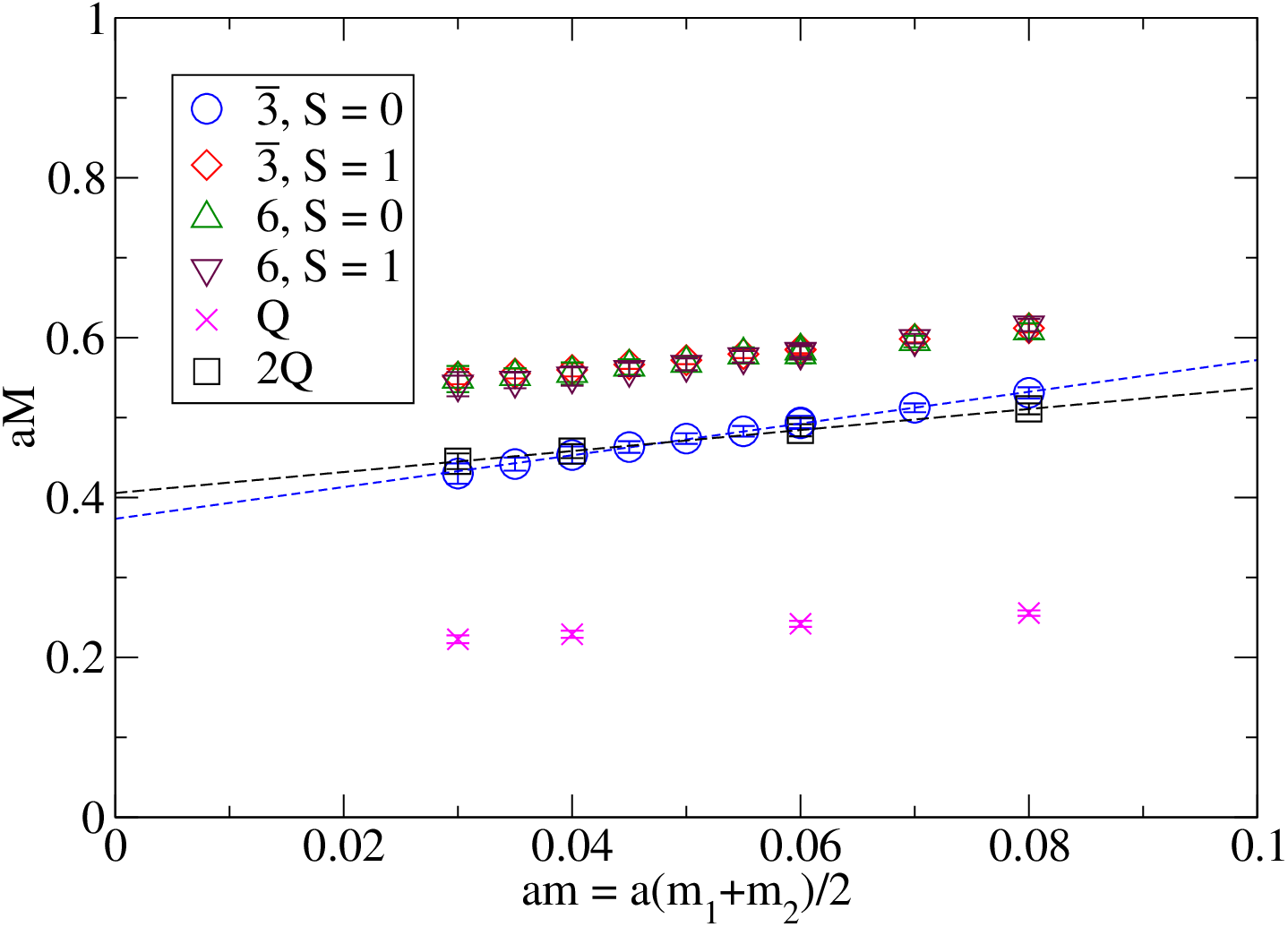}
\caption{Diquark spectrum vs. input quark mass, positive-parity.}
\label{fig-diquarks-12+34}
\efig

\bfig[tp]
\includegraphics*[width=8.3cm,clip]{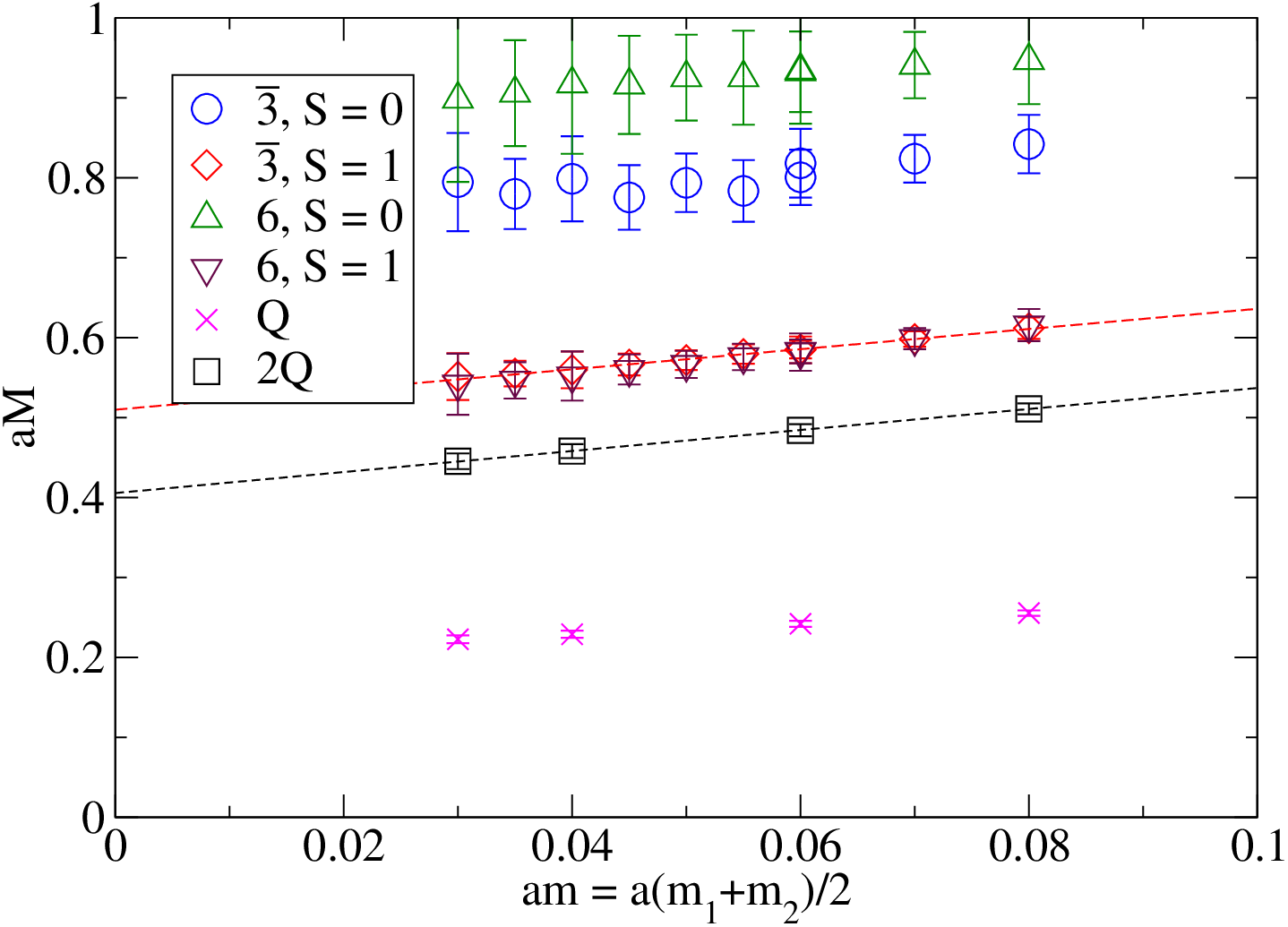}
\caption{Diquark spectrum vs. input quark mass, negative-parity.}
\label{fig-diquarks-13+24}
\efig

\section{Conclusions}
\label{sec-Conclusions}
In this article, we have presented results from quenched lattice QCD
simulations using the overlap Dirac operator on an $18^3 \x 64$ lattice
at $\b=6.0$ and a $14^3 \x 48$ lattice at $\b=5.85$.
We calculated quark propagators with a fixed source point and a variety of
quark mass values for 100 configurations on each lattice, and we have used
them to evaluate meson and baryon observables, quark masses, meson
final-state wave functions, and diquark correlations.

One important result of our work is that the calculation of quark
propagators with the overlap operator to a chosen numerical precision
has been shown to be feasible using available techniques, even when dealing with rough
background gauge configurations. Such a calculation requires projection
of low lying eigenvectors of $H^2=(\g_5D_W)^2$.  Improved
algorithms or smoothing techniques both may help to make the calculation
less computationally demanding~\cite{Durr:2005an}.

Beyond this, our investigation validates the good chiral properties
of the overlap operator and suggests good scaling properties
between $\b=5.85$ and $\b=6$, indicating that the $\b=6$ results
may already be close to the continuum limit.
So far as the actual values of the observables are concerned,
our results suffer from the shortcomings of the quenched approximation.
Nevertheless, from this investigation and others it is clear that it should
be possible to use the overlap operator in
dynamical fermion simulations, at the very least with a
mixed action formulation.  Work in that direction is beginning.
We are also working to extend our diquark results to diquark correlations
in baryons with one heavy and two light quarks, and
we expect to soon publish our results on non-perturbative
renormalization, selected weak matrix elements, and heavy
quark observables.

\begin{acknowledgments}
Work supported in part by US DOE
grants DE-FG02-91ER40676 and DE-AC02-98CH10866, EU HPP contract
HPRN-CT-2002-00311 (EURIDICE), and grant
HPMF-CT-2001-01468.  We thank Boston University and NCSA for use of
their supercomputer facilities.
\end{acknowledgments}

\clearpage
\appendix

\section{The Overlap Formalism and Simulation Details}
\label{sec-Overlap}
The forward and backward
lattice covariant derivatives are defined by
\beq
\del_\mu\y(x) = \fr{1}{a}[U_\mu(x)\y(x+a\muhat)-\y(x)]\;,
\eeq
and
\beq
\del_\mu^*\y(x) = \fr{1}{a}[\y(x)-U_\mu^\dag(x-a\muhat)\y(x-a\muhat)]\;,
\eeq
where $U_\mu(x)$ are the gauge link fields on the lattice.
Let $D_W$ denote the Wilson-Dirac operator
\beq
D_W = \fr{1}{2}\g_\mu(\del_\mu+\del_\mu^*)-\fr{r}{2}a\,\del_\mu^*\del_\mu\;,
\eeq
with $0 < r \le 1$ and $0 < \r < 2r$ (we used $r=1$ in our
calculations) and where $a$ is the lattice spacing.

Neuberger's overlap Dirac operator is then defined as~\cite{Narayanan:1995gw,Neuberger:1998fp}
\beq
D = \fr{\r}{a}(1+V) = \fr{\r}{a}\parens{1+\g_5H\fr{1}{\sr{H^\dag H}}}\;,
\eeq
where
\beq
H = \g_5(D_W - \fr{1}{a}\r)
\eeq
and $\r$ is a parameter that affects the radius of rescaling of eigenvalues of
the overlap operator in the complex plane as compared to the
Wilson-Dirac operator. 

The combination of the Wilson gauge action and the (massive) overlap fermionic
action is
\beqa
S &=& \fr{6}{g_0^2}\sum_P\bracks{1-\fr{1}{6}\Tr(U_P+U_P^\dag)} \nn\\
& & \qquad +\ybar\bracks{\parens{1-\fr{a}{2\r}\,m}D+m}\y\;,
\label{eq-OverlapAction}
\eeqa
where $U_P$ is the Wilson plaquette, $g_0 = \sr{6/\b}$ is the bare
coupling constant, the fermion fields $\y$ and $\ybar$ carry implicit
color, spin, and flavor indices, and $m$ is a diagonal matrix of bare
masses $(m_1,m_2,\dots)$ in flavor space.

The Ginsparg-Wilson relation~\cite{Ginsparg:1982bj}
\beq
\g_5D+D\g_5 = \fr{a}{\r}D\g_5D
\eeq
is satisfied by the fermionic operator of the overlap action, implying
an exact continuous symmetry of the action in the massless
limit~\cite{Luscher:1998pq}. The symmetry can be interpreted as a lattice form of
chiral invariance at finite cutoff,
\beq
\d\y = \hat{\g}_5\y,\ \ \ \d\ybar = \ybar\g_5\;,
\label{eq-ChiralRots}
\eeq
where
\beq
\hat{\g}_5 \defined \g_5\parens{1-\fr{a}{\r}D}\;,
\eeq
which satisfies $\hat{\g}_5^\dag=\hat{\g}_5$ and
$\hat{\g}_5^2=1$. Invariance of the action under non-singlet chiral
transformations, defined including a flavor group generator in
\req{eq-ChiralRots}, forbids mixing among operators of different
chirality~\cite{Hasenfratz:1998jp}. Therefore,
\begin{itemize}
\item no additive quark mass renormalization is required, and the quark
  mass which enters the vector and axial Ward identities is the bare
  parameter $m(a)$;
\item masses and matrix elements are affected only by $\cO(a^2)$ errors
  and no fine-tuned parameters are required to remove $\cO(a)$
  effects;
\item and the chiral condensate (see \rsec{subsec-ChiralCond}) does not require subtractions of
  power divergent terms (in the chiral limit).
\end{itemize}

The non-singlet ``local'' (source and sink at the same point $x$) bilinear
operators we use are defined by
\beq
O_\G(x) = \ybar_1(x)\G\bracks{(1-\fr{a}{2\r}D)\y_2}(x)\;,
\eeq
where $O_\G \in \set{S,P,V_\mu,A_\mu}$ correspond to $\G \in
\set{\id,\g_5,\g_\mu,\g_\mu\g_5}$. (We also use non-local ``extended''
operators; see \rsec{subsec-ExtSinkOps}.) The bilinear operators are subject to
multiplicative renormalization only, i.e.~the corresponding
renormalized operators are
\beq
\hat{O}_\G(x,\mu) = \lim_{a \to 0}Z_\G(a\mu)\,O_\G(x,a)\;,
\eeq
where $Z_\G(a\mu)$ are the appropriate renormalization
constants. Since $S,P$ and $V_\mu,A_\mu$ belong to the same chiral
multiplets, $Z_S=Z_P$ and $Z_V=Z_A$. Also, flavor symmetry imposes
$Z_S=1/Z_m$, with $Z_m$ defined in~\req{eq-massrenorm}.

Using such operators, the following two-point correlation functions
(or ``correlators'') can be formed:
\beq
G_{SS}(t) = \sum_{\v{x}}\<S(\v{x},t)S^c(\v{0},0)\>\;,
\eeq
\beq
G_{PP}(t) = \sum_{\v{x}}\<P(\v{x},t)P^c(\v{0},0)\>\;,
\eeq
\beq
G_{\del_0A_0P}(t) = \sum_{\v{x}}\<\bar{\del}_0A_0(\v{x},t)P^c(\v{0},0)\>\;,
\eeq
\beq
G_{VV}(t) = \sum_{\v{x},i}\<V_i(\v{x},t)V_i^c(\v{0},0)\>\;,
\eeq
where we assume the quarks to be of different flavor, $c$ denotes 
flavor conjugation, i.e.~interchange of the two flavors, and 
$\bar{\del}_0$ is the symmetric lattice derivative in the time
direction. When appropriate, such correlators are symmetrized around
$t=T/2$, where $T$ is the extent of the lattice in the time direction.

Our simulations studied quenched QCD with the Wilson gauge action 
and with Neuberger's overlap Dirac operator for lattice fermions
on two different lattices.  For both simulations, we used
samples of 100 gauge configurations generated by a 6-hit Metropolis 
algorithm with acceptance $\approx 0.5$.  The configurations 
were separated by 10,000 upgrades, after an initial set of 11,000
upgrades for equilibration.

The first simulation was done on an $18^3 \times 64$ lattice, with
$\beta=6$, $\r = 1.4$, and lattice spacing $a^{-1} \approx 2.0\,\GeV$.
We calculated overlap quark propagators 
for a single point source, for all 12 color-spin combinations and 
quark masses $am=0.03, 0.04, 0.06, 0.08, 0.1, 0.25, 0.5$ and $0.75$, 
using a multi-mass solver. 

The second simulation was done on a $14^3 \times 48$ lattice, with  
$\beta=5.85$ and $\r = 1.6$. The second, coarser lattice was chosen to
have roughly the same volume as the finer $18^3 \x 64$ lattice, with lattice 
spacing  $a^{-1} \approx 1.5\,\GeV$, allowing for an investigation of 
scaling effects. For the $14^3 \x 48$ lattice, we calculated 
overlap quark propagators again for a single point
source and all 12 color-spin combinations, and with bare quark 
masses $am=0.03, 0.04, 0.053, 0.08, 0.106, 0.132, 0.33, 0.66$, 
and $0.99$. The largest eight of the nine quark 
masses correspond approximately to the eight quark masses on
the finer lattice.

One question that needs to be addressed when going to a different
coupling is how to set the negative mass parameter $\rho$ of the
overlap operator. A possible and valid
strategy is, of course, to keep the value of $\rho$ constant. One has to
keep in mind, however, that $\rho=1.4$ was chosen to maximize the
locality of the overlap operator at one specific coupling
($\beta=6.0$)~\cite{Hernandez:1998et} and, from a technical
point of view, is not optimal when going to lower values of $\beta$.

We are not restricted, however, to keep $\rho$ fixed. In fact, since a
change in $\rho$ corresponds to an $\cO(a^2)$ redefinition of the overlap
operator (as long as we stay within the 1-fermion sector of the theory),
varying $\rho(a)=\rho_0+f(a)$, where $f(a)$ is a smooth monotonic
function in $a$ with $\lim_{a\rightarrow 0}f(a)=0$, does not change the
continuum limit or $\cO(a^2)$ scaling violations of the theory. It is
very reasonable to assume that choosing $\rho(a)$ by demanding optimum
locality of the resulting overlap operator falls into this class of
variations, and we therefore followed this strategy. Apart from
better locality, that choice has an additional benefit: the
resulting Hermitian Wilson operator is better conditioned and therefore
the number of eigenmodes that need to be treated exactly is smaller,
which is particularly important at the relatively large physical volume
we are working at.

We used two algorithms for numerical implementation of the overlap
operator. The first was a Zolotarev optimal rational 
function~\cite{Neuberger:1998my,Edwards:1998yw,vandenEshof:2002ms}
approximation with 12 poles, as detailed in~\cite{Giusti:2001yw}.
The second algorithm was a Chebyshev polynomial 
approximation~\cite{Neuberger:1998my,Edwards:1998yw,Bunk:1997wj,Hernandez:2000sb}.
In both cases, we performed a Ritz projection~\cite{Kalkreuter:1995mm}
of a certain number $n_l$ of the lowest eigenvectors of
$H^2=(\g_5D_W)^2$, whose contribution
to the propagators was calculated directly.  We used the 
Zolotarev approach for calculating 55 of the quark propagators 
for the $18^3 \x 64$ lattice, but numerical experimentation showed 
the Chebyshev approach to be about 20\% faster than the rational function 
approach. We used the Chebyshev approach for the balance of the propagators on the
$18^3 \x 64$ lattice and for all 100 quark propagators on the $14^3 \x 48$ 
lattice.  

The maximum degree of the Chebyshev polynomials was chosen
so as to achieve a required numerical precision in the range of
eigenvalues left over after the projection.  We emphasize that our method
of implementing the overlap is exact, up to the chosen numerical precision
$\epsilon_1$ in the calculation of $1/\sqrt(H^2)$ and the maximum
residue $\epsilon_2$ in the calculation of the propagators.
No other approximations are involved in the calculation.
We used as convergence criteria $\epsilon_1=1.0 \x 10^{-8}$ and
$\epsilon_2=1.0 \x 10^{-7}$.
For the $18^3 \x 64$ lattice, we found a projection of
$n_l=12$ low eigenvectors to be adequate, leading to a maximum degree
$100\sim 500$ in the expansion of the inverse square root
into Chebyshev polynomials.  However, for the $14^3 \times 48$ lattice
we found that, most likely on account of the increased disorder
of the gauge background, the spectrum of $H^2$ generally contained many more
low lying eigenvalues.  For some configurations, using $n_l=12$, as for the larger lattice,
led to a maximum degree of the Chebyshev polynomials
as high as $\sim 4000$, and in some cases led to loss of convergence.
Therefore, for the $14^3 \times 48$ lattice we increased the number
of projected low eigenvectors to $n_l=40$.
This brought the maximum degree of the Chebyshev polynomials back
into the range $100\sim 500$.

The computations were performed with shared memory Fortran 90 code, optimized
and run on 8, 16, and 32 processor IBM-p690 nodes at BU and NCSA.

\section{Tables}
\label{sec-Tables}
Here we present a comparison between data for $\b=6$ and
$\b=5.85$, as well as results for meson, baryon, and diquark spectra.

\btabstar[htp]
\centering
\caption{Comparison of data for the two lattices.}
\label{tab-scaling}
\brtabu
\btabu{lll}
Quantity & $18^3 \x 64$ lattice & $14^3 \x 48$ lattice \\
\hline
$Z_A$ & 1.5555(47) & 1.4434(18) \\
$\d$ (degenerate) & 0.29(5)  & 0.17(4)  \\
$\d$ (non-degenerate) & 0.18(8)  & 0.22(4)  \\
$a^{-1}$ (Sommer)~\cite{Guagnelli:1998ud,Necco:2001xg} & $2.12\,\GeV$  & $1.61\,\GeV$  \\
$a^{-1}$ (physical planes) & $2.19(6)\,\GeV$  & $1.44(4)\,\GeV$  \\
$a^{-1}$ ($M_\r$) & $1.90(4)\,\GeV$  & $1.28(6)\,\GeV$  \\
$f_K/f_\pi$ & 1.13(4)  & 1.09(4)  \\
$f_{K^*}/f_\r$ & 1.03(6)  & 1.02(10)  \\
$M_{K^*}/M_\r$ & 1.09(5) & 1.07(6)  \\
$Z_S^{\overline{\text{MS}}}(2\,\GeV)$ & 1.44(2)(3)  & 1.48(3)(16) \\
$(m_s+\hat{m})^{\overline{\text{MS}}}(2\,\text{GeV})$ & $105(3)(4)\,\MeV$  &
$102(3)(16)\,\MeV$ \\
$m_s^{\overline{\text{MS}}}(2\,\text{GeV})$ & $101(3)(4)\,\MeV$  &
$98(3)(15)\,\MeV$ \\
$\langle \ybar \y \rangle^{\overline{\text{MS}}}(2\,\GeV)$ & $- [262(5)(9) \,\MeV]^3$  & $- [302(5)(34) \,\MeV]^3$ \\
$r_0(m_s+\hat{m})^{\overline{\text{MS}}}(2\,\text{GeV})$ & $0.265(7)(5)$  & $0.258(7)(3)$ \\
$r_0^3\langle \ybar \y \rangle^{\overline{\text{MS}}}(2\,\GeV)$ & $- [0.684(16)(4)]^3$  & $- [0.78(2)(3)]^3$ \\
\etabu
\ertabu
\etabstar

\btabstar[htp]
\centering
\caption{$18^3 \x 64$ point sink results for $PP$ correlator}
\label{tab-PPpoint}
\brtabu
\btabu{lllll}
$t_{min}$ & $t_{max}$ & $am$ & $aM$ & $a^3Z/(2M)$ \\
\hline       
12 & 32 & 0.030 & 0.2186(26) & 0.00766(64) \\
12 & 32 & 0.040 & 0.2468(22) & 0.00654(48) \\
12 & 32 & 0.060 & 0.2960(17) & 0.00560(32) \\
12 & 32 & 0.080 & 0.3396(14) & 0.00531(25) \\
12 & 32 & 0.100 & 0.3795(12) & 0.00526(21) \\
12 & 32 & 0.250 & 0.6281(9) & 0.00664(16) \\
12 & 32 & 0.500 & 0.9805(7) & 0.01094(21) \\
12 & 32 & 0.750 & 1.3046(9) & 0.01755(34) \\
\etabu
\ertabu
\etabstar

\btabstar[htp]
\centering
\caption{$18^3 \x 64$ point sink results for $PP-SS$ correlator}
\label{tab-PP-SSpoint}
\brtabu
\btabu{lllll}
$t_{min}$ & $t_{max}$ & $am$ & $aM$ & $a^3Z/(2M)$ \\
\hline
12 & 32 & 0.030 & 0.2192(30) & 0.00770(70) \\
12 & 32 & 0.040 & 0.2474(24) & 0.00660(48) \\
12 & 32 & 0.060 & 0.2967(19) & 0.00567(32) \\
12 & 32 & 0.080 & 0.3403(16) & 0.00539(27) \\
12 & 32 & 0.100 & 0.3803(14) & 0.00536(24) \\
12 & 32 & 0.250 & 0.6304(10) & 0.00710(18) \\
12 & 32 & 0.500 & 0.9838(8) & 0.01206(22) \\
12 & 32 & 0.750 & 1.3084(10) & 0.01971(37) \\
\etabu
\ertabu
\etabstar

\btabstar[htp]
\centering
\caption{$18^3 \x 64$ point sink results for $VV$ correlator}
\label{tab-VVpoint}
\brtabu
\btabu{lllll}
$t_{min}$ & $t_{max}$ & $am$ & $aM$ & $a^3Z/(2M)$ \\
\hline
8 & 32 & 0.030 & 0.511(21) & 0.00243(43) \\
8 & 32 & 0.040 & 0.500(16) & 0.00207(30) \\
8 & 32 & 0.060 & 0.500(10) & 0.00183(18) \\
8 & 32 & 0.080 & 0.515(7) & 0.00184(13) \\
8 & 32 & 0.100 & 0.536(6) & 0.00195(11) \\
8 & 32 & 0.250 & 0.728(2) & 0.00340(10) \\
8 & 32 & 0.500 & 1.055(1) & 0.00674(14) \\
8 & 32 & 0.750 & 1.381(1) & 0.01204(24) \\
\etabu
\ertabu
\etabstar

\btabstar[htp]
\centering
\caption{$18^3 \x 64$ extended sink results for $VV$ correlator}
\label{tab-VVextended}
\brtabu
\btabu{llll}
$t_{min}$ & $t_{max}$ & $am$ & $aM$ \\
\hline
4 & 32 & 0.030 & 0.446(13) \\
4 & 32 & 0.040 & 0.453(10) \\
4 & 32 & 0.060 & 0.472(8) \\
4 & 32 & 0.080 & 0.495(6) \\
4 & 32 & 0.100 & 0.519(5) \\
4 & 32 & 0.250 & 0.721(2) \\
4 & 32 & 0.500 & 1.053(1) \\
4 & 32 & 0.750 & 1.380(1) \\
\etabu
\ertabu
\etabstar

\btabstar[htp]
\centering
\caption{$14^3 \x 48$ point sink results for $PP$ correlator}
\label{tab-PPpoint-1448}
\brtabu
\btabu{lllll}
$t_{min}$ & $t_{max}$ & $am$ & $aM$ & $a^3Z/(2M)$ \\
\hline
10 & 24 & 0.030 & 0.2513(29) & 0.0251(13) \\
10 & 24 & 0.040 & 0.2858(25) & 0.0211(10) \\
10 & 24 & 0.053 & 0.3252(22) & 0.0182(8) \\
10 & 24 & 0.080 & 0.3959(21) & 0.0157(6) \\
10 & 24 & 0.106 & 0.4557(20) & 0.0150(6) \\
10 & 24 & 0.132 & 0.5101(19) & 0.0149(5) \\
10 & 24 & 0.330 & 0.8502(13) & 0.0190(5) \\
10 & 24 & 0.660 & 1.3268(11) & 0.0357(9) \\
10 & 24 & 0.990 & 1.6488(43) & 0.0333(17) \\
\etabu
\ertabu
\etabstar

\btabstar[htp]
\centering
\caption{$14^3 \x 48$ point sink results for $PP-SS$ correlator}
\label{tab-PP-SSpoint-1448}
\brtabu
\btabu{lllll}
$t_{min}$ & $t_{max}$ & $am$ & $aM$ & $a^3Z/(2M)$ \\
\hline
10 & 24 & 0.030 & 0.2458(51) & 0.0232(20) \\
10 & 24 & 0.040 & 0.2821(43) & 0.0202(15) \\
10 & 24 & 0.053 & 0.3228(36) & 0.0177(12) \\
10 & 24 & 0.080 & 0.3950(29) & 0.0155(8) \\
10 & 24 & 0.106 & 0.4555(25) & 0.0149(7) \\
10 & 24 & 0.132 & 0.5106(23) & 0.0149(6) \\
10 & 24 & 0.330 & 0.8532(13) & 0.0200(6) \\
10 & 24 & 0.660 & 1.3317(11) & 0.0395(11) \\
10 & 24 & 0.990 & 1.6595(41) & 0.0420(21) \\
\etabu
\ertabu
\etabstar

\btabstar[htp]
\centering
\caption{$14^3 \x 48$ point sink results for $VV$ correlator}
\label{tab-VVpoint-1448}
\brtabu
\btabu{lllll}
$t_{min}$ & $t_{max}$ & $am$ & $aM$ & $a^3Z/(2M)$ \\
\hline
8 & 24 & 0.030 & 0.775(81) & 0.0151(40) \\
8 & 24 & 0.040 & 0.732(54) & 0.0104(22) \\
8 & 24 & 0.053 & 0.701(36) & 0.0076(14) \\
8 & 24 & 0.080 & 0.692(21) & 0.0060(8) \\
8 & 24 & 0.106 & 0.708(15) & 0.0058(6) \\
8 & 24 & 0.132 & 0.732(11) & 0.0058(5) \\
8 & 24 & 0.330 & 0.984(3) & 0.0096(3) \\
8 & 24 & 0.660 & 1.433(2) & 0.0212(5) \\
8 & 24 & 0.990 & 1.857(4) & 0.0463(20) \\
\etabu
\ertabu
\etabstar

\btabstar[htp]
\centering
\caption{$14^3 \x 48$ extended sink results for $VV$ correlator}
\label{tab-VVextended-1448}
\brtabu
\btabu{llll}
$t_{min}$ & $t_{max}$ & $am$ & $aM$ \\
\hline
4 & 24 & 0.030 & 0.650(25) \\
4 & 24 & 0.040 & 0.645(21) \\
4 & 24 & 0.053 & 0.646(17) \\
4 & 24 & 0.080 & 0.664(12) \\
4 & 24 & 0.106 & 0.689(9) \\
4 & 24 & 0.132 & 0.719(8) \\
4 & 24 & 0.330 & 0.979(3) \\
4 & 24 & 0.660 & 1.433(2) \\
4 & 24 & 0.990 & 1.877(4) \\
\etabu
\ertabu
\etabstar

\btabstar[htp]
\centering
\caption{Baryon octet spectrum at $\beta=6.0$ for multiple fitting windows}
\label{tab-octet}
\brtabu
\btabu{llllll}
$am$ & & & $aM$ & & \\
\cline{2-6}
& $6-16$ & $8-14$ & $8-16$ & $8-18$ & $10-16$  \\
\hline
0.030 & 0.650(19) & 0.625(22) & 0.627(23) & 0.632(24) & 0.626(32)  \\
0.040 & 0.680(14) & 0.658(16) & 0.658(17) & 0.662(17) & 0.653(22)  \\
0.060 & 0.7353(97) & 0.717(11) & 0.714(11) & 0.714(11) & 0.705(14)  \\
0.080 & 0.7854(77) & 0.7691(87) & 0.7634(88) & 0.7606(88) & 0.753(11)  \\
0.100 & 0.8327(65) & 0.8172(74) & 0.8103(74) & 0.8052(74) & 0.7991(89)  \\
\etabu
\ertabu
\etabstar

\btabstar[htp]
\centering
\caption{Baryon decuplet spectrum at $\beta=6.0$ for multiple fitting windows}
\label{tab-decuplet}
\brtabu
\btabu{llllll}
$am$ & & & $aM$ & & \\
\cline{2-6}
& $6-16$ & $8-14$ & $8-16$ & $8-18$ & $10-16$  \\
\hline
0.030 & 0.857(26) & 0.785(31) & 0.751(32) & 0.728(34) & 0.643(36)  \\
0.040 & 0.863(21) & 0.805(23) & 0.775(23) & 0.757(24) & 0.694(25)  \\
0.060 & 0.889(16) & 0.846(15) & 0.824(15) & 0.812(16) & 0.774(17)  \\
0.080 & 0.920(12) & 0.885(12) & 0.868(12) & 0.859(12) & 0.832(13)  \\
0.100 & 0.953(10) & 0.9222(94) & 0.9085(96) & 0.901(10) & 0.881(11)  \\
\etabu
\ertabu
\etabstar

\btabstar[htp]
\centering
\caption{Baryon octet spectrum at $\beta=5.85$ for multiple fitting windows}
\label{tab-octet14}
\brtabu
\btabu{llllll}
$am$ & & & $aM$ & & \\
\cline{2-6}
& $6-11$ & $7-10$ & $7-11$ & $7-12$ & $8-11$  \\
\hline
0.030 & 0.805(35) & 0.796(43) & 0.813(48) & 0.817(52) & 0.839(70)  \\
0.040 & 0.840(23) & 0.833(27) & 0.840(30) & 0.838(33) & 0.849(41)  \\
0.053 & 0.879(15) & 0.872(17) & 0.873(19) & 0.868(21) & 0.873(25)  \\
0.080 & 0.9526(94) & 0.946(11) & 0.941(11) & 0.935(12) & 0.935(13)  \\
0.106 & 1.0196(78) & 1.0130(93) & 1.0065(88) & 0.9996(88) & 0.998(10)  \\
0.132 & 1.0842(69) & 1.0772(82) & 1.0700(76) & 1.0630(74) & 1.0602(84)  \\
\etabu
\ertabu
\etabstar

\btabstar[htp]
\centering
\caption{Baryon decuplet spectrum at $\beta=5.85$ for multiple fitting windows}
\label{tab-deculpet14}
\brtabu
\btabu{llllll}
$am$ & & & $aM$ & & \\
\cline{2-6}
& $6-9$ & $7-9$ & $7-10$ & $7-11$ & $8-10$  \\
\hline
0.030 & 1.131(52) & 1.106(77) & 1.123(90) & 1.131(98) & 1.17(17)  \\
0.040 & 1.140(38) & 1.110(53) & 1.118(60) & 1.123(65) & 1.135(94)  \\
0.053 & 1.153(28) & 1.121(37) & 1.121(40) & 1.121(41) & 1.118(54)  \\
0.080 & 1.190(18) & 1.157(22) & 1.148(22) & 1.142(21) & 1.133(26)  \\
0.106 & 1.233(13) & 1.201(15) & 1.189(15) & 1.180(14) & 1.171(16)  \\
0.132 & 1.279(11) & 1.248(12) & 1.235(11) & 1.224(10) & 1.217(12)  \\
\etabu
\ertabu
\etabstar

\btabstar[htp]
\centering
\caption{Degenerate ($m_1=m_2$) diquark spectrum, positive-parity.}
\label{tab-dqspectrum1}
\brtabu
\btabu{llll}
color & spin & $am$ & $aM$ \\
\hline
$\bo{\bar{3}}$ & 0 & 0.030 & 0.430(13) \\
$\bo{\bar{3}}$ & 0 & 0.040 & 0.453(11) \\
$\bo{\bar{3}}$ & 0 & 0.060 & 0.494(8) \\
$\bo{\bar{3}}$ & 0 & 0.080 & 0.531(7) \\
\hline
$\bo{\bar{3}}$ & 1 & 0.030 & 0.551(10) \\
$\bo{\bar{3}}$ & 1 & 0.040 & 0.560(8) \\
$\bo{\bar{3}}$ & 1 & 0.060 & 0.585(6) \\
$\bo{\bar{3}}$ & 1 & 0.080 & 0.612(5) \\
\hline
$\bo{6}$ & 0 & 0.030 & 0.548(16) \\
$\bo{6}$ & 0 & 0.040 & 0.555(14) \\
$\bo{6}$ & 0 & 0.060 & 0.579(11) \\
$\bo{6}$ & 0 & 0.080 & 0.609(10) \\
\hline
$\bo{6}$ & 1 & 0.030 & 0.542(15) \\
$\bo{6}$ & 1 & 0.040 & 0.552(12) \\
$\bo{6}$ & 1 & 0.060 & 0.582(9) \\
$\bo{6}$ & 1 & 0.080 & 0.616(8) \\
\etabu
\ertabu
\etabstar

\btabstar[htp]
\centering
\caption{Degenerate ($m_1 = m_2$) diquark spectrum, negative-parity.}
\label{tab-dqspectrum2}
\brtabu
\btabu{llll}
color & spin & $am$ & $aM$ \\
\hline
$\bo{\bar{3}}$ & 0 & 0.030 & 0.795(61) \\
$\bo{\bar{3}}$ & 0 & 0.040 & 0.799(53) \\
$\bo{\bar{3}}$ & 0 & 0.060 & 0.818(43) \\
$\bo{\bar{3}}$ & 0 & 0.080 & 0.842(37) \\
\hline
$\bo{\bar{3}}$ & 1 & 0.030 & 0.551(29) \\
$\bo{\bar{3}}$ & 1 & 0.040 & 0.560(23) \\
$\bo{\bar{3}}$ & 1 & 0.060 & 0.585(17) \\
$\bo{\bar{3}}$ & 1 & 0.080 & 0.612(14) \\
\hline
$\bo{6}$ & 0 & 0.030 & 0.898(104) \\
$\bo{6}$ & 0 & 0.040 & 0.918(88) \\
$\bo{6}$ & 0 & 0.060 & 0.935(67) \\
$\bo{6}$ & 0 & 0.080 & 0.947(55) \\
\hline
$\bo{6}$ & 1 & 0.030 & 0.542(38) \\
$\bo{6}$ & 1 & 0.040 & 0.552(31) \\
$\bo{6}$ & 1 & 0.060 & 0.582(23) \\
$\bo{6}$ & 1 & 0.080 & 0.616(20) \\
\etabu
\ertabu
\etabstar

\btabstar[htp]
\centering
\caption{Constituent quark mass}
\label{tab-quark}
\brtabu
\btabu{llll}
& $am$ & $aM$ & \\
\hline
& 0.030 & 0.229(5) & \\
& 0.040 & 0.235(5) & \\
& 0.060 & 0.248(4) & \\
& 0.080 & 0.261(4) & \\
\etabu
\ertabu
\etabstar

\clearpage
\bibliography{lighthadron}

\end{document}